\newif \iffull \fulltrue
\newif \ifdraft \draftfalse
\newif \ifanonymous \anonymousfalse

\iffull
\documentclass[acmsmall,screen,authorversion]{acmart}
\else
\documentclass[sigplan,screen]{acmart}
\fi

\setcopyright{rightsretained}
\acmPrice{15.00}
\acmDOI{10.1145/3519939.3523722}
\acmYear{2022}
\copyrightyear{2022}
\acmSubmissionID{pldi22main-p517-p}
\acmISBN{978-1-4503-9265-5/22/06}
\acmConference[PLDI '22]{Proceedings of the 43rd ACM SIGPLAN International Conference on Programming Language Design and Implementation}{June 13--17, 2022}{San Diego, CA, USA}
\acmBooktitle{Proceedings of the 43rd ACM SIGPLAN International Conference on Programming Language Design and Implementation (PLDI '22), June 13--17, 2022, San Diego, CA, USA}

\usepackage{microtype}
\usepackage{amsmath,amsthm}
\usepackage{mathtools}
\usepackage{xcolor}
\usepackage{wasysym}
\usepackage{xspace}
\usepackage{bcprules}
\usepackage{url}
\usepackage{enumitem}
\usepackage{xfrac}

\usepackage{listings}
\usepackage{lstocaml}
\usepackage{lstwhile}
\usepackage{local} 

\newcommand{\blindurl}[1]{
\ifanonymous
\texttt{https://suppressed.for.double.blind}
\else
\url{#1}
\fi
}

\ifdraft
\fi

\begin{document}

\title[Kleene Algebra Modulo Theories]{Kleene Algebra Modulo Theories}
\subtitle{A Framework for Concrete KATs}


\author{Michael Greenberg}
\orcid{0000-0003-0014-7670}
\affiliation{
  \institution{Stevens Institute of Technology}           
  \city{Hoboken}
  \state{NJ}
  \country{USA}  
}
\email{michael@greenberg.science}         

\author{Ryan Beckett}      
\affiliation{
  \institution{Microsoft Research}            
  \city{Redmond}
  \state{WA}
  \country{USA}  
}
\email{Ryan.Beckett@microsoft.com}          

\author{Eric Campbell}
\orcid{0000-0001-5954-2136}
\affiliation{
  \institution{Cornell University}           
  \city{Ithaca}
  \state{NY}
  \country{USA}  
}
\email{ehc86@cornell.edu}         

\begin{abstract}
Kleene algebras with tests (KATs) offer sound, complete, and decidable equational reasoning about regularly structured programs. Interest in KATs has increased greatly since NetKAT demonstrated how well extensions of KATs with domain-specific primitives and extra axioms apply to computer networks.
Unfortunately, extending a KAT to a new domain by adding custom primitives, proving its equational theory sound and complete, and coming up with an efficient implementation is still an expert's task.
Abstruse metatheory is holding back KAT's potential.

We offer a fast path to a ``minimum viable
model'' of a KAT, formally or in code through our framework, \textit{Kleene algebra modulo theories} (KMT).
Given primitives and a notion of state, we can automatically derive a corresponding KAT's semantics, prove its equational theory sound and complete with respect to a tracing semantics (programs are denoted as traces of states), and derive a normalization-based decision procedure for equivalence checking. Our framework is based on \textit{pushback}, a generalization of weakest preconditions that specifies how predicates and actions interact.
We offer several case studies, showing tracing variants of theories from the literature (bitvectors, NetKAT) along with novel compositional theories (products, temporal logic, and sets). We derive new results over \emph{unbounded state}, reasoning about monotonically increasing, unbounded natural numbers. Our OCaml implementation closely matches the theory: users define and compose KATs with the module system.
\end{abstract}

\ccsdesc[500]{Software and its engineering~Formal language definitions}
\ccsdesc[300]{Software and its engineering~Frameworks}
\ccsdesc[300]{Software and its engineering~Formal software verification}
\ccsdesc[300]{Software and its engineering~Correctness}
\ccsdesc[300]{Software and its engineering~Automated static analysis}
\ccsdesc[500]{Theory of computation~Regular languages}

\maketitle


\section{Introduction}
\label{sec:intro}

Kleene algebra with tests (KAT) provides a powerful framework for reasoning about regularly structured programs. Modeling simple programs with while loops and beyond, KATs can handle a variety of analysis tasks~\cite{Angus:2001:KAT:867190, Cohen94hypothesesin, Cohen94lazycaching, Cohen94Concurrency, Kozen:2000:CCO:647482.728262, Barth02equationalverification} and typically enjoy sound, complete, and decidable equational theories.
Interest in KATs has followed their success in networking: NetKAT, a language for programming and verifying Software Defined Networks (SDNs), is a remarkably successful concrete KAT~\cite{Anderson:2014:NSF:2535838.2535862}, followed by many other variations and extensions~\citep{Schlesinger:2014:CNP:2628136.2628157,Foster2016probabilistic,Beckett:2016:TN:2908080.2908108,McClurg:2016:ENP:2908080.2908097,Larsen16wnetkat,Arashloo:2016:SSN:2934872.2934892}.

What's holding back KAT and its decidable equivalence from applying in other domains?
\emph{It's hard to generate useful, concrete instances of KAT.}
But defining concrete KATs remains a challenging task even for KAT experts. To build a custom KAT, one must craft custom domain primitives, derive a collection of new domain-specific axioms, prove the soundness and completeness of the resulting algebra, and implement a decision procedure.
For example, NetKAT's theory and implementation was developed over several papers \cite{Anderson:2014:NSF:2535838.2535862,Foster:2015:CDP:2676726.2677011,Smolka:2015:FCN:2784731.2784761}, following a series of papers that resembled, but did not use, the KAT framework~\cite{DBLP:conf/icfp/FosterHFMRSW11,DBLP:conf/popl/MonsantoFHW12,DBLP:conf/pldi/GuhaRF13,DBLP:conf/sigcomm/ReitblattCGF13}.
A pessimistic analysis concludes that making a domain-specific KAT requires moving to an institution with a KAT expert!

Abstract KAT has not successfully transferred to other domains.
The conventional, abstract approach to KAT leaves actions and predicates abstract, without any domain-specific equations~\cite{Pous:2015:SAL:2676726.2677007,10.1007/978-3-319-66902-1_16,nakamura2015decision,kozen2017coalgebraic}. Abstract KATs can't do domain-specific reasoning. Domain-specific knowledge must be encoded manually as additional equational assumptions, which makes equivalence undecidable in general; decision procedures have limited support for reasoning over domain-specific primitives and axioms~\cite{Cohen94hypothesesin,Kozen2003sa}.
Applying KAT is hard because existing work is too abstract, with challengingly telegraphic completeness proofs: normalization procedures are implicit in the very terse proofs. Such concision makes it hard for a domain expert to adapt KAT to their needs.

Domain-specific KATs will find more general application when it is possible to cheaply build and experiment with them. Our goal is to democratize KATs, offering a general framework for automatically deriving sound, complete, and decidable KATs with tracing semantics for client theories.

By tracing semantics, we mean that programs are denoted as traces of actions and states.
Such a semantics is useful for proving not just end-to-end properties (e.g., does the packet arrive at the correct host?) but also more fine-grained ones (e.g., does the packet traverse the firewall on the way?).

To demonstrate the effectiveness of our approach, we not only reproduce results from the literature (e.g., tracing variants of finite-state KATs, like bit vectors and NetKAT), but we also derive new KATs that have monotonically increasing, unbounded state (e.g., naturals).
The proof obligations of our approach are relatively mild
and our approach is \emph{compositional}: a
client can compose smaller theories to form larger, more interesting
KATs than might be tractable by hand. Our completeness proof corresponds directly to a \emph{modular} equivalence decision procedure; users compose KATs and their decision procedures
from theories specified as OCaml modules.
We offer a fast path to a ``minimum viable
model'' for those wishing to experiment with KATs.


\subsection{What is a KAT?}

\newcommand{\prog}[1]{$\text{P}_\textsf{{#1}}$\xspace}

From a bird's-eye view, a Kleene algebra with tests is a first-order language with loops (the Kleene algebra) and interesting decision making (the tests). 
Formally, a KAT consists of two parts: a Kleene algebra $\langle 0, 1,
\mathord{+}, \mathord{\cdot}, {}^* \rangle$ of ``actions'' with an
embedded Boolean algebra $\langle 0, 1, \mathord{+}, \mathord{\cdot},
\mathord{\neg} \rangle$ of ``predicates''.
KATs subsume While programs: the $1$ is interpreted as skip, $\cdot$ as sequence, $+$ as branching, and ${}^*$ for iteration.\footnote{KATs are more general, though---guarded KAT~\cite{10.1145/3371129} corresponds directly to
While programs, while KAT admits general parallel composition and
iteration.}
%
Simply adding opaque actions and predicates gives us a While-like language, where our domain is simply traces of the actions taken.
For example, if $\alpha$ and $\beta$ are predicates and $\pi$ and $\rho$ are actions, then the KAT term $\alpha \cdot \pi + \neg \alpha \cdot (\beta \cdot \rho)^* \cdot \neg \beta \cdot \pi$ defines a program denoting two kinds of traces: either $\alpha$ holds and we simply run $\pi$, or $\alpha$ doesn't hold, and we run $\rho$ until $\beta$ no longer holds and then run $\pi$.
i.e., the set of traces of the form $\{ \pi, \rho^*\pi \}$. 
Translating the KAT term into a While program, we write:
$\texttt{if } \alpha \texttt{ then } \pi \texttt{ else \{ while } \beta \texttt{ do \{ } \rho \texttt{ \}; } \pi \texttt{ \}}$.
Moving from While to KAT, consider the following program---a simple loop over two natural-valued variables \texttt{i} and \texttt{j}:
\begin{center}
\begin{while}
assume i<50; while (i<100) {i += 1;j += 2}; assert j>100
\end{while}
\end{center}
\noindent 
To model such a program in KAT, one replaces each concrete test or action with an abstract representation. Let the atomic test $\alpha$ represent the test $\texttt{i}<50$, $\beta$ represent $\texttt{i}<100$, and $\gamma$ represent $\texttt{j}>100$; the atomic actions $p$ and $q$  represent the assignments $\texttt{i} := \texttt{i} + 1$ and $\texttt{j} := \texttt{j} + 2$, respectively. We can now write the program as the KAT expression
$\alpha \cdot (\beta \cdot p \cdot q)^* \cdot \neg \beta \cdot \gamma$.
The complete equational theory of KAT makes it possible to reason about program transformations and decide equivalence between KAT terms. For example, KAT's theory that the original loop is equivalent to its unfolding:
%
\[ \alpha \cdot (\beta \cdot p \cdot q)^* \cdot \neg \beta \cdot \gamma 
   ~\equiv~ 
   \alpha \cdot (1 + \beta \cdot p \cdot q \cdot (\beta \cdot p \cdot q)^*) \cdot \neg \beta \cdot \gamma \]
But there's a catch: $\alpha$ and $\beta$ and $p$ and $q$ are abstract.
KATs are na\"{i}vely propositional, with no model of the underlying domain or the semantics of the abstract predicates and actions. For example, the fact that $(\texttt{j} := \texttt{j} + 2 \cdot \texttt{j} > 200) \equiv (\texttt{j} > 198 \cdot \texttt{j} := \texttt{j} + 2)$ does not follow from the KAT axioms and must be added manually to any proof as an equational assumption.
%
%
Yet the ability to reason about the equivalence of programs in the presence of particular domains is critical for reasoning about real programs and domain-specific languages. 
Unfortunately, it remains an expert's task to extend the KAT with new domain-specific axioms, provide new proofs of soundness and completeness, and develop the corresponding implementation~\cite{Anderson:2014:NSF:2535838.2535862,Kozen2014kaeqs,Grathwohl:2014:KB:2603088.2603095,Beckett:2016:TN:2908080.2908108,Arashloo:2016:SSN:2934872.2934892}. 

As an example of such a domain-specific KAT, NetKAT models packet forwarding in computer networks as KAT terms. Devices in a network must drop or permit packets (tests), update packets by modifying their fields (actions), and iteratively pass packets to and from other devices (loops): a network is the logical crossbar $\mathit{in}; (p; t)^*; p; \mathit{out}$, where $p$ is a policy, $t$ models the network topology and $\mathit{in}$ and $\mathit{out}$ are edge predicates.
NetKAT extends KAT with two actions and one predicate: an action to write to packet fields, $f \ASSIGN v$, where we write value $v$ to field $f$ of the current packet; an action $\DUP$, which records a packet in a history log; and a field matching predicate, $f = v$, which determines whether the field $f$ of the current packet is set to the value $v$. 
Each NetKAT program is denoted as a function from a packet history to a set of packet histories. For example, the program: $$\mathsf{dstIP} \ASSIGN 192.168.0.1 \cdot \mathsf{dstPort} \ASSIGN 4747 \cdot \DUP$$ takes a packet history as input, updates the current packet to have a new destination IP address and port, and then records the current packet state.
%
%
The original NetKAT paper
defines a denotational semantics not just for its primitive parts, but for the various KAT operators;
they explicitly restate the KAT equational theory along with custom axioms for the new primitive forms, prove the theory's soundness, and then devise a novel normalization routine to reduce NetKAT to an existing KAT with a known completeness result. 
Later papers~\cite{Foster:2015:CDP:2676726.2677011, Smolka:2015:FCN:2784731.2784761} then developed the NetKAT automata theory used to compile NetKAT programs into forwarding tables and to verify networks. 
NetKAT's power is costly: one must prove metatheorems and develop an implementation---a high bar to meet for those hoping to apply KAT in their domain.

We aim to make it easier to define new KATs. Our theoretical framework and its corresponding implementation allow for quick and easy composition of sound and complete KATs with normalization-based decision procedures when given arbitrary domain-specific theories.
Our framework, which we call Kleene algebras modulo theories (KMT) after the objects it produces, allows us to derive metatheory and implementation for KATs based on a given theory. The KMT framework obviates the need to deeply understand KAT metatheory and implementation for a large class of extensions; a variety of higher-order theories allow language designers to compose new KATs from existing ones, allowing them to rapidly prototype their KAT theories.

{\iffull
\begin{figure}[t!]\small
\begin{minipage}[t]{\iffull.28\else.4\fi\linewidth}
\centering
  \begin{while}
  assume i < 50
  while (i < 100) do
      i := i + 1 
      j := j + 2
  end
  assert j > 100
  \end{while}
\end{minipage}\hfill
\begin{minipage}[t]{\iffull.28\else.4\fi\linewidth}
\centering
  \begin{while} 
  assume 0 $\leq$ j < 4
  while (i < 10) do
      i := i + 1 
      j := (j << 1) + 3
      if i < 5 then
          insert(X, j)
  end
  assert in(X, 9)
  \end{while}
\end{minipage}{\iffull\hfill
\begin{minipage}[t]{.28\linewidth}
\centering
  \begin{while}
  i := 0
  parity := false
  while (true) do 
      odd[i] := parity
      i := i + 1
      parity := !parity
  end
  assert odd[99]
  \end{while}
\end{minipage}
\fi}

\vspace*{-1.2em}

\begin{minipage}[t]{\iffull.28\else.4\fi\linewidth}
\centering
(a) \prog{nat}
\end{minipage}\hfill
\begin{minipage}[t]{\iffull.28\else.4\fi\linewidth}
\centering
(b) \prog{set}
\end{minipage}{\iffull\hfill
\begin{minipage}[t]{.28\linewidth}
\centering
(c) \prog{map}
\end{minipage}\fi}

\vspace*{-1em}
\caption{Example simple while programs.}
\label{fig:example-programs}
\end{figure}
\fi}

{\iffull We offer some cartoons of KMTs here; see \S\ref{sec:casestudies} for technical details. 

\else

\fi}

\newcommand{\ttj}{\texttt{j}\xspace}
{\iffull
Consider \prog{set} (Fig.~\ref{fig:example-programs}b), a program defined over both naturals and a \emph{set} data structure with two operations: insertion and membership tests. The insertion action $\texttt{insert}(x,j)$ inserts the value of an expression ($j$) into a given set ($x$); the membership test $\texttt{in}(x,c)$ determines whether a constant ($c$) is included in a given set ($x$). An axiom characterizing pushback for this theory has the form:
\[ \begin{array}{rcl}
    \mathsf{insert}(x,e) \cdot \mathsf{in}(x,c) 
        &\equiv& 
        ((e=c) + \mathsf{in}(x,c)) \cdot \mathsf{insert}(x,e)  \\
   \end{array}
\]
Our theory of sets works for expressions $e$ taken from another theory, so long as the underlying theory supports tests of the form $e = c$. For example, this would work over the theory of naturals since a test like $j = 10$ can be encoded as $(j > 9) \cdot \neg (j > 10)$.

Finally, \prog{map} (Fig.~\ref{fig:example-programs}c) uses a combination of mutable \emph{boolean} values and a \emph{map} data structure. Just as before, we can craft custom theories for reasoning about each of these types of state. For booleans, we can add actions of the form $b := \true$ and $b := \false$ and tests of the form $b = \true$ and $b = \false$. The axioms are then simple equivalences like $(b := \true \cdot b = \false) \equiv 0$ and $(b := \true \cdot b = \true) \equiv (b := \true)$. 
To model map data structures, we add actions of the form $\texttt{X}[e] := e$ and tests of the form $\texttt{X}[c] = c$. Just as with the set theory, the map theory is parameterized over other theories, which can provide the type of keys and values---here, integers and booleans. In \prog{map}, the \texttt{odd} map tracks whether certain natural numbers are odd or not by storing a boolean into the map's index. A sound axiom characterizing pushback in the theory of maps has the form:
\[ (\texttt{X}[e_1] := e_2 \cdot \texttt{X}[c_1] = c_2) 
   ~\equiv~ 
   (e_1 = c_1 \cdot e_2 = c_2 + \texttt{X}[c_1] = c_2) \cdot \texttt{X}[e_1] := e_2 
\]

Each of the theories we have described so far---naturals, sets, booleans\iffull, and maps\fi---have tests that only examine the \textit{current} state of the program. However, we need not restrict ourselves in this way. Primitive tests can make dynamic decisions or assertions based on any previous state of the program. As an example, consider the theory of past-time, finite-trace linear temporal logic (\LTLf)~\cite{Giacomo2012ltlf,Giacomo14insensitivity}. Linear temporal logic introduces new operators such as: $\LAST a$ (in the last state $a$), $\EVER a$ (in some previous state $a$), and $\ALWAYS a$ (in every state $a$); we use finite-time LTL because finite traces are a reasonable model in most domains modeling programs. 

Finally, we can encode a tracing variant of NetKAT, a system that extends KAT with actions of the form $f \leftarrow v$, where some value $v$ is assigned to one of a finite number of fields $f$, and tests of the form $f = v$ where field $f$ is tested for value $v$. It also includes a number of axioms such as $f \leftarrow v \cdot f = v \equiv f \leftarrow v$. The NetKAT axioms can be captured in our framework with minor changes. Further extending NetKAT to Temporal NetKAT is captured trivially in our framework as an application of the \LTLf theory to NetKAT's theory, deriving Beckett et al.'s~\cite{Beckett:2016:TN:2908080.2908108} completeness result compositionally (in fact, we can strengthen it---see \S\ref{sec:ltlf}).

\fi}

\subsection{An example instance: incrementing naturals}
\label{sec:incnat}

We can model programs like the While program over \texttt{i} and \texttt{j} from earlier by introducing a new client theory for natural numbers (Fig.~\ref{fig:incnat}). First, we extend the KAT syntax with actions $x := n$ and $\INC_x$ (increment $x$) and a new test $x > n$ for variables $x$ and natural number constants $n$. 
Next, we define the client semantics. We fix a set of variables,
$\mathcal{V}$, which range over natural numbers, and the
program state $\sigma$ maps from variables to natural numbers.
Primitive actions and predicates are interpreted over the state $\sigma$
by the $\mathsf{act}$ and $\mathsf{pred}$ functions (where $t$ is a trace of states).

\paragraph*{Proof obligations}

\begin{figure}\small\flushleft

  {\textbf{Syntax}
  \[ \alpha ::= x > n \quad
     \pi ::= \INC_x \ALT x := n \quad
    \sub(x > n) = \{ x > m \mid m \le n \}
  \]
  }%
  
  {\textbf{Semantics}
  \[ n \in \mathbb{N} \qquad
     x \in \mathcal{V} \qquad
     \mathsf{State} = \mathcal{V} \rightarrow \mathbb{N}
  \]
  \[ \pred(x > n,t) = \last(t)(x) > n \]
  \[
    \act(\INC_x, \sigma) = \sigma[x \mapsto \sigma(x)+1] \qquad
    \act(x := n, \sigma) = \sigma[x \mapsto n]
  \]
   }
  
  {\textbf{Weakest precondition}
   \[ \begin{array}{l}
    x := n \cdot (x > m) \WP (n > m) \\
    \INC_y \cdot (x > n) \WP (x > n) \qquad \qquad
    \INC_x \cdot (x > 0) \WP 1 \\
    \INC_x \cdot (x > n) \WP (x > n-1) \text{ when $n \ne 0$} \\
  \end{array} \]}
  
  {\textbf{Axioms}  
  \[ \begin{array}{ll}
    \neg (x > n) \cdot (x > m) \equiv 0 \text{ when $n \le m$}   & \ax{GT-Contra} \\
    x := n \cdot (x > m) \equiv (n > m) \cdot x := n  & \ax{Asgn-GT} \\
    (x > m) \cdot (x > n) \equiv (x > \max(m,n)) & \ax{GT-Min} \\
    \INC_y \cdot (x > n) \equiv (x > n) \cdot \INC_y    & \ax{GT-Comm} \\
    \INC_x \cdot (x > n) \equiv (x > n-1) \cdot \INC_x  \text{ when $n > 0$} & \ax{Inc-GT} \\
    \INC_x \cdot (x > 0) \equiv \INC_x & \ax{Inc-GT-Z} \\
  \end{array} \]}
\vspace*{-1em}
  \caption{$\mathsf{IncNat}$, increasing naturals}
  \label{fig:incnat}
\end{figure}

Our framework takes a \emph{client theory} and produces a KAT, but what must one provide in order to know that the generated KAT is deductively complete, or to derive an implementation?
We require, at a minimum, a description of the theory's \emph{primitive} predicates and actions along with how these apply to some notion of state. We call these parts the \textit{client theory} (Fig.~\ref{fig:incnat}).
The resulting KAT is a \textit{Kleene algebra modulo theory} (KMT).

Our framework hinges on an operation relating predicates and operations called \emph{pushback}. Pushback is a generalization of weakest preconditions, built out of a notion of weakest preconditions for each pair of primitive test and action. Accordingly, client theories must define a weakest preconditions relation $\WP$ along with axioms that are sufficient to justify $\WP$.
The $\WP$ relation provides a way to compute the weakest precondition
for any primitive action and test: we write $\pi \cdot \alpha \WP a$ to mean that $\pi \cdot \alpha \equiv a \cdot \pi$.
For example, the weakest precondition
of $\INC_x \cdot x > n$ is $x > n - 1$ when $n$ is not zero; the weakest preconditoin of $x := n \cdot (x > m)$ is $n > m$, which is statically either $1$ (when the constant $n$ is greater than the constant $m$) or $0$ (otherwise).
The client theory's $\WP$ should have two properties: it should
be sound, (i.e., the resulting expression is equivalent to the
original one); and none of the resulting predicates should be any
bigger than the original predicates, by some measure (see
\S\ref{sec:framework}).
For example, the domain axiom: 
$\INC_x \cdot (x > n) \equiv (x > n - 1) \cdot \INC_x$ 
ensures that weakest preconditions for $\INC_x$ are modeled by the equational theory.
The other axioms are used to justify the remaining weakest 
preconditions that relate other actions and predicates.
Additional axioms that do not involve actions (\ax{Gt-Contra}, \ax{GT-Min}), are included to ensure 
that the predicate fragment of $\mathsf{IncNat}$ is 
complete in isolation.

Formally, the client must provide the following for our normalization
routine (part of completeness):
primitive tests and actions ($\alpha$ and
$\pi$), semantics for those primitives (states $\sigma$ and
functions $\pred$ and $\act$), a function identifying each primitive's subterms ($\sub$), a weakest precondition relation ($\WP$) justified by sound domain axioms ($\equiv$),
restrictions on $\WP$ term size growth.
In addition to these definitions, our client theory incurs a few proof obligations: $\equiv$ must be sound with respect to the semantics; the pushback relation should never push back a term that's larger than the input; the pushback relation should be sound with respect to $\equiv$; and we need a satisfiability checking procedure for a Boolean algebra extended with the primitive predicates.
Given these things, we can construct a sound and complete KAT with a normalization-based equivalence procedure.
For this example, the deductive completeness of the model shown here can be reduced to
Presburger arithmetic.

It was relatively easy to define $\mathsf{IncNat}$, and we get real
power---we've extended KAT with unbounded state. It is sound to add
other operations to $\mathsf{IncNat}$, like scalar multiplication or addition. So long as the operations are monotonically increasing
and invertible, we can still define a $\WP$ and corresponding
axioms.
It is \textit{not} possible, however, to compare two variables
directly with tests like $x = y$---doing so would break the requirement that weakest precondition does not enlarge tests.
Put another way, the test $x=y$ can
encode context-free languages! The non-KMT term $x := 0 \cdot
y := 0; (\INC_x)^* \cdot (\INC_y)^* \cdot x = y$ does balanced increments of $x$ and $y$. For similar
reasons, we cannot add a decrement operation
$\mathrm{dec}_x$. Either of these would let us define counter
machines, leading inevitably to undecidability.

\paragraph*{Implementation}
Users implement KMT's client theories by defining OCaml
modules; users give the types of actions and tests along with
functions for parsing, computing subterms, calculating weakest
preconditions for primitives, mapping predicates to an SMT solver, and deciding
predicate satisfiability (see \S\ref{sec:impl} for
more detail).
%

Our example implementation starts by defining a new, recursive module called \texttt{IncNat}. Recursive modules let the client theory make use of the derived KAT functions and types. For example, the module \texttt{K} on the fifth line gives us a recursive reference to the resulting KMT instantiated with the \texttt{IncNat} theory; such self-reference is key for higher-order theories, which must embed KAT predicates inside of other kinds of predicates (\S\ref{sec:casestudies}).
The client defines two types: tests $a$ and actions $p$. Here, tests are just $x > n$ where variables are \texttt{string}s, and numbers are \texttt{int}s. Actions store the variable being incremented ($\INC_x$); we omit assignment to save space.
\begin{ocaml}
type a = Gt of string * int      (* alpha ::= x > n *)
type p = Increment of string  (* pi    ::= inc x *)
module rec IncNat : THEORY
  (* generated KMT, for recursive use *)
  module K = KAT (IncNat)
  (* extensible parser; pushback; subterms *)
  let parse name es = ... 
  let push_back p a = match (p,a) with
    | (Increment x, Gt (y, j)) when x = y ->
       singleton_set (K.theory (Gt (y, j - 1)))
    | ...
  let rec subterms x = ...
  (* decision procedure for predicates *)
  let satisfiable (a: K.Test.t) = ...
end
\end{ocaml}
\noindent
The first function, \texttt{parse}, allows the library author to extend the KAT parser (if desired) to include new kinds of tests and actions in terms of infix and named operators. 
%
The implementation obligations---syntactic extensions, subterms functions, $\WP$ on primitives, a satisfiability
checker for the test fragment---mirror our formal development. We offer more client theories in \S\ref{sec:casestudies} and more implementation detail in \S\ref{sec:impl}.

\paragraph{Contributions}
We claim the following contributions:
\begin{itemize}[leftmargin=*]
\item A compositional framework for defining KATs and proving their metatheory,
  with a novel development of the normalization procedure used in
  completeness (\S\ref{sec:framework}).
  Completeness yields a decision procedure based on normalization.

{\iffull
\item Several case studies of this framework
  (\S\ref{sec:casestudies}), including a strengthening of Temporal
  NetKAT's completeness result, theories for unbounded state
  (naturals, sets, maps), distributed routing protocols, and, most
  importantly, compositional theories that allow designers to
  experiment new, complex theories. Several of these theories use
  unbounded state (e.g., naturals, sets, and maps), going beyond what
  the state of the art in KAT metatheory is able to accommodate.
  
   \fi}

\item An implementation of KMT (\S\ref{sec:impl})
  mirroring our proofs; deriving an equivalence decision procedure for client
  theories from just a few definitions. Our implementation is efficient
  enough to experiment with small programs
  (\S\ref{sec:eval}).
\end{itemize}
Finally, our framework offers a new way in for those looking to
work with KATs. Researchers comfortable with inductive relations from,
e.g., type theory and semantics, will find a familiar friend in pushback, our
generalization of weakest preconditions---we define it as an inductive
relation.

\paragraph{Notes for KAT experts}
To restate our contributions for readers more deeply familiar with
KAT:
our work is quite different from conventional work on KAT, which tends to focus on abstract, general theories.
KAT's success in NetKAT is portable, but would-be KAT users need help constructing concrete KAT instances.
Our framework is similar to Schematic KAT~\cite{KOZEN20043}. But Schematic KAT is incomplete. Our framework identifies a complete subset of Schematic KATs: tracing semantics and monotonic weakest preconditions.



\section{Case studies}
\label{sec:casestudies}

We define KAT client theories for bitvectors and
networks, as well as higher-order theories for products of theories,
sets, and temporal logic (Fig.~\ref{fig:client}). To give a sense
of the range and power of our framework, we offer these case studies
before the formal details of the framework itself
(\S\ref{sec:framework}).
{\iffull
We start with a simple theory 
(bit vectors in \S\ref{sec:katbbang}), 
building up to unbounded state from naturals (\S\ref{sec:incnat}) 
to sets and maps parameterized over a notion of value and variable (\S\ref{sec:sets}).
As an example of a higher-order theory, we define LTL on finite traces (a/k/a \LTLf; \S\ref{sec:ltlf}), extending the predicate language with temporal operators like $\LAST a$, meaning ``the predicate $a$ holds in the previous state of the trace''.
\fi}
%

\begin{figure*}[tp]
\iffull\footnotesize\fi

  \sidebyside[.35][.64][t]
  {\textbf{Syntax}\centering
  \[ 
    \alpha ::= b=\true \qquad
    \pi ::= b := \true \ALT b := \false
  \]
  \[
    \sub(\alpha) = \{ \alpha \}
  \]
  }
  {
  \textbf{Semantics}
  {\[ b \in \mathcal{B} \qquad
      \mathsf{State} = \mathcal{B} \rightarrow \{ \true, \false \} \]
   \[ \pred(b = \true,t) = \last(t)(b) \qquad
      \act(b := \true, \sigma) = \sigma[b \mapsto \true] \qquad
      \act(b := \false, \sigma) = \sigma[b \mapsto \false]
   \]
  }
  }
  
  \sidebyside[.35][.64][t]
   {
   \textbf{Weakest precondition}
   \[ 
    b := \true  \cdot b = \true  \WP 1 \qquad
    b := \false \cdot b = \true  \WP 0 
   \]
  }
  {
  \textbf{Axioms}
  \[
  (b := \true) \cdot (b = \true) \equiv (b := \true) ~~ \ax{True-True} \qquad
  (b := \false) \cdot (b = \true) \equiv 0 ~~ \ax{False-True} 
  \]
  }

(a) $\mathsf{BitVec}$, theory of bitvectors

  \sidebyside[0.45][0.54][t]
  {\textbf{Syntax}
  {\[ 
    \alpha ::= \client{\alpha_1} \ALT \client{\alpha_2} \qquad
    \pi ::= \client{\pi_1} \ALT \client{\pi_2}
   \]
   \[
    \sub(\alpha_i) = \client{\sub_i(\alpha_i)}
   \]}}
  {\textbf{Semantics}
  {\[ \mathsf{State} = \client{\mathsf{State}_1} \times \client{\mathsf{State}_2} \]
   \[ \pred(\alpha_i,t) = \client{\pred_i(\alpha_i,t_i)} \qquad
      \act(\pi_i, \sigma) = \sigma[\sigma_i \mapsto \client{\act_i(\pi_i, \sigma_i)}]
   \]}}

  \sidebyside[0.45][0.54]
  {\textbf{Weakest precondition extending $\THY_1$ and $\THY_2$}
   \[ \pi_1 \cdot \alpha_2 \WP \alpha_2 \qquad \pi_2 \cdot \alpha_1 \WP \alpha_1 \]
  }  
  {\textbf{Axioms extending $\THY_1$ and $\THY_2$}
    \[ 
    \pi_1 \cdot \alpha_2 \equiv \alpha_2 \cdot \pi_1 \quad \ax{L-R-Comm} \qquad
    \pi_2 \cdot \alpha_1 \equiv \alpha_1 \cdot \pi_2 \quad \ax{R-L-Comm}
   \]}

\centering
(b) $\mathsf{Prod}(\THY_1, \THY_2)$, products of two disjoint theories

  \sidebyside[0.49][0.49][t]
  {  \textbf{Syntax}
    \[ \alpha ::= \IN(x, c) \ALT \client{e=c} \ALT \client{\alpha_e} \qquad
       \pi ::= \INSERT(x, e) \ALT \client{\pi_e}
    \]
    \[ \begin{array}{r@{~}c@{~}l}
    \sub(\IN(x,c)) &=& \{ \IN(x,c) \} \cup \client{\sub(\neg(e = c))} \\
    \sub(e = c) &=& \client{\sub(e = c)} \\
    \sub(\alpha_e) &=& \client{\sub(\alpha_e)} \\
    \end{array} \]
  }
  { \textbf{Semantics}
   \[  c \in \mathcal{C} \qquad
       e \in \mathcal{E} \qquad
       x \in \mathcal{V} \qquad
      \mathsf{State} = (\mathcal{V} \rightarrow \mathcal{P}({\mathcal{C}})) \times (\client{\mathcal{E} \rightarrow \mathcal{C}})
   \]
   \[
      \pred(\IN(x,c),t) = c \in \last(t)_1(x) \qquad
      \pred(\alpha_e, t) = \pred(\alpha_e, t_2)
   \]
   \[ \begin{array}{l@{~~}c@{~~}l}
      \act(\INSERT(x,e), \sigma) &=& \sigma[\sigma_1[x \mapsto \sigma_1(x) \cup \{ \sigma(e) \}]] \\
      \act(\pi_e, \sigma) &=& \sigma[\sigma_2 \mapsto \client{\act(\pi_e, \sigma_2)}] \\
   \end{array} \]}

\flushleft
\sidebyside[0.49][0.49][t]
{\textbf{Weakest precondition extending $\mathcal{E}$}
  \[ \begin{array}{l}
    \INSERT(y,e) \cdot \IN(x,c) \WP \IN(x,c) \\
    \INSERT(x,e) \cdot \IN(x,c) \WP (e = c) + \IN(x,c) \\
    \INSERT(x,e) \cdot \alpha_e \WP \alpha_e \\
  \end{array} \]}
{\textbf{Axioms extending $\mathcal{E}$}
   \[ \begin{array}{r@{~~~}l}
    \INSERT(y,e) \cdot \IN(x,c) \equiv \IN(x,c) \cdot \INSERT(y,e) & \ax{Add-Comm} \\ 
    \INSERT(x,e) \cdot \IN(x,c) \equiv ((e = c) + \IN(x,c)) \cdot \INSERT(x,e) & \ax{Add-In} \\
    \INSERT(x,e) \cdot \alpha_e \equiv \alpha_e \cdot \INSERT(x,e) & \ax{Add-Comm2} \\
  \end{array} \]}

\centering
(c) $\mathsf{Set}(\mathcal{E})$, unbounded sets over expressions

\flushleft
  \sidebyside[0.4][0.59][t]
  {\textbf{Syntax}
   \[ \alpha ::= \LAST \client{a} \ALT \SINCE{\client{a}}{\client{b}} \ALT \client{a} \qquad
      \pi ::= \client{\pi_\THY} 
   \]
   \[ \begin{array}{r@{~}c@{~}l}
      \sub(\LAST a) &=& \{ \LAST a \} \cup \client{\sub(a)} \\ 
      \sub(\SINCE{a}{b}) &=& \{ \SINCE{a}{b} \} \cup \client{\sub(a)} \cup \client{\sub(b)} \\
      && \\
     \multicolumn{3}{c}{\WLAST a \triangleq \neg \LAST \neg a \qquad
     \BACKTO{a}{b} \triangleq \SINCE{a}{b} + \ALWAYS{a}} \\
     \multicolumn{3}{c}{\START \triangleq \neg \LAST 1 \qquad
     \EVER a \triangleq \SINCE{1}{a}  \qquad
     \ALWAYS a \triangleq \neg \EVER \neg a} \\
    \end{array} \]}
{\textbf{Semantics}
   \[ \mathsf{State} = \client{\mathsf{State}_\THY} \]
   \[ \begin{array}{l@{~}c@{~}l}
      \multicolumn{3}{c}{
      \pred(\LAST a,\langle \sigma, l \rangle) = \false \qquad
      \pred(\LAST a, t \langle \sigma, l \rangle) = \client{\pred(a, t)}} \\
      \pred(\SINCE{a}{b},\langle \sigma, l \rangle) &=& \client{\pred(b, \langle \sigma, l \rangle)} \\
      \pred(\SINCE{a}{b}, t \langle \sigma, l \rangle) &=& \client{\pred(b, t \langle \sigma, l \rangle)} \vee (\client{\pred(a, t \langle \sigma, l \rangle)} \wedge \pred(\SINCE{a}{b}, t)) \\
     \end{array} \]
   \[ \act(\pi,\sigma) = \client{\act(\pi, \sigma) } \]}

  \sidebyside[0.4][0.59][t]
  {\textbf{Weakest precondition extending \THY}
  \[ \pi \cdot \LAST a \WP a \]
  \infrule%
         {\client{\pi \cdot a \PBdot_\THY a' \cdot \pi} \andalso \client{\pi \cdot b \PBdot_\THY b' \cdot \pi}}
         {\pi \cdot (\SINCE{a}{b}) \WP b' + a' \cdot (\SINCE{a}{b})}
  \[ \begin{array}{rl}
      a \leq \WLAST a \cdot b ~ \rightarrow ~ a \leq \ALWAYS b  &  \ax{LTL-Induction} \\
    \end{array}
  \]
  } 
  {\textbf{Axioms (extending those of \THY)}  
  \[ \begin{array}{rl}
    \LAST (a \cdot b) \equiv \LAST a \cdot \LAST b  &  \ax{LTL-Last-Dist-Seq} \\
    \LAST (a+b) \equiv \LAST a + \LAST b  &  \ax{LTL-Last-Dist-Plus} \\
    \WLAST 1 \equiv 1  &  \ax{LTL-WLast-One} \\    
    \SINCE{a}{b} \equiv b + a \cdot \LAST (\SINCE{a}{b})  &  \ax{LTL-Since-Unroll} \\
    \neg (\SINCE{a}{b}) \equiv \BACKTO{(\neg b)}{(\neg a \cdot \neg b)}  & \ax{LTL-Not-Since} \\
    \ALWAYS a \le \EVER(\START \cdot a) & \ax{LTL-Finite} \\
  \end{array} \]}

  \centering
  (d) $\LTLf(\THY)$, linear temporal logic on finite traces over an arbitrary theory

\caption{Client theories for KMT; in higher-order theories, we $\highlight{\text{highlight}}$ client obligations.}
\label{fig:client}
\end{figure*}
\newcommand{\refbv}{\ref{fig:client}(a)}
\newcommand{\refprod}{\ref{fig:client}(b)}
\newcommand{\refset}{\ref{fig:client}(c)}
\newcommand{\refltlf}{\ref{fig:client}(d)}

\subsection{Bit vectors}
\label{sec:katbbang}

The simplest KMT is bit vectors: we extend KAT with some finite number of bits, each of which can be set to true or false and tested for their current value (Fig.~\refbv). The theory adds actions $b := \true$ and $b := \false$ for boolean variables $b$, and tests of the form $b = \true$, where $b$ is drawn from some set of names $\mathcal{B}$.

Since our bit vectors are embedded in a KAT, we can use KAT operators to build up encodings on top of bits: $b = \false$ desugars to $\neg (b = \true)$; $\mathsf{flip}~ b$ desugars to $(b = \true \cdot b := \false) + (b = \false \cdot b := \true)$.We could go further and define numeric operators on collections of bits, at the cost of producing larger terms. We are not limited to numbers, of course; once we have bits, we can encode any bounded structure we like.
KAT+B!~\cite{Grathwohl:2014:KB:2603088.2603095} develops a similar theory, but our semantics admit different equalities. KMT uses \textit{trace} semantics, distinguishing $b := \true \cdot b := \true$ and $b := \true$. Even though the final states are equivalent, they produce different traces because they run different actions.
KAT+B!, on the other hand, doesn't distinguish based on the trace of actions, so they find that $(b := \true \cdot b := \true) \equiv (b := \true)$. KMT can't \emph{exactly} model KAT+B!. (We have a similar relationship to NetKAT (\S\ref{sec:netkat}).)
It's difficult to say if one model is `better'---either could be appropriate, depending on the setting. For example, our tracing semantics is useful for answering model-checking-like questions (\S\ref{sec:ltlf}).
%


\subsection{Disjoint products}

Given two client theories, we can combine them into a disjoint product theory, $\mathsf{Prod}(\THY_1, \THY_2)$, where the states are products of the two sub-theory's states and the predicates and actions from $\THY_1$ can't affect $\THY_2$ and vice versa (Fig.~\refprod).
We explicitly give definitions for \pred and \act that defer to the corresponding sub-theory, using $t_i$ to project the trace state to the $i$th component.
It may seem that disjoint products don't give us much, but they in fact allow for us to simulate much more interesting languages in our derived KATs. For example, $\mathsf{Prod}(\mathsf{BitVec}, \mathsf{IncNat})$ allow boolean- or (increasing) natural-valued variables; the product theory lets us directly express things \citet{Kozen2003sa} encoded manually, i.e., loops over boolean and numeric state.

\subsection{Unbounded sets}
\label{sec:sets}

We define a KMT for unbounded sets parameterized on a theory of expressions $\mathcal{E}$ (Fig.~\refset). We also support maps, but we omit them out of space concerns.
The set data type supports just one operation: $\INSERT(x,e)$ adds the value of expression $e$ to set $x$ (we could add $\REMOVE(x,e)$, but we omit it to save space). It also supports a single test: $\IN(x,c)$ checks if the constant $c$ is contained in set $x$.
The idea is that $e \in \mathcal{E}$ refers to expressions with, say, variables $x$ and constants $c$. We allow arbitrary expressions $e$ in some positions and constants $c$ in others.
It's critical that for each constant $c$, the test of expression equality $e = c$ be smaller in our global ordering than the membership test $\IN(x,c)$.
\iffull\ 
(If we allowed expressions in all positions, $\WP$ wouldn't necessarily be non-increasing.)
\fi
For example, we can have sets of naturals by setting $\mathcal{E} ::= n \in \mathbb{N} \ALT i \in \mathcal{V}'$, where our constants $\mathcal{C} = \mathbb{N}$ and $\mathcal{V}'$ is some set of variables distinct from those we use for sets. We can then prove that the term $(\INC_i \cdot \INSERT(x,i))^* \cdot (i > 100) \cdot \IN(x,100)$ is non-empty by pushing tests back (and unrolling the loop 100 times).

To instantiate the $\mathsf{Set}$ theory, we need a few things: expressions $\mathcal{E}$, a subset of \textit{constants} $\mathcal{C} \subseteq \mathcal{E}$, and predicates for testing (in)equality between expressions and constants ($e = c$ and $e \ne c$). Comparing two variables would cause us to accidentally define a counter machine. Our state has two parts: $\sigma_1 : \mathcal{V} \rightarrow \mathcal{P}(\mathcal{C})$ records the current sets for each set in $\mathcal{V}$, while $\sigma_2 : \mathcal{E} \rightarrow \mathcal{C}$ evaluates expressions in each state. Since each state has its own evaluation function, the expression language can have actions that update $\sigma_2$.
\iffull
The set theory's $\sub$ function calls the client theory's $\sub$ function, so all $\IN(x,e)$ formulae must come \textit{later} in the global well ordering than any of those generated by the client theory's $e = c$ or $e \ne c$.
%
\fi

\subsection{Past-time linear temporal logic}
\label{sec:ltlf}

Past-time linear temporal logic on finite traces (\LTLf)~\cite{Baier:2006:PFT:1597538.1597664,Giacomo2012ltlf,Giacomo14insensitivity,CoinductiveLTLf_2016,Beckett:2016:TN:2908080.2908108,Campbell17,DBLP:journals/corr/abs-2107-06045} is a \emph{higher-order theory}: \LTLf extends another theory \THY, with its own predicates and actions. Any \THY test can appear in of \LTLf's temporal predicates (Fig.~\refltlf).

\LTLf adds just two predicates: $\LAST a$, pronounced ``last $a$'', means $a$ held in the prior state; and $\SINCE{a}{b}$, pronounced ``$a$ since $b$'', means $b$ held at some point in the past, and $a$ has held since then. 
There is a slight subtlety around the beginning of time: we say that $\LAST a$ is false at the beginning (what can be true in a state that never happened?), and $\SINCE{a}{b}$ degenerates to $b$ at the beginning of time.
These two predicates suffice to encode the rest of \LTLf; encodings are given below the syntax.
{\iffull The \pred definitions mostly defer to the client theory's definition of \pred (which may recursively reference the \LTLf \pred function), unrolling $\S$ as it goes (\ax{LTL-Since-Unroll}). \fi} Weakest preconditions uses inference rules: to push back $\mathcal{S}$, we unroll $\SINCE{a}{b}$ into $a \cdot \LAST (\SINCE{a}{b}) + b$; pushing last through an action is easy, but pushing back $a$ or $b$ recursively uses the $\PBdot$ judgment from the normalization routine of the KMT framework (Fig.~\ref{fig:normalization}). \iffull Adding these rules leaves our judgments monotonic, and if $\pi \cdot a \PBdot x$, then $x = \sum a_i \pi$ (Lemma~\ref{lem:pbpi}).\fi Our implementation's recursive modules let us use the derived pushback to define weakest preconditions.

The equivalence axioms come from Temporal
NetKAT~\cite{Beckett:2016:TN:2908080.2908108}; the deductive
completeness result for these axioms comes from Campbell's undergraduate thesis~\cite{Campbell17,DBLP:journals/corr/abs-2107-06045}; Ro\c{s}u's proof uses different axioms~\cite{CoinductiveLTLf_2016}.

\iffull
As a use of \LTLf, recall the simple While program from \S\ref{sec:intro}.
We may want to check that, before the last state after the loop, the variable $\ttj$ was always less than or equal to 200. We can capture this with the test $\LAST \ALWAYS (\ttj \leq 200)$. We can use the \LTLf axioms to push tests back through actions; for example, we can rewrite terms using these \LTLf axioms alongside the natural number axioms:
\[
  \begin{array}{@{}c@{}l}
                 & \ttj := \ttj + 2 \cdot \ALWAYS (\ttj \leq 200) \\
    {} \equiv {} & \ttj := \ttj + 2 \cdot (\ttj \leq 200 \cdot \LAST \ALWAYS (\ttj \leq 200)) \\
    {} \equiv {} & (\ttj := \ttj + 2 \cdot \ttj \leq 200) \cdot \LAST \ALWAYS (\ttj \leq 200) \\
    {} \equiv {} & (\ttj \leq 198) \cdot \ttj := \ttj + 2 \cdot \LAST \ALWAYS (\ttj \leq 200) \\
    {} \equiv {} & (\ttj \leq 198) \cdot \ALWAYS (\ttj \leq 200) \cdot \ttj := \ttj + 2 \\
  \end{array}
\]
Pushing the temporal test back through the action reveals that \ttj is never greater than 200 if before the action \ttj was not greater than 198 in the previous state and \ttj never exceeded 200 before the action as well.
The final pushed back test $(j \leq 198) \cdot \ALWAYS (j \leq 200)$ satisfies the theory requirements for pushback not yielding larger tests, since the resulting test is only in terms of the original test and its subterms. Note that we've embedded our theory of naturals into \LTLf: we can generate a complete equational theory for \LTLf over any other complete theory.

The ability to use temporal logic in KAT means that we can model check programs by phrasing model checking questions in terms of program equivalence. For example, for some program $r$, we can check if $r \equiv r \cdot \LAST \ALWAYS (j \leq 200)$. In other words, if there exists some program trace that does not satisfy the test, then it will be filtered---resulting in non-equivalent terms. If the terms are equal, then every trace from $r$ satisfies the test. Similarly, we can test whether $r \cdot \LAST \ALWAYS (j \leq 200)$ is empty---if so, there are \textit{no} satisfying traces.

In addition to model checking, temporal logic is a useful programming
language feature: programs can make dynamic program decisions based on
the past more concisely. Such a feature is
useful for Temporal NetKAT (\S\ref{sec:temporalnetkat} below),
but could also be used for, e.g., regular expressions with lookbehind or even a limited form of back-reference.
\fi

\subsection{Tracing NetKAT}
\label{sec:netkat}

\begin{figure}\small\flushleft
  { \textbf{Syntax}
  \[ \alpha ::= f = v \qquad\quad
    \pi ::= f \ASSIGN v \qquad\quad
    \sub(\alpha) = \{ \alpha \}
  \]}

{ \textbf{Semantics}
  \[ \mathsf{F} = \text{packet fields} \qquad
    \mathsf{V} = \text{packet field values} \qquad
    \mathsf{State} = \mathsf{F} \rightarrow \mathsf{V}
  \]
  \[ \pred(f = v,t) = \last(t).f = v \qquad
     \act(f \ASSIGN v, \sigma) = \sigma[f \mapsto v]
  \]}

  {\textbf{Weakest precondition}
   \[ \begin{array}{r@{~}c@{~}l}
    f \ASSIGN v \cdot f = v\phantom{'}  &\WP& 1 \\
    f \ASSIGN v \cdot f = v' &\WP& 0 \text{ when $v \ne v'$}\\
    f' \ASSIGN v \cdot f = v\phantom{'} &\WP& f = v
  \end{array}\]}

{\textbf{Axioms}
  \[ \begin{array}{rl}
    f \ASSIGN v \cdot f' = v' \equiv {} f' = v' \cdot f \ASSIGN v & \ax{PA-Mod-Comm} \\
    f \ASSIGN v \cdot f = v \equiv f \ASSIGN v & \ax{PA-Mod-Filter} \\
    f = v \cdot f = v' \equiv 0, \text{ if } v \ne v' & \ax{PA-Contra} \\
    \sum_v f = v \equiv 1 & \ax{PA-Match-All}
  \end{array} \]}

\vspace*{-1em}
  \caption{Tracing NetKAT a/k/a NetKAT without \DUP}
  \label{fig:netkat}
\end{figure}

We define NetKAT as a KMT over packets, which we model as functions from packet fields to values (Fig.~\ref{fig:netkat}).
KMT's tracing semantics diverge slightly from NetKAT's:
like KAT+B! (\S\ref{sec:katbbang}; \cite{Grathwohl:2014:KB:2603088.2603095}), NetKAT normally merges adjacent writes. If the policy analysis demands reasoning about the history of packets traversing the network---reasoning, for example, about which routes packets actually take---the programmer must insert \DUP{}s to record relevant moments in time. {\iffull Typically, $\DUP$s are automatically inserted at the topology level, i.e., before a packet enters a switch, we record its state by running \DUP. \fi}
From our perspective, NetKAT very nearly has a tracing semantics, but the traces are selective. If we put an implicit \DUP before \textit{every} field update, NetKAT has our tracing semantics.
{\iffull
The upshot is that our ``tracing NetKAT'' has a slightly different equational theory from conventional NetKAT, rejecting the following NetKAT laws as unsound for tracing semantics:
\[ \begin{array}{rl}
    f = v \cdot f \ASSIGN v \equiv f = v & \ax{PA-Filter-Mod} \\
    f \ASSIGN v \cdot f \ASSIGN v' \equiv f \ASSIGN v' & \ax{PA-Mod-Mod} \\
    f \ASSIGN v \cdot f' \ASSIGN v' \equiv
                       f' \ASSIGN v' \cdot f \ASSIGN v & \ax{PA-Mod-Mod-Comm} \\
\end{array} \]
In principle, one can abstract our semantics' traces to find the more restricted NetKAT traces, but we can't offer any formal support in our framework for abstracted reasoning. Just as for $\mathsf{BitVec}$, It is possible that ideas from Kozen and Mamouras could apply here~\cite{Kozen2014kaeqs}; see \S\ref{sec:related}.
\fi}

\subsection{Temporal NetKAT}
\label{sec:temporalnetkat}

We derive Temporal NetKAT as $\LTLf(\mathrm{NetKAT})$, i.e., \LTLf instantiated over tracing NetKAT; the combination yields precisely the system described in the Temporal NetKAT paper~\cite{Beckett:2016:TN:2908080.2908108}.
Recent proofs of deductive completeness for \LTLf~\cite{Campbell17,DBLP:journals/corr/abs-2107-06045} yield a stronger completeness result---the original work showed completeness only for ``network-wide'' policies, i.e., those with $\START$ at the front.


\section{The KMT framework}
\label{sec:framework}

The rest of our paper describes how our framework takes a client
theory and generates a KAT. We emphasize that you need not understand
the following formalism to use our framework---we do it once and for
all, so you don't have to!
{\iffull We first explain the structure of our framework for defining a KAT in terms of a client theory. \fi}
In figures, we $\client{\text{highlight}}$
what the client theory must provide.

We derive a KAT \THYkat (Fig.~\ref{fig:semantics}) from a client
theory \THY, where \THY has two \emph{primitive} parts---predicates $\alpha \in \THY_\alpha$ and
actions $\pi \in \THY_\pi$.
Lifting $\THY_\alpha$ to a Boolean algebra yields $\THYpred \subseteq \THYkat$, where $\THYkat$ is the KAT that embeds the client theory.

A client theory must provide:
(1) primitives $\alpha$ and $\pi$;
(2) a notion of state and semantics for those primitives on that state;
(3) theory-specific axioms of KAT equivalences that should hold, in terms of $\alpha$ and $\pi$ ($\equiv_\THY$);
(4) a weakest precondition operation $\WP$ that relates each $\alpha$ and $\pi$;
and (5) a satisfiability checker for the theory's predicates, i.e., for $\THYpred$.
(See \S\ref{sec:impl} for details on how these are provided.)

Our framework provides results for \THYkat in a pay-as-you-go fashion:
given just the state and an interpretation for the predicates and actions of \THY, we derive a tracing semantics for \THYkat (\S\ref{sec:semantics});
if the axioms of \THY are sound with respect to the tracing semantics, then \THYkat is sound (\S\ref{sec:soundness});
if the axioms of \THY are complete with respect to our semantics and $\WP$ satisfies some ordering requirements, then \THYkat has a complete equational theory (\S\ref{sec:completeness});
and finally, with just a bit of code defining the structure of \THY and deciding the predicate theory $\THYpred$, we can derive a decision procedure for equivalence (\S\ref{sec:impl}) using the normalization routine from completeness (\S\ref{sec:completeness}).

The key to our general, parameterized proof is a novel
\textit{pushback} operation that generalizes weakest preconditions (\S\ref{sec:pushback}): given an
understanding of how to push primitive predicates back to the front of
a term, we can normalize terms for our
completeness proof (\S\ref{sec:completeness}).


%
%

\subsection{Semantics}
\label{sec:semantics}

The first step in turning the client theory \THY into a KAT is to define a semantics (Fig.~\ref{fig:semantics}).
We can give any KAT a \textit{tracing semantics}: the meaning of a term is a trace $t$, which is a non-empty list of log entries $l$.
Each \textit{log entry} records a state $\sigma$ and (in all but the initial state) a primitive action $\pi$.
The client assigns meaning to predicates and actions by defining a set of states $\mathsf{State}$ and two functions: one to determine whether a predicate holds ($\pred$) and another to determine an action's effects ($\act$).
To run a \THYkat term on a state $\sigma$, we start with an initial state $\langle \sigma, \bot \rangle$; when we're done, we'll have a set of traces of the form $\langle \sigma_0, \bot \rangle  \langle \sigma_1, \pi_1 \rangle  \dots$, where $\sigma_i = \act(\pi_i, \sigma_{i-1})$ for $i > 0$.
(A similar semantics shows up in Kozen's application of KAT to static analysis~\cite{Kozen2003sa}.)

\begin{figure*}\small
  \sidebyside[.45][.45][t]
  {\hdr{Predicate syntax}{}
  \vspace*{-1em}
  \[ \begin{array}{rclr}
    a,b &::= & 0         & \textit{additive identity} \\
        &\ALT& 1         & \textit{multiplicative identity} \\
        &\ALT& \neg a    & \textit{negation} \\
        &\ALT& a + b     & \textit{disjunction} \\
        &\ALT& a \cdot b & \textit{conjunction} \\
        &\ALT& \client{\alpha} & \client{\textit{primitive predicates (\,$\THY_\alpha$)}} \\[.5em]
  \end{array} \]}
  {\hdr{Action syntax}{}
  \vspace*{-1em}
   \[ \begin{array}{rclr}
    p,q &::= & a         & \textit{embedded predicates} \\
        &\ALT& p + q     & \textit{parallel composition} \\
        &\ALT& p \cdot q & \textit{sequential composition} \\
        &\ALT& p^*       & \textit{Kleene star} \\
        &\ALT& \client{\pi} & \client{\textit{primitive actions (\,$\THY_\pi$)}}
  \end{array} \]}

  \hdr{Trace definitions}{}
  \vspace*{-1em}
  \sidebyside[.5][.48][t]
  {\[ \begin{array}{rclcl}
    \client{\sigma} &\in& \client{\mathsf{State}} && \\
    l &\in& \mathsf{Log} &::=& \langle \sigma, \bot \rangle \ALT \langle \sigma, \pi \rangle \\
    t &\in& \mathsf{Trace} &=& \mathsf{Log}^+
  \end{array}\]}
  {\[ \client{\begin{array}{rcl}
    \pred &:& \THY_\alpha \times \mathsf{Trace} \rightarrow \{ \true, \false \} \\
    \act  &:& \THY_\pi \times \mathsf{State} \rightarrow \mathsf{State} 
  \end{array}} \]}

  \hdr{Tracing semantics}{\hfill \fbox{$\denot{-} : \THYkat \rightarrow \mathsf{Trace} \rightarrow \mathcal{P}(\mathsf{Trace})$}}
  \vspace*{-1em}
  \sidebyside[.52][.45][t]
  {\[ \begin{array}{rcl}
    \denot{0}(t)         &=& \emptyset \\
    \denot{1}(t)         &=& \{ t \} \\
    \denot{\alpha}(t)    &=& \{ t \mid \client{\pred(\alpha, t)} = \true \} \\
    \denot{\neg a}(t)    &=& \{ t \mid \denot{a}(t) = \emptyset \} \\
    \denot{\pi}(t)       &=& \{ t \langle \sigma', \pi \rangle \mid \sigma' = \client{\act(\pi, \last(t))} \} \\
    \denot{p + q}(t)     &=& \denot{p}(t) \cup \denot{q}(t) \\
    \end{array} \]}
  {\[ \begin{array}{rcl}
      (f \bullet g)(t) &=& \bigcup_{t' \in f(t)} g(t') \\
\iffull      && \\ \fi
f^0(t)    = \{ t \}  && f^{i+1}(t) = (f \bullet f^i)(t) \\
\iffull      && \\ \fi
      \last(\dots \langle \sigma, \_ \rangle) &=& \sigma \\
      && \\
    \denot{p \cdot q}(t) &=& (\denot{p} \bullet \denot{q})(t) \\
    \denot{p^*}(t)       &=& \bigcup_{0 \le i} \denot{p}^i(t)
    \end{array}\]}


\hdr{Axioms (\textsc{KA} = Kleene algebra; \textsc{BA} = Boolean algebra)}{}
\vspace*{-1em}
  \sidebyside[.45][.54][t]
   {\[ \begin{array}{rl}
    p + (q + r)  \equiv (p + q) + r  &  \ax{KA-Plus-Assoc} \\
    p + q  \equiv q + p  &  \ax{KA-Plus-Comm} \\
    p + 0  \equiv p  &  \ax{KA-Plus-Zero} \\
    p + p  \equiv p  &  \ax{KA-Plus-Idem} \\
    p \cdot (q \cdot r)  \equiv (p \cdot q) \cdot r  &  \ax{KA-Seq-Assoc} \\
    1 \cdot p  \equiv p  &  \ax{KA-Seq-One} \\
    p \cdot 1  \equiv p  &  \ax{KA-One-Seq} \\
    p \cdot (q + r)  \equiv p \cdot q + p \cdot r  &  \ax{KA-Dist-L} \\
    (p + q) \cdot r  \equiv p \cdot r + q \cdot r  &  \ax{KA-Dist-R} \\
    0 \cdot p  \equiv 0  &  \ax{KA-Zero-Seq} \\
    p \cdot 0  \equiv 0  &  \ax{KA-Seq-Zero} \\
    1 + p \cdot p^*  \equiv p* & \ax{KA-Unroll-L} \\
    1 + p^* \cdot p  \equiv p* & \ax{KA-Unroll-R} \\
    q + p \cdot r \leq r ~ \rightarrow ~ p^* \cdot q \leq r  &  \ax{KA-LFP-L} \\
    p + q \cdot r \leq q ~ \rightarrow ~ p \cdot r^* \leq q  &  \ax{KA-LFP-R} \\
  \end{array} \]}
  {\[ \begin{array}{rl}
    a + (b \cdot c) \equiv (a + b) \cdot (a + c) & \ax{BA-Plus-Dist} \\
    a + 1  \equiv 1  &  \ax{BA-Plus-One} \\
    a + \neg a  \equiv 1  &  \ax{BA-Excl-Mid} \\
    a \cdot b  \equiv b \cdot a  &  \ax{BA-Seq-Comm} \\
    a \cdot  \neg a  \equiv 0  &  \ax{BA-Contra} \\
    a \cdot a  \equiv a  &  \ax{BA-Seq-Idem} \\
    & \\
    \multicolumn{2}{@{}l@{}}{\textbf{Consequences}} \\
    (p + q)^* \equiv p^* \cdot (q \cdot p^*)^* & \ax{Denesting} \\
    p \cdot a \equiv b \cdot p ~ \leftrightarrow ~  p \cdot \neg a \equiv \neg b \cdot p & \ax{Pushback-Neg} \\
    p \cdot (q \cdot p)^* \equiv (p \cdot q)^* \cdot p & \ax{Sliding} \\
    p \cdot a \equiv a \cdot q + r \rightarrow p^* \cdot a \equiv (a + p^* \cdot r) \cdot q^* & \ax{Star-Inv} \\
    \iffull
    \begin{array}{r}
    p \cdot a \equiv a \cdot q + r \rightarrow \\
    p \cdot a \cdot (p \cdot a)^* \equiv (a \cdot q + r) \cdot (q + r)^*
    \end{array}
    \fi
    & \ax{Star-Expand} \\
    & \\
    \multicolumn{2}{c}{p \le q \Leftrightarrow p + q \equiv q} \\
  \end{array} \]}

\vspace*{-1em}
  \caption{Semantics and equational theory for \THYkat}
  \label{fig:semantics}
\end{figure*}

{\iffull
A reader new to KATs should compare this definition with that of NetKAT or Temporal NetKAT~\cite{Anderson:2014:NSF:2535838.2535862,Beckett:2016:TN:2908080.2908108}: defined recursively over the syntax, the denotation function collapses predicates and actions into a single semantics, using Kleisli composition (written $\bullet$) to give meaning to sequence and an infinite union and exponentiation (written $-^n$) to give meaning to Kleene star.
We've generalized the way that predicates and actions work, though, deferring to two functions that must be defined by the client theory: \pred and \act.
\fi}

The client's $\pred$ function takes a primitive predicate $\alpha$ and a trace --- predicates can examine the entire trace --- returning true or false.
When the $\pred$ function returns $\true$, we return the singleton set holding our input trace; when $\pred$ returns $\false$, we return the empty set.
\iffull\
(Composite predicates follow this same pattern, always returning either a singleton set holding their input trace or the empty set (Lemma~\ref{lem:predsingle}).)
\fi
It's acceptable for $\pred$ to recursively call the denotational semantics (e.g., \S\ref{sec:temporalnetkat}), though we have skipped the formal detail here.

The client's $\act$ function takes a primitive action $\pi$ and the last state in the trace, returning a new state. Whatever new state comes out is recorded in the trace along with $\pi$.

\subsection{Soundness}
\label{sec:soundness}

Proving the equational theory sound relative to our tracing semantics is easy: we depend on the client's $\act$ and $\pred$ functions, and none of our KAT axioms refer to primitives (Fig.~\ref{fig:semantics}).
Our soundness proof requires that the client theory's equations be sound in our tracing semantics.

{\iffull
\fi}

\begin{theorem}[Soundness of \THYkat relative to \THY]
  \label{thm:soundness}
  If $p \equiv_\THY q
  \Rightarrow \denot{p} = \denot{q}$ then $p \equiv q \Rightarrow \denot{p} = \denot{q}$.
  \begin{proof}
    By induction on the derivation of $p \equiv q$.
    {\iffull\ (See Theorem~\ref{fullthm:soundness}.) \fi}
   \end{proof}
\end{theorem}

\noindent
If the client theory is buggy, i.e., the axioms are unsound, then we can offer no guarantees about \THY at all.
For the duration of \S\ref{sec:framework}, we assume that any equations \THY adds are sound and,
so, \THYkat is sound by Theorem~\ref{thm:soundness}.

\subsection{Normalization via pushback}
\label{sec:normalization}

In order to prove completeness (\S\ref{sec:completeness}), we
reduce our KAT terms to a more manageable subset of \emph{normal
forms}. Normalization happens via a generalization of weakest
preconditions; we use a \emph{pushback} operation to translate a
term $p$ into an equivalent term of the form $\sum a_i \cdot m_i$
where each $m_i$ does not contain any tests. The
client theory's completeness result on the $a_i$ then reduces the
completeness of our language to an existing result for Kleene
algebra on the $m_i$.

The client theory
\THY must provide two things for our normalization procedure: (1) a way to extract subterms from predicates, which orders
  predicates for the termination measure on
  normalization (\S\ref{sec:ordering}); and (2) weakest preconditions for primitives (\S\ref{sec:pushback}).
Once we've defined our normalization procedure, we can use it prove
completeness (\S\ref{sec:completeness}).
  
\subsubsection{Normalization and the maximal subterm ordering}
\label{sec:ordering}

Our normalization algorithm works by ``pushing back'' predicates to the front of a term until we reach a normal form with \textit{all} predicates at the front. 
%
%
The pushback algorithm's termination measure is complex: pushing a predicate back may not eliminate it; pushing test $a$ back through $\pi$ may yield $\sum a_i \cdot \pi$ where each of the $a_i$ copies some subterm of $a$---and there may be \textit{many} such copies!

Let the set of \textit{restricted actions} $\RA$ be the subset of
$\THYkat$ where the only test is $1$. Let the metavariables $m$, $n$, and $l$ to denote elements of $\RA$.
Let the set of \textit{normal forms} $\THYnf$ be a set of pairs of tests $a_i \in \THYpred$ and restricted actions $m_i \in \RA$.
Let the metavariables $t$, $u$, $v$, $w$, $x$, $y$, and $z$ to denote elements of $\THYnf$; we typically write these sets as sums, i.e., $x = \sum_{i=1}^k a_i \cdot m_i$ means $x = \{ (a_1,m_1), (a_2,m_2), \dots, (a_k,m_k) \}$. The sum notation is convenient, but normal forms must really be treated as sets---there should be no duplicated terms in the sum.
We write $\sum_i a_i$ to denote the normal form $\sum_i a_i \cdot 1$.
{\iffull
We will call a normal form $\textit{vacuous}$ when it is the empty set
(i.e., the empty sum, which we interpret conventionally as $0$) or
when all of its tests are $0$.
\fi}
The set of normal forms, $\THYnf$, is closed over parallel composition by
simply joining the sums.
The fundamental challenge in our normalization method is to define
sequential composition and Kleene star on $\THYnf$.

{\iffull
\begin{figure*}[t]\small
   \hdr{Sequences and tests}{
     \hfill \fbox{$\seqs : \THYpred \rightarrow \mathcal{P}(\THYpred)$}
     \quad \fbox{$\seqs : \mathcal{P}(\THYpred) \rightarrow \mathcal{P}(\THYpred)$}
     \quad \fbox{$\tests : \THYnf \rightarrow \mathcal{P}(\THYpred)$}}

   \sidebyside[.45][.45][t]
   {\[ \begin{array}{rcl}
     \seqs(a \cdot b) &=& \seqs(a) \cup \seqs(b) \\
     \seqs(a) &=& \{ a \} \\[1em]
   \end{array} \]}
   {\[ \begin{array}{rcl}
     \seqs(A) &=& \bigcup_{a \in A} \seqs(a) \\
     \tests(\sum a_i \cdot m_i) &=& \{ 1 \} \cup \bigcup_{a_i \in \sum a_i} \{ a_i \}
   \end{array} \]}

   \hdr{Subterms}{
     \hfill \fbox{$\sub : \THYpred \rightarrow \mathcal{P}(\THYpred)$} 
     \quad \fbox{$\client{\sub_\THY : \THY_\alpha \rightarrow \mathcal{P}(\THYpred)}$}
     \quad \fbox{$\sub : \mathcal{P}(\THYpred) \rightarrow \mathcal{P}(\THYpred)$}}

   \sidebyside[.4][.58][t]
   {\[ \begin{array}{rcl}
     \sub(0) &=& \{ 0 \} \\
     \sub(1) &=& \{ 0, 1 \} \\
     \sub(\alpha) &=& \{ 0, 1, \alpha \} \cup \sub_\THY(\alpha) \\
   \end{array} \]}
   {\[ \begin{array}{rcl}
     \sub(\neg a) &=& \{ 0, 1 \} \cup \sub(a) \cup \{ \neg b \mid b \in \sub(a) \} \\
     \sub(a + b) &=& \{ a + b \} \cup \sub(a) \cup \sub(b) \\
     \sub(a \cdot b) &=& \{ a \cdot b \} \cup \sub(a) \cup \sub(b)
     \end{array} \]}
   \[ \sub(A) = \bigcup_{a \in A} \sub(a) \]

   \hdr{Maximal tests}{
     \hfill \fbox{$\mt : \mathcal{P}(\THYpred) \rightarrow \mathcal{P}(\THYpred)$}
     \quad \fbox{$\mt : \THYnf \rightarrow \mathcal{P}(\THYpred)$}}

   \[ \mt(A) = \{ b \in \seqs(A) \mid \forall c \in \seqs(A), ~ c \ne b \Rightarrow b \not\in \sub(c) \} 
      \qquad 
      \mt(x) = \mt(\tests(x)) \]

   \hdr{Maximal subterm ordering}{
     \hfill \fbox{$\mathord{\preceq}, \mathord{\prec}, \mathord{\approx} \subseteq \THYnf \times \THYnf$}}
   \centering

   \[ 
     x \preceq y \Longleftrightarrow \sub(\mt(x)) \subseteq \sub(\mt(y))
     \qquad
     x \prec y   \Longleftrightarrow \sub(\mt(x)) \subsetneq \sub(\mt(y))
   \]
   \[ x \approx y  \Longleftrightarrow x \preceq y \wedge y \preceq x \]

\vspace*{-1em}
  \caption{Maximal tests and the maximal subterm ordering}
  \label{fig:subterm}
\end{figure*}
\fi}

{\iffull
The definitions for the maximal subterm ordering are complex
(Fig.~\ref{fig:subterm}), but the intuition is: \seqs gets all the
tests out of a predicate; \tests gets all the predicates out of a
normal form; \sub gets subterms; \mt gets ``maximal'' tests that cover
a whole set of tests; we lift \mt to work on normal forms by
extracting all possible tests; the relation $x \preceq y$ means that
$y$'s maximal tests include all of $x$'s maximal tests. Maximal tests
indicate which test to push back next in order to make progress
towards normalization. For example, the subterms of $\EVER x > 1$ are
defined by the client theory (\S\ref{sec:ltlf}) as $\{ \EVER x >
1, x > 1, x > 0, 1, 0 \}$, which represents the possible tests that
might be generated pushing back $\EVER x > 1$; the maximal tests of
$\EVER x > 1$ are just $\{ \EVER x > 1 \}$; the maximal tests of the
set $\{\EVER x > 1, x > 0, y > 6\}$ are $\{ \EVER x > 1, y > 6 \}$
since these tests are not subterms of any other test. Therefore, we
can choose to push back either of $\EVER x > 1$ or $y > 6$ next and
know that we will continue making progress towards normalization.

\fi}

{\iffull

\begin{figure}[t]
  \hdr{}{\hfill \fbox{$\nnf : \THYpred \rightarrow \THYpred$}}

  \sidebyside[.45][.45][t]
    {\[ \begin{array}{rcl}
        \nnf(0) &=& 0 \\
        \nnf(1) &=& 1 \\
        \nnf(\alpha) &=& \alpha \\ 
        \nnf(a + b) &=& \nnf(a) + \nnf(b) \\
        \nnf(a \cdot b) &=& \nnf(a) \cdot \nnf(b) \\
      \end{array} \]}
    {\[ \begin{array}{rcl}
        \nnf(\neg 0) &=& 1 \\
        \nnf(\neg 1) &=& 0 \\
        \nnf(\neg \alpha) &=& \neg \alpha \\
        \nnf(\neg \neg a) &=& \nnf(a) \\
        \nnf(\neg (a + b)) &=& \nnf(\neg a) \cdot \nnf(\neg b) \\
        \nnf(\neg (a \cdot b)) &=& \nnf(\neg a) + \nnf(\neg b) 
      \end{array} \]}

  \caption{Negation normal form}
  \label{fig:nnf}
\end{figure}

To handle negation, we translate predicates into a \emph{negation
normal form} where only primitive predicates $\alpha$ can be negated
(Fig.~\ref{fig:nnf}). The translation $\nnf$ uses De Morgan's laws
to push negations inwards.
These possibly negated predicates are commonly called ``atoms''.
In our setting, it is important that negation normal form is monotonic
in the maximal subterm ordering ($\preceq$;
Lemma~\ref{lem:nnfmonotonic}).

We can take a normal form $x$ and \textit{split} it around a maximal test $a \in \mt(x)$ such that we have a pair of normal forms: $a \cdot y + z$, where both $y$ and $z$ are smaller than $x$ in our ordering, because $a$ (1) appears at the front of $y$ and (2) doesn't appear in $z$ at all. 
\fi}
\begin{lemma}[Splitting]
  \label{lem:splitting}
  If $a \in \mt(x)$, then there exist $y$ and $z$ such that $x \equiv
  a \cdot y + z$ and $y \prec x$ and $z \prec x$.
  {\iffull
  \begin{proof}
     See Lemma~\ref{fulllem:splitting}.
  \end{proof}
  \fi}
\end{lemma}

\subsubsection{Pushback}
\label{sec:pushback}

Normalization requires that the client theory's
\textit{weakest preconditions} respect the subterm ordering.

\begin{definition}[Weakest preconditions]
  \label{def:wp}
  The client theory's
  \emph{weakest precondition} operation is a
  relation $\mathord{\WP} \subseteq \THY_\pi \times \THY_\alpha
  \times \mathcal{P}(\THYpred)$, where $\THY_\pi$ are the primitive actions and $\THY_\alpha$ are the primitive predicates of $\THY$.
  \WP need not be a function, but we require $\forall \pi \alpha \exists A, (\pi,\alpha,A) \in \mathord{\WP}$.
  We write 
  $\pi \cdot \alpha \WP \sum a_i \cdot \pi$ and read it as
  ``$\alpha$ pushes back through $\pi$ to yield $\sum a_i \cdot \pi$'' (the second $\pi$ is purely notational).
  We require that if $\pi \cdot \alpha \WP \{ a_1, \dots, a_k \} \cdot \pi$, then $\pi \cdot \alpha
  \equiv \sum_{i=1}^k a_i \cdot \pi$, and $a_i \preceq \alpha$.
\end{definition}
\noindent
Given the client theory's weakest-precondition relation \WP, we define
a normalization procedure for \THYkat
by extending the client's \WP relation to a more general
\emph{pushback} relation, \PB (Fig.~\ref{fig:normalization}).
The client's $\WP$ relation need not be a function, nor do the $a_i$
need to be obviously related to $\alpha$ or $\pi$ in any way. Even when
the $\WP$ relation is a function, the $\PB$ relation generally won't be.
While $\WP$ computes the classical weakest precondition, the $\PB$
relations are different: when pushing back we 
\emph{change the program itself}---not normally an option for weakest preconditions (see \S\ref{sec:related}).

\begin{figure*}[tp]\small
  \hdr{Normalization}{\hfill \fbox{$p \norm x$}}

  \threesidebyside
    {\infrule[Pred]{}{a \norm a}}
    {\infrule[Act]{}{\pi \norm 1 \cdot \pi}}
    {\infrule[Par]{p \norm x \qquad q \norm y}{p + q \norm x + y}} \\[1em]

  \sidebyside[.56][.4]
    {\infrule[Seq]{p \norm x \qquad q \norm y \qquad x \cdot y \PBJ z}{p \cdot q \norm z}} 
    {\infrule[Star]{p \norm x \qquad x^* \PBstar y}{p^* \norm y}}

  \hdr{Pushback}{\hfill \fbox{$x \cdot y \PBJ z$} \hfill \fbox{$m \cdot a \PBdot y$} \hfill \fbox{$m \cdot x \PBR y$} \hfill \fbox{$x \cdot a \PBT y$}}

  \vspace*{0.25em}

  \threesidebyside[0.49][0.22][0.22]
    {\infrule[Join]
      {m_i \cdot b_j \PBdot x_{ij}}
      {(\sum_i a_i \cdot m_i) \cdot (\sum_j b_j \cdot n_j) \PBJ \sum_i \sum_j a_i \cdot x_{ij} \cdot n_j}
    }
    {\infrule[SeqZero]
      {}
      {m \cdot 0 \PBdot 0}}
    {\infrule[SeqOne]
      {}
      {m \cdot 1 \PBdot 1 \cdot m}} \\[1em]

  \sidebyside
    {\infrule[SeqSeqTest]
      {m \cdot a \PBdot y \qquad
       y \cdot b \PBT z}
      {m \cdot (a \cdot b) \PBdot z}}
    {\infrule[SeqSeqAction]
      {n \cdot a \PBdot x \qquad
       m \cdot x \PBR y}
      {(m \cdot n) \cdot a \PBdot y}} \\[1em]

  \sidebyside
    {\infrule[SeqParTest]
       {m \cdot a \PBdot x \qquad
        m \cdot b \PBdot y}
       {m \cdot (a + b) \PBdot x + y}}
    {\infrule[SeqParAction]
       {m \cdot a \PBdot x \qquad
        n \cdot a \PBdot y}
       {(m + n) \cdot a \PBdot x + y}} \\[1em]

  \sidebyside[.35][.6]
    {\infrule[Prim]
       {\client{\pi \cdot \alpha \WP \{ a_1, \dots \}}}
       {\pi \cdot \alpha \PBdot \sum_i a_i \cdot \pi}}
    {\infrule[PrimNeg]
       {\pi \cdot a \PBdot \sum_i a_i \cdot \pi \andalso
         \nnf(\neg(\sum_i a_i)) = b}
       {\pi \cdot \neg a \PBdot b \cdot \pi}} \\[1em]

  \sidebyside[.45][.54][t]
  {\infrule[SeqStarSmaller]
     {m \cdot a \PBdot x \qquad
      x \prec a \\
      m^* \cdot x \PBR y}
     {m^* \cdot a \PBdot a + y}} 
  {\infrule[SeqStarInv]
    {m \cdot a \PBdot a \cdot t + u \qquad
     m^* \cdot u \PBR x \\
     t^* \PBstar y \qquad
     x \cdot y \PBJ z}
    {m^* \cdot a \PBdot a \cdot y + z}} \\[1em]

  \sidebyside[.43][.56][t]
  {\infrule[Restricted]
    {m \cdot a_i \PBdot x_i}
    {m \cdot \sum_i a_i \cdot n_i \PBR \sum_i x_i \cdot n_i}}
  {\infrule[Test]
    {m_i \cdot a \PBdot \sum_j b_{ij} \cdot m_{ij}}
    {(\sum_i a_i \cdot m_i) \cdot a \PBT \sum_i \sum_j a_i \cdot b_{ij} \cdot m_{ij}}}

  \hdr{Normalization of star}{\hfill \fbox{$x^* \PBstar y$}}

  \sidebyside[.3][.69][t]
  {\infrule[StarZero]
    {}
    {0^* \PBstar 1}}
  {\infrule[Slide]
    {x \prec a \qquad
     x \cdot a \PBT y \qquad
     y^* \PBstar y' \qquad
     y' \cdot x \PBJ z}
    {(a \cdot x)^* \PBstar 1 + a \cdot z}} \\[1em]

  \sidebyside[.4][.59][t]
  {\infrule[Expand]
    {x \not\prec a \qquad
     x \cdot a \PBT a \cdot t + u \\
     (t + u)^* \PBstar y \qquad
     y \cdot x \PBJ z}
    {(a \cdot x)^* \PBstar 1 + a \cdot z}}
  {\infrule[Denest]
    {a \not\in \mathsf{mt}(z) \qquad
     y \not\equiv 0 \qquad
     y^* \PBstar y' \\
     x \cdot y' \PBJ x' \qquad
     (a \cdot x')^* \PBstar z \qquad
     y' \cdot z \PBJ z'}
    {(a \cdot x + y)^* \PBstar z'}}

\vspace*{-1em}
  \caption{Normalization for \THYkat}
  \label{fig:normalization}
\end{figure*}

The top-level normalization routine is the syntax-directed $p \norm x$ relation (Fig.~\ref{fig:normalization}), which takes a term $p$ and produces a normal form $x = \sum_i a_i m_i$.
Most syntactic forms are easy to normalize:
predicates are already normal forms (\rn{Pred}); primitive actions
$\pi$ have single-summand normal forms where the
predicate is $1$ (\rn{Act}); parallel composition of two normal
forms means just joining the sums (\rn{Par}).
But sequence and Kleene star are harder: we define judgments using \PB to lift these operations to normal forms (\rn{Seq}, \rn{Star}).

For sequences, we can recursively take $p \cdot q$ and normalize $p$ into $x = \sum a_i \cdot m_i$ and $q$ into $y = \sum b_j \cdot n_j$. To combine $x$ and $y$, we can concatenate and rearrange the normal forms to get $\sum_{i,j} a_i \cdot m_i \cdot b_j \cdot n_j$. \iffull If we can push $b_j$ back through $m_i$ to find some new normal form $\sum c_k \cdot l_k$, then $\sum_{i,j,k} a_i \cdot c_k \cdot l_k \cdot n_j$ is a normal form (\rn{Join}); we \fi write $x \cdot y \PBJ z$ to mean that the concatenation of $x$ and $y$ is equivalent to the normal form $z$---the $\cdot$ is suggestive notation, here and elsewhere. 

For Kleene star, we can take $p^*$ and normalize $p$ into $x = \sum a_i \cdot m_i$, but $x^*$ isn't a normal form---we need to somehow move all of the tests out of the star and to the front. We do so with the $\PBstar$ relation (Fig.~\ref{fig:normalization}), writing $x^* \PBstar y$ to mean that the Kleene star of $x$ is equivalent to the normal form $y$---the ${}^*$ on the left is again suggestive notation.
The $\PBstar$ relation is more subtle than \PBJ. Depending on how $x$ splits (Lemma~\ref{lem:splitting}), there are four possibilities:
if $x = 0$, then $0^* \equiv 1$ (\rn{StarZero});
if $x$ splits into $a \cdot x'$, then we can either use the KAT sliding lemma \iffull\ (Lemma~\ref{lem:katsliding})\fi to pull the test out when $a$ is strictly the largest test in $x$ (\rn{Slide}) or by using the KAT expansion lemma \iffull\ (Lemma~\ref{lem:katexpand}) otherwise \fi (\rn{Expand});
if $x$ splits into $a \cdot x' + z$, we use the KAT denesting lemma \iffull\ (Lemma~\ref{lem:katdenesting})\fi to pull $a$ out before continuing recursively (\rn{Denest}).
\rn{SeqStarSmaller} and \rn{SeqStarInv} push a test ($a$) back through a star ($m^*$).
Both rules work by unrolling the loop.
In the simple case (\rn{SeqStarSmaller}), the resulting test at the front is strictly smaller than $a$, and we can generate a normal form directly.
When pushing the test back doesn't shrink $a$, we use \ax{Star-Inv} to divide the term into parts with the maximal test ($a \cdot t$) and without it ($u$).

The bulk of the pushback's work happens in the \PBdot relation, which pushes a test back through a restricted action; \PBR and \PBT use \PBdot to push tests back through other forms.
{\iffull
We write $m \cdot a \PBdot y$ to mean that pushing the test $a$ back through restricted action $m$ yields the equivalent normal form $y$. The \PBdot relation works by analyzing both the action and the test.
The client theory's \WP relation is used in \PBdot when we try to push a primitive predicate $\alpha$ through a primitive action $\pi$ (\rn{Prim}); all other KAT predicates can be handled by rules matching on the action or predicate structure, deferring to other \PB relations.
\fi}
To handle negation, the function $\nnf${}puts predicates in \textit{negation normal form}, where negations only appear on primitive predicates\iffull\ (Fig.~\ref{fig:nnf})\fi, using De Morgan's laws. \iffull \ax{Pushback-Neg} justifies the \rn{PrimNeg} case (\ax{Pushback-Neg}); we use $\nnf$ to respect the maximal subterm ordering (Lemma~\ref{lem:nnfmonotonic}).\fi

We show that our notion of pushback is correct in two steps.
First we prove that pushback is partially correct, i.e., if we can form a derivation in the pushback relations, the right-hand sides are equivalent to the left-hand-sides (Theorem~\ref{thm:pushbacksound}).
Then we show that the mutually recursive tangle of our \PB relations always terminates (Theorem~\ref{thm:pushbackexist}) \iffull, which makes extensive use of our subterm ordering lemma (Lemma~\ref{lem:nforder}) and splitting (Lemma~\ref{lem:splitting}))\fi.

\begin{theorem}[Pushback soundness]
  \label{thm:pushbacksound}
  {\iffull
  ~ \\
  \begin{enumerate}
  \item If $x \cdot y \PBJ z'$ then $x \cdot y \equiv z'$.
  \item If $x^* \PBstar y$ then $x^* \equiv y$.
  \item If $m \cdot a \PBdot y$ then $m \cdot a \equiv y$.
  \item If $m \cdot x \PBR y$ then $m \cdot x \equiv y$.
  \item If $x \cdot a \PBT y$ then $x \cdot a \equiv y$.
  \end{enumerate}
  \fi}
  \begin{proof}
    By simultaneous induction on the derivations (Theorem~\ref{fullthm:pushbacksound}).
  \end{proof}
\end{theorem}
\noindent
Finally, we
show that every left-hand side of each pushback relation has a
corresponding right-hand side. We \emph{haven't} proved that the
pushback relation is functional---there could be many different choices of maximal tests to push back.
\begin{theorem}[Pushback existence]
  \label{thm:pushbackexist}
  {\iffull
  For all $x$ and $m$ and $a$:
  \begin{enumerate}
  \item For all $y$ and $z$, if $x \preceq z$ and $y
    \preceq z$ then there exists some $z' \preceq z$ such that $x
    \cdot y \PBJ z'$.
  \item There exists a $y \preceq x$ such
    that $x^* \PBstar y$.
  \item There exists some $y \preceq
    a$ such that $m \cdot a \PBdot y$.
  \item There exists a $y \preceq x$
    such that $m \cdot x \PBR y$.
  \item If $x \preceq z$ and $a \preceq z$ then there
    exists a $y \preceq z$ such that $x \cdot a \PBT y$.
  \end{enumerate}
  \fi}
  \begin{proof}
    By induction on the lexicographical order of: 
    the subterm ordering ($\prec$); 
    the size of $x$\iffull\ (for (\ref{exj}), (\ref{exstar}), (\ref{exr}), and (\ref{ext}))\fi; 
    the size of $m$\iffull\ (for (\ref{exdot}) and (\ref{exr}))\fi; and
    the size of $a$\iffull\ (for (\ref{exdot}))\fi.
    Cases first split (Lemma~\ref{lem:splitting}) to show that
    derivations exist; subterm ordering congruence finds orderings to
    apply the IH\iffull\
    (Theorem~\ref{fullthm:pushbackexist}).\fi
  \end{proof}
\end{theorem}

\noindent
With pushback in hand, we show that every term has an equivalent normal form.

\begin{corollary}[Normal forms]
  \label{cor:nf}
  For all $p \in \THYkat$, there exists a normal form $x$ such that
  $p \norm x$ and that $p \equiv x$.
  \begin{proof}
    By induction on $p$\iffull\ (Corollary~\ref{fullcor:nf})\fi.
  \end{proof}  
\end{corollary}

\noindent
The \PB relations and these two proofs are one of the contributions of this paper: it is the first time that a KAT normalization procedure has been given as a distinct procedure, rather than hiding inside of normal forms used in completeness proofs.
Temporal NetKAT, which introduced pushback, proved Theorems~\ref{thm:pushbacksound} and~\ref{thm:pushbackexist} as a single theorem, without any explicit normalization or pushback relation.

If the client's $\WP$ doesn't obey the axioms, then normalization may produce garbage.
If the client's $\WP$ is sound but disrespects the global ordering, then normalization may not terminate, but any results it does produce will be correct.

\subsection{Completeness}
\label{sec:completeness}

We prove completeness relative to our tracing semantics---if $\denot{p} = \denot{q}$ then $p \equiv q$---by normalizing $p$ and $q$ and comparing the resulting terms.
Like other completeness proofs, ours uses the completeness of Kleene algebra (KA) as its foundation: the set of possible traces of actions performed for a restricted (test-free) action in our denotational semantics is a regular language, and so the KA axioms are sound and complete for it. In order to relate our denotational semantics to regular languages, we define the regular interpretation of restricted actions $m \in \RA$ in the conventional way and then relate our denotational semantics to the regular interpretation\iffull\  (Fig.~\ref{fig:regularaction})\fi. 
\iffull
Readers familiar with NetKAT's completeness proof may notice that we've omitted the language model and gone straight to the regular interpretation. We're able to shorten our proof because our tracing semantics is more directly relatable to its regular interpretation, and because our completeness proof separately defers to the client theory's decision procedure for the predicates at the front.
\fi
Our normalization routine only uses the KAT axioms and doesn't rely on any property of our tracing semantics. We conjecture that one could prove a similar completeness result and derive a similar decision procedure with a merging, non-tracing semantics, like in NetKAT or KAT+B!~\cite{Anderson:2014:NSF:2535838.2535862,Grathwohl:2014:KB:2603088.2603095}. 

We use several KAT theorems in our completeness proof (Fig.~\ref{fig:semantics}, Consequences), the most complex being star expansion (\ax{Star-Expand})~\cite{Beckett:2016:TN:2908080.2908108}. \ax{Pushback-Neg} is a novel generalization of a theorem of Cohen and Kozen~\cite{Kozen:1997:KAT:256167.256195,DBLP:journals/tocl/CohenK00}\iffull that $b \cdot p \equiv p \cdot
b \leftrightarrow b \cdot p \cdot \neg b + \neg b \cdot p \cdot
b \equiv 0$\fi.

\begin{theorem}[Completeness]
  \label{thm:completeness}
  If the emptiness of \THY predicates is decidable, then if $\denot{p} = \denot{q}$ then $p \equiv q$.
  \begin{proof}
    There must exist normal forms $x$ and $y$ such that $p \norm x$
    and $q \norm y$ and $p \equiv x$ and $q \equiv y$
    (Corollary~\ref{cor:nf}); by soundness (Theorem~\ref{thm:soundness}), we can find that
    $\denot{p} = \denot{x}$ and $\denot{q} = \denot{y}$, so it must be
    the case that $\denot{x} = \denot{y}$.
    We will show that $x \equiv y$ to transitively
    prove that $p \equiv q$.
    We have $x = \sum_i a_i \cdot m_i$ and $y = \sum_j b_j \cdot
    n_j$. In principle, we ought to be able to match up each of the
    $a_i$ with one of the $b_j$ and then check to see whether $m_i$ is
    equivalent to $n_j$ (by appealing to the completeness on Kleene
    algebra). But we can't simply do a syntactic matching---we could
    have $a_i$ and $b_j$ that are in effect equivalent, but not
    obviously so. Worse still, we could have $a_i$ and $a_{i'}$ equivalent!
    We need to perform two steps of disambiguation: first
    each normal form's predicates must be unambiguous locally, and then 
    the predicates must be pairwise comparable between the two normal forms.

    To construct independently unambiguous normal forms, we explode
    our normal form $x$ into a disjoint form $\hat{x}$, where we test
    each possible combination of the predicates $a_i$ and run the actions corresponding to
    the true predicates, i.e., $m_i$ gets run precisely when $a_i$ is
    true:
    \[ \left. \begin{array}{rl}
          \hat{x} 
      = & \phantom{\neg} a_1 \cdot \phantom{\neg} a_2 \cdot \ldots \cdot \phantom{\neg} a_n \cdot (m_1 + m_2 + \ldots + m_n) \\
      + & \neg a_1 \cdot \phantom{\neg} a_2 \cdot \ldots \cdot \phantom{\neg} a_n \cdot (m_2 + \ldots + m_n) \\
      + & \phantom{\neg} a_1 \cdot \neg a_2 \cdot \ldots \cdot \phantom{\neg} a_n \cdot (m_1 + \ldots + m_n) \\
      + & \dots \\
      + & \neg a_1 \cdot \neg a_2 \cdot \ldots \cdot \phantom{\neg} a_n \cdot m_n \\
      + & \neg a_1 \cdot \neg a_2 \cdot \ldots \cdot \neg a_n \cdot 0 \\
    \end{array} \right\} \begin{array}{@{}c@{}} 2^n \\ \text{terms} \end{array}
    \]
    and similarly for $\hat{y}$. We can find $x \equiv \hat{x}$ via
    distributivity (\ax{BA-Plus-Dist}), commutativity
    (\ax{KA-Plus-Comm}, \ax{BA-Seq-Comm}) and the excluded middle
    (\ax{BA-Excl-Mid}).

    Observe that the sum of all of the predicates in $\hat{x}$ and
    $\hat{y}$ are respectively equivalent to $1$, since it enumerates all
    possible combinations of each $a_i$
    (\ax{BA-Plus-Dist}, \ax{BA-Excl-Mid}); i.e., if $\hat{x} = \sum_i
    c_i \cdot l_i$ and $\hat{y} = \sum_j d_j \cdot m_j$, then $\sum_i
    c_i \equiv 1$ and $\sum_j d_j \equiv 1$.
    We can take advantage of exhaustiveness of these sums to translate
    the locally disjoint but syntactically unequal predicates in each
    $\hat{x}$ and $\hat{y}$ to a single set of predicates on both,
    which allows us to do a \textit{syntactic} comparison on each of
    the predicates. Let $\ddot{x}$ and $\ddot{y}$ be the extension of
    $\hat{x}$ and $\hat{y}$ with the tests from the other form, giving
    us $\ddot{x} = \sum_{i,j} c_i \cdot d_j \cdot l_i$ and $\ddot{y}
    = \sum_{i,j} c_i \cdot d_j \cdot m_j$. Extending the normal forms
    to be disjoint between the two normal forms is still provably
    equivalent using commutativity (\ax{BA-Seq-Comm}) and the
    exhaustiveness above (\ax{KA-Seq-One}).
        
    Now that each of the predicates are syntactically uniform and
    disjoint, we can proceed to compare the commands. But there is one
    final risk: what if the $c_i \cdot d_j \equiv 0$? Then $l_i$ and
    $m_j$ could safely be different. Since we can check
    predicates of $\THY$ for emptiness, we can
    eliminate those cases where the expanded tests at the front of
    $\ddot{x}$ and $\ddot{y}$ are equivalent to zero, which is sound
    by the client theory's completeness and zero-cancellation
    (\ax{KA-Zero-Seq} and \ax{KA-Seq-Zero}).
    If one normal form is empty, the other one must be empty as well.
    
    Finally, we can defer to deductive completeness for KA to find
    proofs that the commands are equivalent. To use KA's completeness to
    get a proof over commands, we have to show that if our commands
    have equal denotations in our semantics, then they will also have
    equal denotations in the KA semantics.  We've done exactly this by
    showing that restricted actions have regular interpretations:
    because the zero-canceled $\ddot{x}$ and $\ddot{y}$ are provably
    equivalent, soundness guarantees that their denotations are
    equal. Since their tests are pairwise disjoint, if their
    denotations are equal, it must be that any non-canceled
    commands are equal, which means that \iffull each $\lbl$ of
    these commands must be equal---and so $\R(l_i) = \R(m_j)$
    (Lemma~\ref{lem:regularsound})\fi.  By the deductive completeness of
    KA, we know that $\mathrm{KA} \vdash l_i \equiv m_j$. Since $\THYkat$ includes the KA axioms,
    $l_i \equiv m_j$; we have $c_i \cdot d_j \equiv c_i \cdot d_j$ by reflexivity,
    and so $\ddot{x} \equiv \ddot{y}$. By
    transitivity, we can see that $\hat{x} \equiv \hat{y}$ and so $x
    \equiv y$ and---finally!---$p \equiv q$. \qedhere
  \end{proof}
\end{theorem}

\noindent
Our completeness proof relies on the client theory's decision
procedure for satisfiability of $\THYpred$ terms. If the client
theory's axioms are incomplete or this decision procedure is buggy,
then the derived completeness proof may not be correct.
With a broken decision procedure, the terms $\hat{x}$/$\hat{y}$ and
$\ddot{x}$/$\ddot{y}$ might not actually be unambiguous, and so the
output of the decision procedure would be garbage.


\section{Implementation}
\label{sec:impl}

Our formalism corresponds directly to our OCaml library.\footnote{\blindurl{https://github.com/mgree/kmt}}
Implementing a client theory means defining a module with
the \code{THEORY} signature, lightly abridged:
\begin{ocaml}
module type THEORY = sig
  module A : CollectionType (* predicates *)
  module P : CollectionType (* actions *)
  (* recursive knot for KAT from A and P *)
  module Test with type t = A.t pred
  module Term with type t = (A.t, P.t) kat
  module K : KAT_IMPL
    with module A = A and module P = P
     and module Test = Test and module Term = Term
  (* lightweight extension to parser *)
  val parse : string -> expr list -> (A.t, P.t) either
  (* WP relation *)
  val push_back : P.t -> A.t -> Test.t set
  (* ordering *)
  val subterms : A.t -> Test.t set
  (* optional routines for optimization *)
  val simplify_not : A.t -> Test.t option
  val simplify_and : A.t -> A.t -> Test.t option
  val simplify_or : A.t -> A.t -> Test.t option
  val merge : P.t -> P.t -> P.t
  val reduce : A.t -> P.t -> P.t option
  (* satisfiability checker and z3 encoding *)
  val satisfiable : Test.t -> bool
  val variable : P.t -> string
  val variable_test : A.t -> string
  val create_z3_var : string * A.t -> Z3.context ->
                      Z3.Solver.solver -> Z3.Expr.expr
  val theory_to_z3_expr : A.t -> Z3.context ->
                         Z3.Expr.expr StrMap.t -> Z3.Expr.expr
end
\end{ocaml}

\noindent
\S\ref{sec:incnat}
summarizes the high-level idea and sketches a library
implementation for the theory of increasing natural numbers.
To use a higher-order theory like products, one
need only instantiate the appropriate modules in the
library:
\begin{ocaml}
module P = Product(IncNat)(Boolean)
module D = Decide(P) 
let a = P.K.parse "y<1;(a=F + a=T; inc(y));y>0" in 
let b = P.K.parse "y<1;a=T;inc(y)" in 
assert (D.equivalent a b)
\end{ocaml}
The module \texttt{P} instantiates \texttt{Product} over our theories of incrementing naturals and booleans; the module \texttt{D} gives a way to normalize terms based on the completeness proof: it defines the normalization procedure along with the decision procedure \code{equivalent}. Users of the library can combine these modules to perform any number of tasks such as compilation, verification, inference, and so on. For example, checking language equivalence is then simply a matter of reading in KMT terms and calling the normalization-based equivalence checker.
Our command-line tool works with
these theories; given KMT terms in some supported theory as input, it partitions
them into equivalence classes.


\iffull\subsection{Optimizations}\fi

Our implementation uses several optimizations. The
three most prominent are (1) hash-consing all KAT terms to ensure fast
set operations, (2) lazy construction and exploration of word automata
when checking actions for equivalence, and (3) domain-specific satisfiability checking for some theories.

\iffull
Our hash-consing constructors are \emph{smart} constructors, automatically rewriting common identities (e.g., constructing $p \cdot 1$ will simply return $p$; constructing $(p^*)^*$ will simply return $p^*$). Client theories can extend our smart constructors to witness theory-specific identities. 
These optimizations are partly responsible for the speed of our normalization routine (when it avoids the costly \ax{Denest} case).
To decide word equivalence, we use the Hopcroft and
Karp algorithm~\cite{HopcroftKarp71} on implicit automata using the
Brzozowski derivative~\cite{10.1145/321239.321249} to generate the
transition relation on-the-fly.

Client theories can implement custom solvers or rely on Z3
embeddings---custom solvers are typically faster.
We've implemented a few of these domain-specific optimizations:
the satisfiability procedure for \texttt{IncNat} makes a heuristic
decision between using our incomplete custom solver or
Z3~\cite{DeMoura:2008:ZES:1792734.1792766}---our solver is much faster
on its restricted domain.
\fi


\section{Evaluation}
\label{sec:eval}

\begin{figure*}[t]

  \begin{tabular}{lcc}
    \textbf{Benchmark}     & \THY & \textbf{Time to check equivalence} \\
    \hline \hline
    $a^* \nequiv a$ (for random arithmetic predicate $a$)                     & $\mathbb{N}$ & 0.034s \\ \hline
    $\INC_x^*; x > 10 \equiv \INC_x^*;\INC_x^*; x > 10$                        & $\mathbb{N}$ & <0.001s \\ \hline
    $\INC_x^*; x > 3; \INC_y^*; y > 3 \equiv \INC_x^*; \INC_y^*; x > 3; y > 3$ & $\mathbb{N}$ & <0.001s \\ \hline
    $x=\false; ( \mathsf{flip}~x ; \mathsf{flip}~x )^* \equiv 
     ( \mathsf{flip}~x ; \mathsf{flip}~x )^* ; x=\false$                      & $\mathcal{B}$ & <0.001s \\ \hline 
    $\begin{array}{cl}
            & w:=\false; x:=\true; y:=\false; z:=\false; \\
            & (\mathsf{if}~(w=\true + x=\true + y=\true + z=\true) ~\mathsf{then}~ a:=\true~\mathsf{else}~ a:=\false) \\
     \equiv & w:=\false; x:=\true; y:=\false; z:=\false; \\
            & (\mathsf{if}~(w=\true + x=\true) + (y=\true + z=\true) ~\mathsf{then}~ a:=\true~\mathsf{else}~ a:=\false) \\
     \end{array}$                                                            & $\mathcal{B}$ & <0.001s \\ \hline
    $\begin{array}{cl}
           & y<1; a=\true; \INC_y; (1 + b=\true; \INC_y); (1 + c=\true; \INC_y); y>2  \\
    \equiv & y<1; a=\true; b=\true; c=\true; \INC_y; \INC_y; \INC_y
    \end{array}$                                                             & $\mathbb{N} \times \mathcal{B}$ & 0.309s \\ \hline 
    $(\mathsf{flip}~x + \mathsf{flip}~y + \mathsf{flip}~z)^* \equiv (\mathsf{flip}~x + \mathsf{flip}~y + \mathsf{flip}~z)^*$ & $\mathbb{B}$ & >30s (timeout) \\
  \end{tabular}

  \caption{Implementation microbenchmarks. We timeout at 30s because waiting longer is unreasonable for a prototyping tool.}
  \label{fig:microbenchmarks}
\end{figure*}

We evaluated KMT on a collection of microbenchmarks exercising concrete KAT features (Fig.~\ref{fig:microbenchmarks}). For example, the second-to-last example does population count in a theory combining naturals and booleans: if a counter $y$ is above a certain threshold, then the booleans $a$, $b$, and $c$ must have been set to true.
Our tool is usable for exploration---enough to decide whether to pursue any particular KAT.

Our normalization-based decision procedure is very fast in many cases. This is likely due to a combination of hash-consing and smart constructors that rewrite complex terms into simpler ones when possible, and the fact that, unlike previous KAT-based normalization proofs (e.g., \cite{Anderson:2014:NSF:2535838.2535862, Kozen2003sa}) our normalization proof does not require splitting predicates into all possible ``complete tests.'' However, our decision procedure does very poorly on examples where there is a sum nested inside of a Kleene star, i.e., $(p+q)^*$. The final, bit-flipping benchmark is one such example---it flips three boolean variables in some arbitrary order. In this case the normalization-based decision procedure repeatedly invokes the \ax{Denest} rewriting rule, which greatly increases the size of the term on each invocation.
Consider the simpler loop, which only flips from false to true: $(x_1=\false;x_1:=\true + \dots + x_n=\false;x_n:=\true)^*$.
With $n = 1$, there are 4 disjunctions in the locally unambiguous form; with $n = 2$, there are 16; with $n = 3$, there are 512; with $n = 4$, there are 65,536. The normal forms grow in $O(2^{2^n})$, which quickly becomes intractable in space and time.


\section{Related work}
\label{sec:related}

Kozen and Mamouras's Kleene algebra with equations~\cite{Kozen2014kaeqs} is perhaps the most closely related work: they also devise a framework for proving extensions of KAT sound and complete.
Our works share a similar genesis: Kleene algebra with equations generalizes the NetKAT completeness proof (and then reconstructs it); our work generalizes the Temporal NetKAT completeness proof (and then reconstructs it---while also developing several other, novel KATs).
Both their work and ours use rewriting to find normal forms and prove deductive completeness. Their rewriting systems work on mixed sequences of actions and predicates, but they can only delete these sequences wholesale or replace them with a single primitive action or predicate; our rewriting system's pushback operation only works on predicates (since the tracing semantics preserves the order of actions), but pushback isn't restricted to producing at most a single primitive predicate. Each framework can do things the other cannot. Kozen and Mamouras can accommodate equations that combine actions, like those that eliminate redundant writes in KAT+B! and NetKAT~\cite{Grathwohl:2014:KB:2603088.2603095,Anderson:2014:NSF:2535838.2535862}; 
we can accommodate more complex predicates and their interaction with actions, like those found in Temporal NetKAT~\cite{Beckett:2016:TN:2908080.2908108} or those produced by the compositional theories (\S\ref{sec:casestudies}).
A tracing semantics occurs in previous work on KAT as well \cite{Kozen2003sa, Gabbay:2011:FNS:1987171.1987202}.
Selective tracing (\`a la NetKAT's \DUP) offers more control over which traces are considered equivalent; our pushback offers more flexibility for how actions and predicates interact. It may be possible to build a hybrid framework, with ideas from both.

Kozen studies KATs with arbitrary equations $x := e$~\cite{KOZEN20043}, also called Schematic KAT, where $e$ comes from arbitrary first-order structures over a fixed signature $\Sigma$. He has a pushback-like axiom $x := e \cdot p \equiv p[\sfrac{x}{e}] \cdot x := e$. Arbitrary first-order structures over $\Sigma$'s theory are much more expressive than anything we can handle---the pushback may or may not decrease in size, depending on $\Sigma$; KATs over such theories are generally undecidable. We, on the other hand, are able to offer pay-as-you-go results for soundness and completeness as well as an implementations for deciding equivalence---but only for first-order structures that admit a non-increasing weakest precondition.
Other extensions of KAT often give up on decidabililty, too.
Larsen et al.~\cite{Larsen16wnetkat} allow comparison of variables, leading to an incomplete theory. They are, able, however, to decide emptiness of an entire expression.

Coalgebra provides a general framework for reasoning about state-based systems~\cite{Silva10Coalgebra,Rutten:1996:UCT:869662,kozen2017coalgebraic}, which has proven useful in the development of automata theory for KAT extensions. Although we do not explicitly develop the connection in this paper, we've developed an automata theoretic decision procedure for KMT that uses tools similar to those used in coalgebraic approaches, and one could perhaps adapt our theory and implementation to that setting.
In principle, we ought to be able to combine ideas from the two schemes into a single, even more general framework that supports complex actions \textit{and} predicates.

Symkat is a powerful decision procedure for symbolic KAT, but doesn't work in our concrete setting~\cite{Pous:2015:SAL:2676726.2677007}. It's possible to give symkat extra equations, and it can solve some equivalences that KMT can, but it can't handle, e.g., commutativity in general. Knotical uses KAT to model program traces for trace refinement~\cite{DBLP:journals/pacmpl/AntonopoulosKL19}. Our tracing semantics may be particularly well adapted for them, though they could generate KAT equations that fall outside of KMT's capabilities.

Smolka et al.~\cite{10.1145/3371129} find an almost linear algorithm for checking equivalence of \emph{guarded} KAT terms ($O(n \cdot \alpha(n))$, where $\alpha$ is the inverse Ackermann function), i.e., terms which use $\mathsf{if}$ and $\mathsf{while}$ instead of $+$ and ${}^*$, respectively. Their guarded KAT is completely abstract (i.e., actions are purely symbolic), while our KMTs are completely concrete (i.e., actions affect a clearly defined notion of state). 


Our work is loosely related to Satisfiability Modulo Theories (SMT) ~\cite{DeMoura:2011:SMT:1995376.1995394}. Both aim to create an extensible framework where custom theories can be combined~\cite{Nelson:1979:SCD:357073.357079} and used to increase the expressiveness and power~\cite{Stump2001ADP} of the underlying technique (SAT vs. KA). {\iffull However, the specifics vary greatly---while SMT is used to reason about the formula satisfiability, KMT is used to reason about how program structure interacts with tests. \fi}
Some of our KMT theories implement satisfiability checking by calling out to Z3~\cite{DeMoura:2008:ZES:1792734.1792766}.


The pushback requirement detailed in this paper is closely related to the classical notion of weakest precondition~\cite{Barnett:2005:WUP:1108792.1108813, Dijkstra:1975:GCN:360933.360975, DBLP:journals/corr/Santosa15}.
The pushback operation isn't quite a generalization of weakest preconditions
because the various \PB relations can change the program itself.
Automatic weakest precondition generation is generally limited in the presence of loops in while-programs. While there has been much work on loop invariant inference~\cite{DBLP:journals/corr/abs-0909-0884, Furia2010, DBLP:journals/corr/GaleottiFMFZ14, Kong:2010:AIQ:1947873.1947904,Sharma:2014:ICI:2735050.2735057,Padhi:2016:DPI:2908080.2908099}, the problem remains undecidable in most cases; however, the pushback restrictions of ``growth'' of terms makes it possible for us to automatically lift the weakest precondition generation to loops in KAT. In fact, this is exactly what the normalization proof does when lifting tests out of the Kleene star operator. 

The core technique we discuss here was first developed in Beckett et
al.'s work on Temporal
NetKAT~\cite{Beckett:2016:TN:2908080.2908108}. Our work 
significantly extends that work: our normalization proof is explicit, rather than implicit; we separate proofs of correctness and termination of normalization;
our treatment of negation is improved; we prove a new KAT theorem (\ax{Pushback-Neg});
KMT is a general \emph{framework} for proving completeness, while the Temporal NetKAT development is specialized to a particular instance;
and Temporal NetKAT proof achieves limited completeness because of its limited understanding of \LTLf; we are able to achieve a more general result~\cite{Campbell17,DBLP:journals/corr/abs-2107-06045}.
Beckett et al.~handles compilation to forwarding decision
diagrams~\cite{Smolka:2015:FCN:2784731.2784761}, while our
presentation doesn't discuss compilation.


%
%


\section{Conclusion}

Kleene algebra modulo theories (KMT) is a new framework for extending
Kleene algebra with tests with the addition of actions and predicates
in a custom domain. KMT uses an operation that pushes tests back
through actions to go from a decidable client theory to a
domain-specific KMT. Derived KMTs are sound and complete with respect
to a tracing semantics; we derive a decision procedure
in an implementation that mirrors our formalism. The KMT
framework captures common use cases and can reproduce \emph{by mere
  composition} several results from the literature as well as several new results: we offer theories for
bitvectors~\cite{Grathwohl:2014:KB:2603088.2603095}, natural numbers,
unbounded sets\iffull and maps\fi,
networks~\cite{Anderson:2014:NSF:2535838.2535862}, and temporal
logic~\cite{Beckett:2016:TN:2908080.2908108}. Our ability to reason about unbounded state is novel.
Our decision procedure follows our proof; automata-theoretic/coinductive approaches would be more
efficient. Our approach isn't inherently limited to tracing semantics, as alternative regular
interpretations could merge actions (as in KAT+B!, NetKAT, and Kleene
algebra with
equations~\cite{Anderson:2014:NSF:2535838.2535862,Grathwohl:2014:KB:2603088.2603095,Kozen2014kaeqs});
future work could develop a relational semantics.

\begin{acks}

Dave Walker and Aarti Gupta provided valuable advice.
Ryan Beckett was supported by NSF CNS award 1703493.
The first two authors did preliminary work at Princeton University.
The first and last authors did later work at Pomona College.
Justin Hsu and Eric Koskinen provided advice and encouragement.
Colin Gordon helped shepherd the paper; we thank the PLDI reviewers
for helping improve the paper.

\end{acks}

\bibliographystyle{ACM-Reference-Format}
\bibliography{genkat}

{\iffull
\appendix
\section{Soundness proofs}

\begin{lemma}[Kleisli composition is associative]
  \label{lem:kleisliassoc}
  $\denot{p} \bullet (\denot{q} \bullet \denot{r}) = (\denot{p} \bullet \denot{q}) \bullet \denot{r}$.
  \begin{proof}
    We compute:
      \[ \begin{array}{rl}
            \denot{p \cdot (q \cdot r)}(t) 
        & = \bigcup_{t' \in \denot{p}(t)} \denot{q \cdot r}(t') \\
        & = \bigcup_{t' \in \denot{p}(t)} \bigcup_{t'' \in \denot{q}(t')} \denot{r}(t'') \\
        & = \bigcup_{t'' \in \bigcup_{t' \in \denot{p}(t)} \denot{q}(t')} \denot{r}(t'') \\
        & = \bigcup_{t'' \in \denot{p \cdot q}(t)} \denot{r}(t'') \\
        & = \denot{(p \cdot q) \cdot r}(t)
      \end{array} \]
      \qedhere
  \end{proof}
\end{lemma}

\begin{lemma}[Exponentiation commutes]
  \label{lem:expcommute}
  $\denot{p}^{i+1} = \denot{p}^i \bullet \denot{p}$
  \begin{proof}
    By induction on $i$.
    When $i = 0$, both yield $\denot{p}$.
    In the inductive case, we compute:
    \[ \begin{array}{rl}
          \denot{p}^{i+2}
      = & \denot{p} \bullet \denot{p}^{i+1} \\
      = & \denot{p} \bullet (\denot{p}^i \bullet \denot{p}) \text{\quad by the IH} \\
      = & (\denot{p} \bullet \denot{p}^i) \bullet \denot{p} \text{\quad by Lemma~\ref{lem:kleisliassoc}} \\
      = & (\denot{p}^{i+1}) \bullet \denot{p} \text{\quad by Lemma~\ref{lem:kleisliassoc}} \\
    \end{array} \]
    \qedhere
  \end{proof}
\end{lemma}

\begin{lemma}[Predicates produce singleton or empty sets]
  \label{lem:predsingle}
  $\denot{a}(t) \subseteq \{ t \}$.
  \begin{proof}
    By induction on $a$, leaving $t$ general.
    \begin{proofcases}
    \item[($a=0$)] We have $\denot{a}(t) = \emptyset$.
    \item[($a=1$)] We have $\denot{a}(t) = \{ t \}$.
    \item[($a=\alpha$)] If $\pred(\alpha, t) = \true$, then our output trace is $\{ t \}$; otherwise, it is $\emptyset$.
    \item[($a=\neg b$)] We have $\denot{\neg a}(t) = \{ t \mid \denot{b}(t) = \emptyset \}$. By the IH, $\denot{b}(t)$ is either $\emptyset$ (in which case we get $\{ t \}$ as our output) or $\{ t \}$ (in which case we get $\emptyset$).
    \item[($a=b + c$)] By the IHs.
    \item[($a=b \cdot c$)] We get $\denot{b \cdot c}(t) = \bigcup_{t' \ in \denot{b}(t)} \denot{c}(t')$. By the IH on $b$, we know that $b$ yields either the set $\{ t \}$ or the emptyset; by the IH on $c$, we find the same. \qedhere
    \end{proofcases}
  \end{proof}
\end{lemma}

\begin{theorem}[Soundness of \THYkat relative to \THY]
  \label{fullthm:soundness}
  If $p \equiv_\THY q
  \Rightarrow \denot{p} = \denot{q}$ then $p \equiv q \Rightarrow \denot{p} = \denot{q}$.
  \begin{proof}
    By induction on the derivation of $p \equiv q$.
    \begin{proofcases}

    \item[(\ax{KA-Plus-Assoc})] We have $p + (q + r) \equiv (p + q) + r$; by associativity of union.
    \item[(\ax{KA-Plus-Comm})] We have $p + q  \equiv q + p$; by commutativity of union.
    \item[(\ax{KA-Plus-Zero})] We have $p + 0 \equiv p$; immediate,
      since $\denot{0}(t) = \emptyset$.
    \item[(\ax{KA-Plus-Idem})] By idempotence of union $p + p \equiv p$.
    \item[(\ax{KA-Seq-Assoc})] We have $p \cdot (q \cdot r) \equiv (p
      \cdot q) \cdot r$; by Lemma~\ref{lem:kleisliassoc}.
    \item[(\ax{KA-Seq-One})] We have $1 \cdot p \equiv p$; immediate, since $\denot{1}(t) = \{ t \}$.
    \item[(\ax{KA-One-Seq})] We have $p \cdot 1 \equiv p$; immediate, since $\denot{1}(t) = \{ t \}$.
    \item[(\ax{KA-Dist-L})] We have $p \cdot (q + r) \equiv p \cdot q + p \cdot r$; we compute:
      \[ \begin{array}{rl}
            \denot{p \cdot (q + r)}(t)
        & = \bigcup_{t' \in \denot{p}(t)} \denot{q + r}(t') \\
        & = \bigcup_{t' \in \denot{p}(t)} \denot{q}(t') \cup \denot{r}(t') \\
        & = \bigcup_{t' \in \denot{p}(t)} \denot{q}(t') \cup \bigcup_{t' \in \denot{p}(t)} \denot{r}(t') \\
        & = \denot{p \cdot q}(t) \cup \denot{p \cdot r}(t) \\
        & = \denot{p \cdot q + p \cdot r}(t)
      \end{array} \]
    \item[(\ax{KA-Dist-R})]
      As for \ax{KA-Dist-L}. We have $(p + q) \cdot r \equiv p \cdot r + q \cdot r$; we compute:
      \[ \begin{array}{rl}
            \denot{(p + q) \cdot r}(t)
        & = \bigcup_{t' \in \denot{p + q}(t)} \denot{r}(t') \\
        & = \bigcup_{t' \in \denot{p}(t) \cup \denot{q}(t)} \denot{r}(t') \\
        & = \bigcup_{t' \in \denot{p}(t)} \denot{r}(t') \cup \bigcup_{t' \in \denot{q}(t)} \denot{r}(t') \\
        & = \denot{p \cdot r}(t) \cup \denot{q \cdot r}(t) \\
        & = \denot{p \cdot r + q \cdot r}(t)
      \end{array} \]
    \item[(\ax{KA-Zero-Seq})] We have $0 \cdot p \equiv 0$; immediate, since $\denot{0}(t) = \emptyset$.
    \item[(\ax{KA-Seq-Zero})] We have $p \cdot 0 \equiv 0$; immediate, since $\denot{0}(t) = \emptyset$.
    \item[(\ax{KA-Unroll-L})] We have $p^* \equiv 1 + p \cdot
      p^*$. We compute:
      \[ \begin{array}{rl}
            \denot{p^*}(t)
        & = \bigcup_{0 \le i} \denot{p}^i(t) \\
        & = \denot{1}(t) \cup \bigcup_{1 \le i} \denot{p}^i(t) \\
        & = \denot{1}(t) \cup \denot{p}(t) \cup \bigcup_{2 \le i} \denot{p}^i(t)) \\
        & = \denot{1}(t) \cup (\denot{p} \bullet \denot{1})(t) \cup \bigcup_{1 \le i} (\denot{p} \bullet \denot{p}^i)(t) \\
        & = \denot{1}(t) \cup (\denot{p} \bullet \denot{1})(t') \cup (\denot{p} \bullet \bigcup_{1 \le i} \denot{p}^i)(t) \\
        & = \denot{1}(t) \cup (\denot{p} \bullet \bigcup_{0 \le i} \denot{p}^i)(t) \\
        & = \denot{1}(t) \cup \denot{p \cdot p^*}(t)) \\
        & = \denot{1 + p \cdot p^*}(t)
      \end{array} \]

    \item[(\ax{KA-Unroll-R})] 
      As for \ax{KA-Unroll-L}. We have $p^* \equiv 1 + p^* \cdot p$. We compute, using Lemma~\ref{lem:expcommute} to unroll the exponential in the other direction:
      \[ \begin{array}{rl}
            \denot{p^*}(t)
        & = \bigcup_{0 \le i} \denot{p}^i(t) \\
        & = \denot{1}(t) \cup \bigcup_{1 \le i} \denot{p}^i(t) \\
        & = \denot{1}(t) \cup \denot{p}(t) \cup \bigcup_{2 \le i} \denot{p}^i(t) \\
        & = \denot{1}(t) \cup \denot{p}(t) \cup \bigcup_{1 \le i} (\denot{p}^i \bullet \denot{p})(t') \text{\quad by Lemma~\ref{lem:expcommute}} \\
        & = \denot{1}(t) \cup (\denot{1} \bullet \denot{p})(t) \cup \bigcup_{1 \le i} (\denot{p}^i \bullet \denot{p})(t') \\
        & = \denot{1}(t) \cup \bigcup_{0 \le i} (\denot{p}^i \bullet \denot{p})(t') \\
        & = \denot{1}(t) \cup (\bigcup_{0 \le i} \denot{p}^i \bullet \denot{p})(t') \\
        & = \denot{1}(t) \cup \denot{p^* \cdot p}(t) \\
        & = \denot{1 + p^* \cdot p}(t)
      \end{array} \]

    \item[(\ax{KA-LFP-L})] We have $p^* \cdot q \le r$, i.e., $p^* \cdot q + r \equiv r$. By the IH, we know that $\denot{q}(t) \cup (\denot{p} \bullet \denot{r})(t) \cup \denot{r}(t) = \denot{r}(t)$. We show, by induction on $i$, that $(\denot{p}^i \bullet \denot{q})(t) \cup \denot{r}(t) = \denot{r}(t)$.
      \begin{proofcases}[leftmargin=1cm]
      \item[($i=0$)] We compute:
        \[ \begin{array}{r@{~}l}
           & (\denot{p}^0 \bullet \denot{q})(t) \cup \denot{r}(t) \\
          =& (\denot{1} \bullet \denot{q})(t) \cup \denot{r}(t) \\
          =&                    \denot{q}(t) \cup \denot{r}(t) \\
          =&                    \denot{q}(t) \cup (\denot{q}(t) \cup (\denot{p} \bullet \denot{r})(t) \cup \denot{r}) \quad \text{by the outer IH} \\
          =& \denot{q}(t) \cup (\denot{p} \cdot \denot{r})(t) \cup \denot{r}(t) \\
          =& \denot{r}(t) \quad \text{by the outer IH again}
        \end{array} \]

      \item[($i=i'+1$)] We compute:
        \[ \begin{array}{r@{~}l}
           & (\denot{p}^{i'+1} \bullet \denot{q})(t) \cup \denot{r}(t) \\
          =& (\denot{p} \bullet \denot{p}^{i'} \bullet \denot{q})(t) \cup \denot{r}(t) \\
          =& (\denot{p} \bullet \denot{p}^{i'} \bullet \denot{q})(t) \cup (\denot{q}(t) \cup (\denot{p} \bullet \denot{r})(t) \cup \denot{r}(t)) \quad \text{by the outer IH} \\
          =& \bigcup_{t' \in \denot{p}(t)} (\bigcup_{t'' \in \denot{p}^{i'}(t')} \denot{q}(t') \cup \denot{r}(t')) \cup (\denot{q}(t) \cup \denot{r}(t)) \\
          =& (\denot{p} \bullet \denot{r})(t) \cup (\denot{q}(t) \cup \denot{r}(t)) \quad \text{by the inner IH} \\
          =& \denot{r}(t) \quad \text{by the outer IH again}
        \end{array} \]
      \end{proofcases}

      So, finally, we have:
      \[ \denot{p^* \cdot q + r}(t) = (\bigcup_{0 \le i} \denot{p}^i \bullet \denot{q})(t) \cup \denot{r}(t) = \bigcup_{0 \le i} (\denot{p}^i \bullet \denot{q})(t) \cup \denot{r}(t)) 
         = \bigcup_{0 \le i} \denot{r}(t) = \denot{r}(t) \]

    \item[(\ax{KA-LFP-R})] 
      As for \ax{KA-LFP-L}.
      We have $p \cdot r^* \le q$, i.e., $p \cdot r^* + q \equiv q$. By the IH, we know that $\denot{p}(t) \cup (\denot{q} \bullet \denot{r})(t) \cup \denot{q}(t) = \denot{q}(t)$. We show, by induction on $i$, that $(\denot{p} \bullet \denot{r}^i)(t) \cup \denot{q}(t) = \denot{q}(t)$.
      \begin{proofcases}[leftmargin=1cm]
      \item[($i=0$)] We compute:
        \[ \begin{array}{r@{~}l}
           & (\denot{p} \bullet \denot{r}^0)(t) \cup \denot{q}(t) \\
          =& (\denot{p} \bullet \denot{1})(t) \cup \denot{q}(t) \\
          =& \denot{p}(t) \cup \denot{q}(t) \\
          =& \denot{p}(t) \cup \denot{p}(t) \cup (\denot{q} \bullet \denot{r})(t) \cup \denot{q}(t) \quad \text{by the outer IH} \\
          =& \denot{p}(t) \cup (\denot{q} \bullet \denot{r})(t) \cup \denot{q}(t) \\
          =& \denot{q}(t)
        \end{array} \]

      \item[($i=i'+1$)] We compute:
        \[ \begin{array}{r@{~}l}
           & (\denot{p} \bullet \denot{r}^{i'+1})(t') \cup \denot{q}(t) \\
          =& (\denot{p} \bullet \denot{r}^{i'} \bullet \denot{r})(t) \cup \denot{q}(t) \quad \text{by Lemma~\ref{lem:expcommute}} \\
          =& (\denot{p} \bullet \denot{r}^{i'} \bullet \denot{r})(t) \cup \denot{p}(t) \cup (\denot{q} \bullet \denot{r})(t) \cup \denot{q}(t) \quad \text{by the outer IH} \\
          =& \bigcup_{t' \in \denot{p}(t)} \bigcup_{t'' \in \denot{r}^{i'}(t') \cup \denot{q}(t)} \denot{r}(t'') \cup \denot{p}(t) \cup \denot{q}(t)  \\
          =& \bigcup_{t' \in \bigcup_{t'' \in \denot{p}(t)} \denot{r}^{i'}(t'') \cup \denot{q}(t)} \denot{r}(t') \cup \denot{p}(t) \cup \denot{q}(t)  \\
          =& (\denot{q} \bullet \denot{r})(t) \cup \denot{p}(t) \cup \denot{q}(t) \quad \text{by the inner IH} \\
          =& \denot{q}(t) \quad \text{by the inner IH again}
        \end{array} \]
      \end{proofcases}

      So, finally, we have:
      \[ 
      \denot{p \cdot r^* + q}(t) =
      (\denot{p} \bullet \bigcup_{0 \le i} \denot{r}^i)(t) \cup \denot{q}(t) =
      \bigcup_{0 \le i} (\denot{p} \bullet \denot{r}^i)(t) \cup \denot{q}(t)) =
      \bigcup_{0 \le i} \denot{q}(t) =
      \denot{q}(t)
      \]

    \item[(\ax{BA-Plus-Dist})] We have $a + (b \cdot c) \equiv (a + b)
      \cdot (a + c)$.  We have $\denot{a + (b \cdot c)}(t) =
      \denot{a}(t) \cup (\denot{b} \bullet \denot{c})(t)$.  By
      Lemma~\ref{lem:predsingle}, we know that each of these
      denotations produces either $\{ t \}$ or $\emptyset$, where
      $\cup$ is disjunction and $\bullet$ is conjunction. By distributivity of these operations.
    \item[(\ax{BA-Plus-One})] We have $a + 1 \equiv 1$; we have this
      directly by Lemma~\ref{lem:predsingle}.
    \item[(\ax{BA-Excl-Mid})] We have $a + \neg a \equiv 1$; we have
      this directly by Lemma~\ref{lem:predsingle} and the definition
      of negation.
    \item[(\ax{BA-Seq-Comm})] $a \cdot b \equiv b \cdot a$; we have
      this directly by Lemma~\ref{lem:predsingle} and unfolding the
      union.
    \item[(\ax{BA-Contra})] We have $a \cdot \neg a \equiv 0$; we have
      this directly by Lemma~\ref{lem:predsingle} and the definition
      of negation.
    \item[(\ax{BA-Seq-Idem})] $a \cdot a \equiv a$; we have this
      directly by Lemma~\ref{lem:predsingle} and unfolding the union. \qedhere
    \end{proofcases}
  \end{proof}
\end{theorem}

\section{Normalization proofs}

\begin{lemma}[Terms are subterms of themselves]
  \label{lem:asuba}
  $a \in \sub(a)$
  \begin{proof}
    By induction on $a$. All cases are immediate except for $\neg a$, which uses the IH. \qedhere
  \end{proof}
\end{lemma}

\begin{lemma}[0 is a subterm of all terms]
  \label{lem:0suba}
  $0 \in \sub(a)$
  \begin{proof}
    By induction on $a$. The cases for $0$, $1$, and $\alpha$ are
    immediate; the rest of the cases follow by the IH.       \qedhere

  \end{proof}
\end{lemma}

\begin{lemma}[Maximal tests are tests]
  \label{lem:mtinseqs}
  $\mt(A) \subseteq \seqs(A)$ for all sets of tests $A$.
  \begin{proof}
    We have by definition:
    \[ \begin{array}{rcl}
      \mt(A) &=& \{ b \in \seqs(A) \mid \forall c \in \seqs(A), ~ c \ne b \Rightarrow b \not\in \sub(c) \} \\
             &\subseteq& \seqs(A)
    \end{array} \]
          \qedhere

  \end{proof}
\end{lemma}

\begin{lemma}[Maximal tests contain all tests]
  \label{lem:seqsinmt}
  $\seqs(A) \subseteq \sub(\mt(A))$ for all sets of tests $A$.
  \begin{proof}
    Let an $a \in \seqs(A)$ be given; we must show that $a \in
    \sub(\mt(A))$.
    If $a \in \mt(A)$, then $a \in \sub(\mt(A))$
    (Lemma~\ref{lem:asuba}).
    If $a \not\in \mt(A)$, then there must exist a $b \in \mt(A)$ such
    that $a \in \sub(b)$. But in that case, $a \in \sub(b) \cup
    \bigcup_{a \in \mt(A) \setminus \{ b \}} \sub(\mt(a))$, so $a \in
    \sub(\mt(A))$.       \qedhere

  \end{proof}
\end{lemma}

\begin{lemma}[\seqs distributes over union]
  \label{lem:seqsunion}
  $\seqs(A \cup B) = \seqs(A) \cup \seqs(B)$
  \begin{proof}
    We compute:
    \[ \begin{array}{rcl}
          \seqs(A \cup B) 
      &=& \bigcup_{c \in A \cup B} \seqs(c) \\
      &=& \bigcup_{c \in A} \seqs(c) \cup \bigcup_{c \in B} \seqs(c) \\
      &=& \seqs(A) \cup \seqs(B)
    \end{array} \]
      \qedhere
  \end{proof}
\end{lemma}

\begin{lemma}[\seqs is idempotent]
  \label{lem:seqsidempotent}
  $\seqs(a) = \seqs(\seqs(a))$
  \begin{proof}
    By induction on $a$.
    \begin{proofcases}[leftmargin=3cm]
    \item[($a=b \cdot c$)] We compute:
      \[ \begin{array}{rcl@{\qquad}r}
            \seqs(\seqs(b \cdot c))
        &=& \seqs(\seqs(b) \cup \seqs(c)) & \\
        &=& \seqs(\seqs(b)) \cup \seqs(\seqs(c)) & \text{(by the IH)} \\
        &=& \seqs(b) \cup \seqs(c) & \\
        &=& \seqs(b \cdot c) &
      \end{array} \]
    \item[($a=0,1,\alpha,\neg b,b + c$)] We compute:
      \[ \seqs(a) 
        = \{ a \} 
        = \seqs(a)
        = \bigcup_{a \in \{ a \}} \seqs(a)
        = \seqs(\{ a \}) 
        = \seqs(\seqs(a))
      \]
      \qedhere
    \end{proofcases}
  \end{proof}  
\end{lemma}

We can lift Lemma~\ref{lem:seqsidempotent} to sets of terms,
as well.

\begin{lemma}[Sequence extraction]
  \label{lem:seqsequiv}
  If $\seqs(a) = \{ a_1, \dots, a_k \}$ then $a \equiv a_1 \cdot \ldots
  \cdot a_k$.
  \begin{proof}
    By induction on $a$. The only interesting case is when $a=b \cdot c$.
    \begin{proofcases}
    \item[($a=b \cdot c$)] We have: \[ \{ a_1, \dots, a_k \} =
      \seqs(a) = \seqs(b \cdot c) = \seqs(b) \cup \seqs(c). \]
      Furthermore, $\seqs(b)$ (resp. $\seqs(c)$) is equal to some
      subset of the $a_i \in \seqs(a)$, such that $\seqs(b) \cup
      \seqs(c) = \seqs(a)$.
      By the IH, we know that $b \equiv \Pi_{b_i \in \seqs(b)} b_i$
      and $c \equiv \Pi_{c_i \in \seqs(c)} c_i$, so we have:
      \[ \begin{array}{r@{}lr}
        a \equiv {} & b \cdot c & \\
          \equiv {} & \left( \prod_{b_i \in \seqs(b)} b_i \right) \cdot \left( \prod_{b_i \in \seqs(b)} b_i \right) & \text{(\ax{BA-Seq-Idem})} \\
          \equiv {} & \prod_{a_i \in \seqs(b) \cup \seqs(c)} a_i & \text{(\ax{BA-Seq-Comm})} \\
          \equiv {} & \prod_{i=1}^k a_i &
      \end{array} \]
    \item[($a=0,1,\alpha,\neg b,b+c$)] Immediate by reflexivity, since
      $\seqs(a) = \{ a \}$. \qedhere
    \end{proofcases}
  \end{proof}
\end{lemma}

\begin{corollary}[Maximal tests are invariant over tests]
  \label{cor:mtseqs}
  $\mt(A) = \mt(\seqs(A))$
  \begin{proof}
    We compute:
    \[ \begin{array}{rcl}
      \mt(A) &=& \{ b \in \seqs(A) \mid \forall c \in \seqs(A), c \neq b \Rightarrow b \not\in \sub(c) \} \\
      \multicolumn{3}{r}{\text{(Lemma~\ref{lem:seqsidempotent})}} \\
             &=& \{ b \in \seqs(\seqs(A)) \mid \forall c \in \seqs(\seqs(A)), c \neq b \Rightarrow b \not\in \sub(c) \} \\
             &=& \mt(\seqs(A))
    \end{array} \]
    \qedhere
  \end{proof}
\end{corollary}

\begin{lemma}[Subterms are closed under subterms]
  \label{lem:subsub}
  If $a \in \sub(b)$ then $\sub(a) \subseteq \sub(b)$.
  \begin{proof}
    By induction on $b$, letting some $a \in \sub(b)$ be given.
    \begin{proofcases}
    \item[($b=0$)] We have $\sub(0) = \{ 0 \}$, so it must be that $a
      = 0$ and $\sub(a) = \sub(b)$.
    \item[($b=1$)] We have $\sub(0) = \{ 0, 1 \}$; either $a = 0$ (and
      so $\sub(a) = \{ 0 \} \subseteq \sub(1)$) or $a = 1$ (and so
      $\sub(a) = \sub(b)$).
    \item[($b=\alpha$)] Immediate, since $\sub_\THY$ is well behaved.
    \item[($b=\neg c$)] $a$ is either in $\sub(c)$ or $a = \neg d$ and
      $d \in \sub(c)$. We can use the IH either way.
    \item[($b=c + d$)] We have $\sub(b) = \{c + d\} \cup \sub(c) \cup
      \sub{d}$. If $a$ is in the first set, we have $a = b$ and we're
      done immediately. If $a$ is in the second set, we have $\sub(a)
      \subseteq \sub(c)$ by the IH, and $\sub(c)$ is clearly a subset
      of $\sub(b)$. If $a$ is in the third set, we similarly have
      $\sub(a) \subseteq \sub(d) \subseteq \sub(b)$.
    \item[($b=c \cdot d$)] We have $\sub(b) = \{c \cdot d\} \cup
      \sub(c) \cup \sub{d}$. If $a$ is in the first set, we have $a =
      b$ and we're done immediately. If $a$ is in the second set, we
      have $\sub(a) \subseteq \sub(c)$ by the IH, and $\sub(c)$ is
      clearly a subset of $\sub(b)$. If $a$ is in the third set, we
      similarly have $\sub(a) \subseteq \sub(d) \subseteq \sub(b)$. \qedhere
    \end{proofcases}
  \end{proof}
\end{lemma}

\begin{lemma}[Subterms decrease in size]
  \label{lem:subsize}
  If $a \in \sub(b)$, then either $a \in \{ 0, 1, b \}$ or $a$ comes
  before $b$ in the global well ordering.
  \begin{proof}
    By induction on $b$.
    \begin{proofcases}
    \item[($b=0$)] Immediate, since $\sub(b) = \{ 0 \}$.
    \item[($b=1$)] Immediate, since $\sub(b) = \{ 0, 1 \}$.
    \item[($b=\alpha$)] By the assumption that $\sub_\THY(\alpha)$ is well behaved.
    \item[($b=\neg c$)] Either $a = \neg c$---and we're done
      immediately, or $a \ne \neg c$, so $a$ is a possibly negated
      subterm of $c$. In the latter case, we're done by the IH.
    \item[($b=c + d$)] Either $a = c + d$---and we're done
      immediately, or $a \ne c + d$, and so $a \in \sub(c) \cup
      \sub(d)$. In the latter case, we're done by the IH.
    \item[($b=c \cdot d$)] Either $a = c \cdot d$---and we're done
      immediately, or $a \ne c \cdot d$, and so $a \in \sub(c) \cup
      \sub(d)$. In the latter case, we're done by the IH. \qedhere
    \end{proofcases}
  \end{proof}
\end{lemma}

\begin{lemma}[Maximal tests always exist]
  \label{lem:mtexsts}
  If $A$ is a non-empty set of tests, then $\mt(A) \ne \emptyset$.
  \begin{proof}
    We must show there exists at least one term in $\mt(A)$. 
    
    If $\seqs(A) = \{ a \}$, then $a$ is a maximal test. If $\seqs(A)
    = \{ 0, 1 \}$, then $1$ is a maximal test. If $\seqs(A) = \{ 0, 1,
    \alpha \}$, then $\alpha$ is a maximal test. If $\seqs(A)$ isn't
    any of those, then let $a \seqs A$ be the term that comes last in
    the well ordering on predicates. 

    To see why $a \in mt(A)$, suppose (for a contradiction) we have $b
    \in \mt(A)$ such that $b \ne a$ and $a \in \sub(b)$. By
    Lemma~\ref{lem:subsize}, either $a \in \{ 0, 1, b \}$ or $a$ comes
    before $b$ in the global well ordering. We've ruled out the first
    two cases above. If $a = b$, then we're fine---$a$ is a maximal
    test. But if $a$ comes before $b$ in the well ordering, we've
    reached a contradiction, since we selected $a$ as the term which
    comes \textit{latest} in the well ordering. \qedhere
  \end{proof}
\end{lemma}

As a corollary, note that a maximal test exists even for vacuous
normal forms, where $\mt(x) = \{ 0 \}$ when $x$ is vacuous.

\begin{lemma}[Maximal tests generate subterms]
  \label{lem:submtseqs}
  $\sub(\mt(A)) = \bigcup_{a \in \seqs(A)} \sub(a)$
  \begin{proof}
    Since $\mt(A) \subseteq \seqs(A)$ (Lemma~\ref{lem:mtinseqs}), we
    can restate our goal as:
    \[ \sub(\mt(A)) = \bigcup_{a \in \mt(A)} \sub(a) \cup \bigcup_{a \in \seqs(A) \setminus \mt(A)} \sub(a) \]
    We have $\sub(\mt(A)) = \bigcup_{a \in \mt(A)} \sub(a)$ by
    definition; it remains to see that the latter union is subsumed by
    the former; but we have $\seqs(A) \subseteq \sub(\mt(A))$ by
    Lemma~\ref{lem:seqsinmt}. \qedhere
  \end{proof}
\end{lemma}

\begin{lemma}[Union distributes over maximal tests]
  \label{lem:mtuniondist}
  ~ \\
  $\sub(\mt(A \cup B)) = \sub(\mt(A)) \cup \sub(\mt(B))$
  \begin{proof}
    We compute:
    \[ \begin{array}{rl@{\quad}r}
          \sub(\mt(A \cup B))
      = & \bigcup_{a \in \seqs(A \cup B)} \sub(a) & \text{(Lemma~\ref{lem:submtseqs})} \\
      = & \bigcup_{a \in \seqs(A) \cup \seqs(B)} \sub(a) & \\
      = & \left[ \bigcup_{a \in \seqs(A)} a \right] \cup \left[ \bigcup_{b \in \seqs(B)} \sub(b) \right] & \\
      = & \sub(\mt(A)) \cup \sub(\mt(B)) & \text{(Lemma~\ref{lem:submtseqs})}
    \end{array} \]
        \qedhere    
  \end{proof}
\end{lemma}

\begin{lemma}[Maximal tests are monotonic]
  \label{lem:mtmonotonic}
  If $A \subseteq B$ then $\sub(\mt(A)) \subseteq \sub(\mt(B))$.
  \begin{proof}
    We have $\sub(\mt(B)) = \sub(\mt(A \cup B)) = \sub(\mt(A)) \cup
    \sub(\mt(B))$ (by Lemma~\ref{lem:mtuniondist}). \qedhere
  \end{proof}
\end{lemma}

\begin{corollary}[Sequences of maximal tests]
  \label{cor:seqmt}
  $\sub(\mt(a \cdot b)) = \sub(\mt(a)) \cup \sub(\mt(b))$
  \begin{proof}
    \[ \begin{array}{rl@{\quad}r}
        &  \sub(\mt(c \cdot d)) & \\
      = & \sub(\mt(\seqs(c \cdot d))) & \text{(Corollary~\ref{cor:mtseqs})} \\
      = & \sub(\mt(\seqs(c) \cup \seqs(d))) & \\
      = & \sub(\mt(\seqs(c))) \cup \sub(\mt(\seqs(d))) & \text{(distributivity; Lemma~\ref{lem:mtuniondist})} \\
      = & \sub(\mt(c)) \cup \sub(\mt(d)) & \text{(Corollary~\ref{cor:mtseqs})}
    \end{array} \]
    \qedhere
  \end{proof}
\end{corollary}

\begin{definition}[Negation normal form]
  \label{def:nnf}
  The \textit{negation normal form} of a term $p$ is a term $p'$
  such that $p \equiv p'$ and negations occur only on primitive predicates in $p'$.
\end{definition}

\begin{lemma}[Terms are equivalent to their negation-normal forms]
  \label{lem:nnfsound}
  $\nnf(p) \equiv p$ and $\nnf(p)$ is in negation normal form.
  \begin{proof}
    By induction on the size of $p$. The only interesting case is when $p=\neg a$; we go by cases on $a$.
    \begin{proofcases}
    \item[($p=0$)] Immediate.
    \item[($p=1$)] Immediate.
    \item[($p=\alpha$)] Immediate.
    \item[($p=\pi$)] Immediate.
    \item[($p=\neg a$)] By cases on $a$.
      \begin{proofcases}[leftmargin=0.3cm]
      \item[($a=0$)] We have $\neg 0 \equiv 1$ immediately, and the
        latter is clearly negation free.
      \item[($a=1$)] We have $\neg 1 \equiv 0$; as above.
      \item[($a=\alpha$)] We have $\neg alpha$, which is in normal form.
      \item[($a=b + c$)] We have $\neg (b + c) \equiv \neg b \cdot
        \neg c$ as a consequence of \ax{BA-Excl-Mid} and soundness
        (Theorem~\ref{thm:soundness}). By the IH on $\neg b$ and $\neg
        c$, we find that $\nnf(\neg b) \equiv \neg b$ and $\nnf(\neg
        c) \equiv \neg c$---where the left-hand sides are negation
        normal. So transitively, we have $\neg (b + c) \equiv \nnf(\neg
        b) \cdot \nnf(\neg c)$, and the latter is negation normal.
      \item[($a=b \cdot c$)] We have $\neg (b \cdot c) \equiv \neg b +
        \neg c$ as a consequence of \ax{BA-Excl-Mid} and soundness
        (Theorem~\ref{thm:soundness}). By the IH on $\neg b$ and $\neg
        c$, we find that $\nnf(\neg b) \equiv \neg b$ and $\nnf(\neg
        c) \equiv \neg c$---where the left-hand sides are negation
        normal. So transitively, we have $\neg (b \cdot c) \equiv
        \nnf(\neg b) + \nnf(\neg c)$, and the latter is negation normal.
      \end{proofcases}
    \item[($p=q + r$)] By the IHs on $q$ and $r$.
    \item[($p=q \cdot r$)] By the IHs on $q$ and $r$.
    \item[($p=q^*$)] By the IH on $q$. \qedhere
    \end{proofcases}
  \end{proof}
\end{lemma}

\begin{lemma}[Negation normal form is monotonic]
  \label{lem:nnfmonotonic}
  If $a \preceq b$ then $\nnf(\neg a) \preceq \neg b$.
  \begin{proof}
    By induction on $a$.
    \begin{proofcases}
    \item[($a=0$)] We have $\nnf(\neg 0) = 1$ and $1 \preceq \neg b$ by definition.
    \item[($a=1$)] We have $\nnf(\neg 1) = 0$ and $0 \preceq \neg b$ by definition.
    \item[($a=\alpha$)] We have $\nnf(\neg \alpha) = \neg \alpha$;
      since $a \preceq b$, it must be that $\alpha \in \sub(\mt(b))$,
      so $\neg \alpha \in \sub(\mt(\neg b))$.  We have $\alpha \in \sub(\neg b)$, since
      $\alpha \in \sub(b)$.
    \item[($a=\neg c$)] We have $\nnf(\neg \neg c) = \nnf(c)$; since
      $c \in \sub(a)$ and $a \preceq b$, it must be that $c \in
      \sub(\mt(b))$, so $\nnf(c) \preceq \neg b$ by the IH.
    \item[($a=c + d$)] We have $\nnf(\neg (c + d)) = \nnf(\neg c)
      \cdot \nnf(\neg d)$; since $c$ and $d$ are subterms of $a$ and
      $a \preceq b$, $\neg c$ and $\neg d$ must be in $\sub(\mt(\neg
      b))$, and we are done by the IHs.
    \item[($a=c \cdot d$)] We have $\nnf(\neg (c \cdot d)) = \nnf(\neg
      c) + \nnf(\neg d)$; since $c$ and $d$ are subterms of $a$ and $a
      \preceq b$, $\neg c$ and $\neg d$ must be in $\sub(\mt(\neg
      b))$, and we are done by the IHs. \qedhere
    \end{proofcases}
  \end{proof}
\end{lemma}

\begin{lemma}[Normal form ordering]
  \label{lem:nforder}
  For all tests $a, b, c$ and normal forms $x, y, z$, the following
  inequalities hold:
  \begin{enumerate}
  \item $a \preceq a \cdot b$ (extension);
  \item if $a \in \tests(x)$, then $a \preceq x$ (subsumption);
  \item \label{nforder:nftests} $x \approx \sum_{a \in \tests(x)} a$ (equivalence);
  \item if $x \preceq x'$ and $y \preceq y'$, then $x + y \preceq x' + y'$ (normal-form parallel congruence);
  \item if $x + y \preceq z$, then $x \preceq z$ and $y \preceq z$ (inversion);
  \item if $a \preceq a'$ and $b \preceq b'$, then $a \cdot b \preceq a' \cdot b'$ (test sequence congruence);
  \item if $a \preceq x$ and $b \preceq x$ then $a \cdot b \preceq x$ (test bounding);
  \item if $a \preceq b$ and $x \preceq c$ then $a \cdot x \preceq b \cdot c$ (mixed sequence congruence);
  \item if $a \preceq b$ then $\nnf(\neg a) \preceq \neg b$ (negation normal-form monotonic).
  \end{enumerate}
  Each of the above equalities also hold replacing $\preceq$ with
  $\prec$, excluding the equivalence (\ref{nforder:nftests}).
  {\iffull
  \begin{proof}
    We prove each properly independently and in turn.
    Each property can be proved using the foregoing lemmas and set-theoretic reasoning.
    \begin{enumerate}
    \item We must show that $a \preceq a \cdot b$ (extension); we compute:
      \[ \begin{array}{rl@{\quad}r}  
          \sub(\mt(a)) 
        = & \sub(\mt(\seqs(a)) & \text{(Corollary~\ref{cor:mtseqs})} \\
        \subseteq & \sub(\mt(\seqs(a))) \cup \sub(\mt(\seqs(b))) & \\
        = & \sub(\mt(\seqs(a) \cup \seqs(b))) & \\
        \multicolumn{3}{r}{\text{(distributivity; Lemma~\ref{lem:mtuniondist})}} \\
        = & \sub(\mt(\seqs(a \cdot b))) \\
        = & \sub(\mt(a \cdot b)) & \text{(Corollary~\ref{cor:mtseqs})}
      \end{array} \]

    \item We must show that if $a \in \tests(x)$, then $a \preceq x$
      (subsumption). We have $\sub(\mt(\{a\})) \subseteq
      \sub(\mt(\tests(x)))$ by monotonicity (Lemma~\ref{lem:mtmonotonic})
      immediately.

    \item We must show that $x \approx \sum_{a \in \tests(x)} a$
      (equivalence). Let $x = \sum a_i \cdot m_i$, and recall that
      $\sum_{a in \tests(x)} a$ really denotes the normal form
      $\sum_{a \in \tests(x)} a \cdot 1$. We compute:
      \[ \begin{array}{rl}
            \sub(\mt(x)) 
        = & \sub(\mt(\tests(x))) \\
        = & \sub(\mt(\{ a_i \})) \\
        = & \sub(\mt(\bigcup_{a \in \tests(x)} a)) \\
        = & \sub(\mt(\tests(\sum_{a \in tests(x)} a \cdot 1))) \\
        = & \sub(\mt(\sum_{a \in tests(x)} a)) \\
      \end{array} \]

    \item We must show that if $x \preceq x'$ and $y \preceq y'$, then
      $x + y \preceq x' + y'$ (normal-form parallel congruence).
      Unfolding definitions, we find $\sub(\mt(x)) \subseteq
      \sub(\mt(x'))$ and $\sub(\mt(y)) \subseteq \sub(\mt(y'))$.
      We compute:
      \[ \begin{array}{rl@{\quad}r}
          &  \sub(\mt(x + y)) \\
        = & \sub(\mt(\tests(x + y))) & \\
        = & \sub(\mt(\tests(x) \cup \tests(y))) & \\
        = & \sub(\mt(\tests(x))) \cup \sub(\mt(\tests(y))) & \text{(distributivity; Lemma~\ref{lem:mtuniondist})} \\
        \subseteq & \sub(\mt(\tests(x'))) \cup \sub(\mt(\tests(y'))) & \text{(assumptions)} \\
        = & \sub(\mt(\tests(x') \cup \tests(y'))) & \text{(distributivity; Lemma~\ref{lem:mtuniondist})} \\
        = & \sub(\mt(x' + y'))
      \end{array} \] 

    \item We must show that if $x + y \preceq z$, then $x \preceq z$
      and $y \preceq z$ (inversion).  We have $\sub(\mt(x + y)) =
      \sub(\mt(x)) \cup \sub(\mt(y))$ by distributivity
      (Lemma~\ref{lem:mtuniondist}). Since we've assumed $\sub(\mt(x +
      y)) \subseteq \sub(\mt(z))$, we must have $\sub(\mt(x))
      \subseteq \sub(\mt(z))$ (and similarly for $y$).

    \item We must show that if $a \preceq a'$ and $b \preceq b'$, then
      $a \cdot b \preceq a' \cdot b'$ (test sequence congruence).
      Unfolding our assumptions, we have $\sub(\mt(a)) \subseteq
      \sub(\mt(a'))$ and $\sub(\mt(b)) \subseteq \sub(\mt(b'))$.
      We can compute:
      \[ \begin{array}{rl@{\quad}{r}}
            \sub(\mt(a \cdot b))
        = & \sub(\mt(a)) \cup \sub(\mt(b)) & \text{(Corollary~\ref{cor:seqmt})} \\
        \subseteq & \sub(\mt(a')) \cup \sub(\mt(b')) & \\
        = & \sub(\mt(a' \cdot b')) & \text{(Corollary~\ref{cor:seqmt})}
      \end{array} \]      

    \item We must show that if $a \preceq x$ and $b \preceq x$ then $a
      \cdot b \preceq x$ (test bounding). Immediate by
      Corollary~\ref{cor:seqmt}.

    \item We must show that if $a \preceq b$ and $x \preceq c$ then $a
      \cdot x \preceq b \cdot c$ (mixed sequence congruence).
      We compute:
      \[ \begin{array}{rl@{\quad}r}
            \sub(\mt(a \cdot x))
        = & \sub(\mt(\tests(\sum a \cdot a_i \cdot m_i))) & \\
        = & \sub(\mt(\{ a \} \cup \{ a_i \})) & \\
        = & \sub(\mt(a)) \cup \sub(\mt(x)) & \text{(Corollary~\ref{cor:seqmt})} \\
        \subseteq & \sub(\mt(b)) \cup \sub(\mt(c)) & \\
        = & \sub(\mt(b \cdot c)) &  \text{(Corollary~\ref{cor:seqmt})}
      \end{array} \]

    \item A restatement of Lemma~\ref{lem:nnfmonotonic}. \qedhere
    \end{enumerate}
  \end{proof}
  \fi}
\end{lemma}

\begin{lemma}[Test sequence split]
  \label{lem:testsplit}
  If $a \in \mt(c)$ then $c \equiv a \cdot b$ for some $b \prec c$.
  \begin{proof}
    We have $a \in \seqs(c)$ by definition. Suppose $\seqs(c) = \{ a,
    c_1, \dots, c_k \}$. By sequence extraction, we have $c \equiv a
    \cdot c_1 \cdot \dots \cdot c_k$ (Lemma~\ref{lem:seqsequiv}).  So
    let $b = c_1 \cdot \dots \cdot c_k$; we must show $b \prec c$,
    i.e., $\sub(\mt(b)) \subsetneq \sub(\mt(c))$. Note that $\{ c_1,
    \dots, c_k \} = \seqs(b)$. We find:
    \[ \begin{array}{rcl}
      \sub(\mt(b)) &\subsetneq& \sub(\mt(c)) \\
      & \Updownarrow & \multicolumn{1}{r}{\text{(Corollary~\ref{cor:mtseqs})}} \\
      \sub(\mt(\seqs(b))) &\subsetneq& \sub(\mt(\seqs(c))) \\
      & \Updownarrow & \\
      \sub(\mt(\{c_1, \dots, c_k \})) &\subsetneq & \sub(\mt(\{a, c_1, \dots, c_k \})) \\
      & \Updownarrow & \multicolumn{1}{r}{\text{\qquad\qquad\qquad\qquad (distributivity; Lemma~\ref{lem:mtuniondist})}} \\ 
      \bigcup_{i=1}^k \sub(\mt(\{c_i\})) &\subsetneq & \sub(\mt(a)) \cup \bigcup_{i=1}^k \sub(\mt(\{c_i\}))
    \end{array} \]
    Since $a \in \mt(c)$, we know that $a \not\in
    \sub(\mt(c_i))$ for all $i$. But terms are subterms of themselves
    (Lemma~\ref{lem:asuba}), so $a \in \sub(a) = \sub(\mt(a))$. \qedhere
  \end{proof}
\end{lemma}

\begin{lemma}[Maximal test inequality]
  \label{lem:mtprec}
  If $a \in \mt(y)$ and $x \preceq y$ then either $a \in \mt(x)$ or $x
  \prec y$.
  \begin{proof}
    Since $a \in \mt(y)$, we have $a \in \sub(\mt(y))$. Since $x
    \preceq y$, we know that $\sub(\mt(x)) \subseteq \sub(\mt(y))$. We
    go by cases on whether or not $a \in \mt(x)$:
    \begin{proofcases}
    \item[($a \in \mt(x)$)] We are done immediately.
    \item[($a \not\in \mt(x)$)] In this case, we show that $a \not\in
      \sub(\mt(x))$ and therefore $x \prec y$. Suppose, for a
      contradiction, that $a \in \sub(\mt(x))$. Since $a \not\in
      \mt(x)$, there must exist some $b \in \sub(\mt(x))$ where $a \in
      \sub(b)$. But since $x \preceq y$, we must also have $b \in
      \sub(\mt(y))$... and so it couldn't be that case that $a \in
      \mt(y))$. We can conclude that it must, then, be the case that
      $a \not\in \sub(\mt(x))$ and so $x \prec y$. \qedhere
    \end{proofcases}
  \end{proof}
\end{lemma}

\begin{lemma}[Splitting]
  \label{fulllem:splitting}
  If $a \in \mt(x)$, then there exist $y$ and $z$ such that $x \equiv
  a \cdot y + z$ and $y \prec x$ and $z \prec x$.
  {\iffull
  \begin{proof}
    Suppose $x = \sum_{i=1}^k c_i \cdot m_i$. We have $a \in \mt(x)$,
    so, in particular:
    \[ a \in \seqs(\tests(x)) 
    = \seqs(\tests(\sum_{i=1}^k c_i \cdot m_i)) 
    = \seqs(\{ c_1,\dots,c_k \}) 
    = \bigcup_{i=1}^k \seqs(c_i). \]
    That is, $a \in \seqs(c_i)$ for at least one $i$. We can, without
    loss of generality, rearrange $x$ into two sums, where the first
    $j$ elements have $a$ in them but the rest don't, i.e., $x \equiv
    \sum_{i=1}^j c_i \cdot m_i + \sum_{i=j+1}^k c_i \cdot m_i$ where
    $a \in \seqs(c_i)$ for $1 \le i \le j$ but $a \not\in \seqs(c_i)$
    for $j + 1 \le i \le k$.
    By subsumption (Lemma~\ref{lem:nforder}), we have $c_i \preceq x$.
    Since $a \in \mt(x)$, it must be that $a \in \mt(c_i)$ for $1 \le
    i \le j$ (instantiating Lemma~\ref{lem:mtprec} with the normal
    form $c_i \cdot 1$).
    By test sequence splitting (Lemma~\ref{lem:testsplit}), we find
    that $c_i \equiv a \cdot b_i$ with $b_i \prec c_i \preceq x$ for
    $1 \le i \le j$, as well.

    We are finally ready to produce $y$ and $z$: they are the first
    $j$ tests with $a$ removed and the remaining tests which never had
    $a$, respectively. Formally, let $y = \sum_{i=1}^j b_i \cdot m_i$;
    we immediately have that $a \cdot y \equiv \sum_{i=1}^j c_i \cdot
    m_i$; let $z = \sum_{i=j+1}^k c_i \cdot m_i$. We can conclude that
    $x \equiv a \cdot y + z$.

    It remains to be seen that $y \prec x$ and $z \prec x$. The
    argument is the same for both; presenting it for $y$, we have $a
    \not\in \seqs(y)$ (because of sequence splitting), so $a \not\in
    \sub(\mt(y))$. But we assumed $a \in \mt(x)$, so $a \in
    \sub(\mt(x))$, and therefore $y \prec x$. The argument for $z$ is
    nearly identical but needs no recourse to sequence splitting---we
    never had any $a \in \seqs(c_i)$ for $j+1 \le i \le k$. \qedhere
  \end{proof}
  \fi}
\end{lemma}

To elucidate the way \PBdot handles structure, suppose we have the term $(\pi_1 + \pi_2) \cdot (\alpha_1 + \alpha_2)$. One of two rules could apply: we could split up the tests and push them through individually (\rn{SeqParTest}), or we could split up the actions and push the tests through together (\rn{SeqParAction}). It doesn't particularly matter which we do first: the next step will almost certainly be the other rule, and in any case the results will be equivalent from the perspective of our equational theory.
It \textit{could} be the case that choosing a one rule over another could give us a smaller term, which might yield a more efficient normalization procedure. Similarly, a given normal form may have more than one maximal test---and therefore be splittable in more than one way (Lemma~\ref{lem:splitting})---and it may be that different splits produce more or less efficient terms. We haven't yet studied differing strategies for pushback.

\begin{lemma}[Sliding]
  \label{lem:katsliding}
  $p \cdot (q \cdot p)^* \equiv (p \cdot q)^* \cdot p$.
  \begin{proof} Following Kozen~\cite{DBLP:journals/iandc/Kozen94}, as
    a corollary of a related result: if $p \cdot x \equiv x \cdot q$ then $p^* \cdot x \equiv x \cdot q^*$.
    We prove this separate property by mutual inclusion.

    \begin{proofcases}[leftmargin=1cm]

    \item[($\Rightarrow$)]
      We use \ax{KA-LFP-L} with $p = p$ and $q = x$ and $r = x \cdot q^*$. We must show
      that $x + p \cdot x \cdot q^* \le x \cdot q^*$ to find $p^* \cdot x \le x \cdot q^*$.

      If $p \cdot q \le x \cdot q$ then $p \cdot x \cdot
      q^* \le x \cdot q \cdot q^*$ by monotonicity.
      We have $x + x \cdot q \cdot q^* \le x \cdot q^*$ by \ax{KA-Unroll-L} and \ax{KA-Plus-Idem}.
      Therefore $x + p \cdot x \cdot q^* \le x + x \cdot q \cdot
      q^* \le x \cdot q^*$, as desired.

    \item[($\Leftarrow$)] This case is symmetric to the first,
      using \ax{-R} rules instead of \ax{-L} rules.
      We apply \ax{KA-LFP-R} with $p = x$ and $r = q$ and $q =
      p^* \cdot x$. We must show $x + p^* \cdot x \cdot q \le
      p^* \cdot x$ to find $x \cdot q^* \le p^* \cdot x$.

      If $x \cdot q \le p \cdot x$, then $p^* \cdot x \cdot q \le
      p^* \cdot p \cdot x$ by monotonicity.
      We have $x + p^* \cdot p \cdot x \le p^* \cdot x$ by \ax{KA-Unroll-R} and \ax{KA-Plus-Idem}.
      Therefore $x + p^* \cdot x \cdot q \le x + p^* \cdot p \cdot
       x \le p^* \cdot x$, as desired.
     \end{proofcases}

    We can now find sliding by letting $p = p \cdot q$ and $x = p$ and
    $q = q \cdot p$ in the above, i.e., we have the premise $p \cdot
    q \cdot p \equiv p \cdot q \cdot p$ by reflexivity, and so
    $(p \cdot q)^* \cdot p \equiv p \cdot (q \cdot p)^*$. \qedhere
    \end{proof}
\end{lemma}

\begin{lemma}[Denesting]
  \label{lem:katdenesting}
  $(p + q)^* \equiv p^* \cdot (q \cdot p^*)^*$.
  \begin{proof} Following Kozen~\cite{DBLP:journals/iandc/Kozen94}, we
    do the proof by mutual inclusion. The proof is surprisingly
    challenging, so we include it here.
  \begin{proofcases}[leftmargin=1cm]

  \item[($\Rightarrow$)] To show $(p + q)^* \le a^* \cdot (b \cdot
  a^*)^*$, we apply induction with $q=1$ and $r=p^* \cdot (q \cdot
  p^*)^*$ (to show $(a+b)^* \cdot 1 \le r$).  We must show that $1 +
  (p+q) \cdot p^*\cdot(q \cdot p^*)^* \le p^* \cdot (q \cdot
  p^*)^*$. We do so in several parts, working our way there in five steps.

  First, we observe that $1 \le p^* \cdot (q \cdot p^*)^*$ (A) because:
  \[ \begin{array}{rlr}
             & 1 + p^* \cdot (q \cdot p^*)^* & \\
      \equiv & 1 + (1 + p \cdot p^* \cdot (q \cdot p^*)^*) & \ax{KA-Unroll-L} \\
      \equiv & 1 + p \cdot p^* \cdot (q \cdot p^*)^* & \ax{KA-Plus-Assoc}, \ax{KA-Plus-Idem} \\
      \equiv & p^* \cdot (q \cdot p^*)^* & \ax{KA-Unroll-L} \\
     \end{array} \]

  Next, $p \cdot p^* \cdot (q \cdot p^*)^* \le p^* \cdot (q \cdot p^*)^*$ (B) because:
  \[ \begin{array}{rlr}
            & p \cdot p^* \cdot (q \cdot p^*)^* + p^* \cdot (q \cdot p^*)^* & \\
     \equiv & p \cdot p^* \cdot (q \cdot p^*)^* + 1 + p \cdot p^* \cdot (q \cdot p^*)^* & \ax{KA-Unroll-L} \\
     \equiv & 1 + p \cdot p^* \cdot (q \cdot p^*)^* &  \ax{KA-Plus-Idem}  \\ 
     \equiv & p^* \cdot (q \cdot p^*)^* & \ax{KA-Unroll-L} \\
     \end{array} \]

  We have $q \cdot p^* \cdot (q \cdot p^*)^* \le (q \cdot p^*)^*$ because:
  \[ \begin{array}{rlr}
            & q \cdot p^* \cdot (q \cdot p^*)^* + (q \cdot p^*)^* & \\
     \equiv & q \cdot p^* \cdot (q \cdot p^*)^* + 1 + q \cdot p^* \cdot (q \cdot p^*)^* & \ax{KA-Unroll-L}\\
     \equiv & 1 + q \cdot p^* \cdot (q \cdot p^*)^* & \ax{KA-Plus-Idem} \\
     \equiv & (q \cdot p^*)^* & \ax{KA-Unroll-L} \\
     \end{array} \]

  Further, $(q \cdot p^*)^* \le p^* \cdot (q \cdot p^*)^*$ because:
  \[ \begin{array}{rlr}
            & (q \cdot p^*)^* + p^* \cdot (q \cdot p^*)^* & \\
     \equiv & (q \cdot p^*)^* + 1 \cdot (q \cdot p^*)^* + p \cdot p^* \cdot (q \cdot p^*)^* & \ax{KA-Unroll-L}, \ax{KA-Dist-R} \\
     \equiv & 1 \cdot (q \cdot p^*)^* + p \cdot p^* \cdot (q \cdot p^*)^* & \ax{KA-Plus-Idem} \\
     \equiv & p^* \cdot (q \cdot p^*)^* & \ax{KA-Unroll-L} \\
     \end{array} \]

  Finally, $q \cdot p^* \cdot (q \cdot p^*)^* \le a^* \cdot (q \cdot
  p^*)^*$ (C) by transitivity with the last two results.

  Now we can find that
  \[ 1 + (p + q)p^* \cdot (q \cdot p^*)^* \le
     1 + p \cdot p^* \cdot (q \cdot p^*)^* + q \cdot p^* \cdot (q \cdot p^*)^* \le
     p^* \cdot (q \cdot p^*)^* \]
  because:
  \[ \begin{array}{rlr}
             & 1 + (p + q)p^* \cdot (q \cdot p^*)^* + 1 + p \cdot p^* \cdot (q \cdot p^*)^* + q \cdot p^* \cdot (q \cdot p^*)^* & \\
      \equiv & 1 + p \cdot p^* \cdot (q \cdot p^*)^* + q \cdot p^* \cdot (q \cdot p^*)^* + p \cdot p^* \cdot (q \cdot p^*)^* + q \cdot p^* \cdot (q \cdot p^*)^* & \ax{KA-Plus-Idem} \\
      \equiv & 1 + p \cdot p^* \cdot (q \cdot p^*)^* + q \cdot p^* \cdot (q \cdot p^*)^* & \ax{KA-Plus-Idem} \\
     \end{array} \]
  because, finally:
  \[ \begin{array}{rlr}
             & 1 + p \cdot p^* \cdot (q \cdot p^*)^* + q \cdot p^* \cdot (q \cdot p^*)^* + p^* \cdot (q \cdot p^*)^* & \\
      \equiv & p^* \cdot (q \cdot p^*)^* + p \cdot p^* \cdot (q \cdot p^*)^* + q \cdot p^* \cdot (q \cdot p^*)^* & \text{(A)} \\
      \equiv & p^* \cdot (q \cdot p^*)^* + q \cdot p^* \cdot (q \cdot p^*)^* & \text{(B)} \\
      \equiv & p^* \cdot (q \cdot p^*)^* & \text{(C)} \\
     \end{array} \]

   \item[($\Leftarrow$)] To show $p^* \cdot (q \cdot p^*)^* \le (p +
   q)^*((p + q)(p + q)^*)^*$, we have first that $p \le p + q$ and
   $q \le p + q$, and so $p + q \le (p + q)^*$.
   And so, by monotonicity $p^* \cdot (q \cdot p^*)^* \le (p + q)^*((p + q)(p + q)^*)^*$.
   We can then find that $(p + q)^*\cdot ((p + q) \cdot (p + q)^*)^* \le (p + q)^* \cdot ((p + q)^*)^*$
   because:
   \[ \begin{array}{rlr}
              & (p + q)(p + q)^* + (p + q)^* & \\
       \equiv & p \cdot (p + q)^* + q \cdot (p + q)^* + (p + q)^* & \ax{KA-Dist-R} \\
       \equiv & p \cdot (p + q)^* + q \cdot (p + q)^* + 1 + (p + q)(p + q)^* & \ax{KA-Unroll-L} \\
       \equiv & p \cdot (p + q)^* + q \cdot (p + q)^* + 1 + p \cdot (p + q)^* + q \cdot (p + q)^* & \ax{KA-Dist-R} \\
       \equiv & 1 + p \cdot (p + q)^* + q \cdot (p + q)^* & \ax{KA-Plus-Idem} \\
       \equiv & 1 + (p + q) \cdot (p + q)^* & \ax{KA-Dist-R} \\
       \equiv & (p + q)^* & \ax{KA-Unroll-L} \\
      \end{array} \]
    But we also have $(p + q)^* \cdot ((p + q)^*)^* \le (p + q)^*$
    because:
    \[ \begin{array}{rlr}
               & (p + q)^* \cdot ((p + q)^*)^* + (p + q)^* & \\
        \equiv & (p + q)^* \cdot (p + q)^* + (p + q)^* & \text{because $(x^*)^* = x^*$} \\
        \equiv & (p + q)^* + (p + q)^* & \text{because $x^*x^* = x^*$} \\
        \equiv & (p + q)^* & \ax{KA-Plus-Idem} \\
        \end{array} \]
        \qedhere
    \end{proofcases}
  \end{proof}
\end{lemma}

\begin{lemma}[Star invariant]
  \label{lem:katstarinv}
  If $p \cdot a \equiv a \cdot q + r$ then $p^* \cdot a \equiv (a +
  p^* \cdot r) \cdot q^*$.
  \begin{proof}
    We show two implications using $\le$ to derive the equality.

    \begin{proofcases}[leftmargin=1cm]
      \item[($\Rightarrow$)] We want to show $p^*;a \le (a + p^*;y);x^*$.

        We know that $q + pr \le r \implies p^*q \le r$ by the
        induction axiom \ax{KA-LFP-L}, so we can instantiate it with
        $p$ as $p$ and $q$ as $a$ and $r$ as $(a + p^*;y);x^*$. We find:
        \[ \begin{array}{rcl}
          a + p;(a + p^*;y);x^* &\le& (a + p^*;y);x^* \\
          a + p;a;x^* + p;p^*;y;x^* &\le& (a + p^*;y);x^* \\
          a + p;a;x^* + p;p^*;y;x^* + (a + p^*;y);x^* &=& (a + p^*;y);x^* \\
          a + p;a;x^* + p;p^*;y;x^* + a;x^* + p^*;y;x^* &=& (a + p^*;y);x^* \\
          (a + a;x^* + p;a;x^*) + (p;p^*;y;x^* + p^*;y;x^*) &=& (a + p^*;y);x^* \\
          (a;x^* + p;a;x^*) + (p;p^*;y;x^* + p^*;y;x^*) &=& (a + p^*;y);x^* \\
          (1+p);a;x^* + (1+p);p^*;y;x^* &=& (a + p^*;y);x^* \\
          a;x^* + p^*;y;x^* &=& (a + p^*;y);x^* \\
          (a + p^*;y);x^* &=& (a + p^*;y);x^* \\
        \end{array} \]

      \item[($\Leftarrow$)] We can to show $(a + p^*;y);x^* \le p^*;a$
        We can apply the other induction axiom (\ax{KA-LFP-R}), $q + r;p
        \le r \implies q;p^* \le r$, with $p = x$ and $q = (a + p^*;y)$
        and $r = p^*;a$.
        We find:
        \[ \begin{array}{rcl}
          (a + p^*;y) + (p^*;a);x &\le& p^*;a \\
          a + p^*;y + p^*;a;x + p^*;a &=& p^*;a \\
          a + p^*;(a;x + y + a) &=& p^*;a \\
          a + p^*;(p;a + a) &=& p^*;a \\
          a + p^*;(a;(p + 1)) &=& p^*;a \\
          a + p^*;a &=& p^*;a \\
          p^*;a &=& p^*;a \\
        \end{array} \]
        \qedhere
    \end{proofcases}
  \end{proof}
\end{lemma}

\begin{lemma}[Star expansion]
  \label{lem:katexpand}
  If $p \cdot a \equiv a \cdot q + r$ then $p \cdot a \cdot (p \cdot
  a)^* \equiv (a \cdot q + r) \cdot (q + r)^*$.
  \begin{proof}
    First we observe that $p;a;(p;a)^*$ is equivalent to $(p;a)^*;p;a$
    (apply \ax{KA-SLIDING} twice).
    We show two implications using $\le$ to derive the equality.

    \begin{proofcases}[leftmargin=1cm]
      \item[($\Rightarrow$)] We want to show $(p;a)^*;p;a \le (a;x +
        y);(x+y)^*$.

        We know that $q + pr \le r \implies p^*q \le r$ by the
        induction axiom \ax{KA-LFP-L}, so we can instantiate it with
        $p$ and $q$ as $p;a$ and $r$ as $(a;x + y);(x+y)^*$. We find:
        \[ \begin{array}{rcl}
          p;a + p;a;(a;x + y);(x+y)^* &\le& (a;x + y);(x+y)^* \\
          p;a + (p;a;x + p;a;y);(x+y)^* &\le& (a;x + y);(x+y)^* \\
          (a;x + y) + ((a;x + y);x + (a;x + y);y);(x+y)^* &\le& (a;x + y);(x+y)^* \\
          (a;x + y) + (a;x + y);(x+y);(x+y)^* &\le& (a;x + y);(x+y)^* \\
          (a;x + y);(1 + (x+y);(x+y)^*) &\le& (a;x + y);(x+y)^* \\
          (a;x + y);(x+y)^* &\le& (a;x + y);(x+y)^*
        \end{array} \]

      \item[($\Leftarrow$))] We can to show $(a;x + y);(x+y)^* \le p;a;(p;a)^*$
        We can apply the other induction axiom (\ax{KA-LFP-R}), $q + r;p
        \le r \implies q;p^* \le r$, with $p = x+y$ and $q = a;x + y$
        and $r = p;a(p;a)^*$.
        We find:
        \[ \begin{array}{rcl}
          (a;x + y) + p;a;(p;a)^*;(x+y) &\le& (p;a)^*;p;a \\
          p;a + p;a;(p;a)^*;(x+y) &\le& (p;a)^*;p;a \\
          p;a + p;a;(p;a)^*;(x+y) &\le& p;a + (p;a)^*;p;a;p;a \\
          p;a + p;a;(p;a)^*;(x+y) &\le& p;a + (p;a)^*;p;a;(a;x + y) \\
          p;a + p;a;(p;a)^*;(x+y) &\le& p;a + (p;a)^*;(a;x + y);(x+y) \\
          p;a + p;a;(p;a)^*;(x+y) &\le& p;a + (p;a)^*;(p;a);(x+y)
        \end{array} \]
        \qedhere
    \end{proofcases}

  \end{proof}
\end{lemma}

\begin{lemma}[Pushback through primitive actions]
  \label{lem:pbpi}
  Pushing a test back through a primitive action leaves the primitive
  action intact, i.e., if $\pi \cdot a \PBdot x$ or $(\sum b_i \cdot
  \pi) \cdot a \PBT x$, then $x = \sum a_i \cdot \pi$.
  \begin{proof}
    By induction on the derivation rule used.
    \begin{proofcases}
    \item[(\rn{SeqZero})] Immediate---$x$ is the empty sum.
    \item[(\rn{SeqOne})] By definition.
    \item[(\rn{SeqSeqTest})] By the IHs.
    \item[(\rn{SeqSeqAction})] Contradictory---$m \cdot n$ isn't primitive.
    \item[(\rn{SeqParTest})] By the IHs.
    \item[(\rn{SeqParAction})] Contradictory---$m + n$ isn't primitive.
    \item[(\rn{Prim})] By definition.
    \item[(\rn{PrimNeg})] By definition.
    \item[(\rn{SeqStarSmaller})] Contradictory---$m^*$ isn't primitive.
    \item[(\rn{SeqStarInv})] Contradictory---$m^*$ isn't primitive.
    \item[(\rn{Test})] By the IH. \qedhere
    \end{proofcases}
  \end{proof}
\end{lemma}

\begin{theorem}[Pushback soundness]
  \label{fullthm:pushbacksound}
  {\iffull
  ~ \\
  \begin{enumerate}
  \item \label{pbj} If $x \cdot y \PBJ z'$ then $x \cdot y \equiv z'$.
  \item \label{pbstar} If $x^* \PBstar y$ then $x^* \equiv y$.
  \item \label{pbdot} If $m \cdot a \PBdot y$ then $m \cdot a \equiv y$.
  \item \label{pbr} If $m \cdot x \PBR y$ then $m \cdot x \equiv y$.
  \item \label{pbt} If $x \cdot a \PBT y$ then $x \cdot a \equiv y$.
  \end{enumerate}
  \else Each of \PB relations' left is equivalent to its
  right, e.g., if $x^* \PBstar y$ then $x^* \equiv y$. \fi}
  \begin{proof}
    By simultaneous induction on the derivations. 
    {\iffull Cases are grouped by judgment. \else Most cases follow by the IH and axioms, with a few relying on KAT theorems like sliding, denesting, star expansion~\cite{Beckett:2016:TN:2908080.2908108}, and pushback negation \iffull\ (Lemma~\ref{lem:pushbackneg})\else (Fig.~\ref{fig:semantics}, Consequences)\fi. \qedhere \fi}

    {\iffull    
    
    \paragraph*{Sequential composition of normal forms ($x \cdot y \PBJ z$)}
    \begin{proofcases}
    \item[(\rn{Join})] We have $x = \sum_{i=1}^k a_i \cdot m_i$ and $y
      = \sum_{j=1}^l b_j \cdot n_j$.  By the IH on (\ref{pbdot}), each $m_i \cdot b_j
      \PBdot x_{ij}$. 
      We compute:
      \[ \begin{array}{rl@{\qquad}r}
               & x \cdot y & \\
        \equiv & \left[ \sum_{i=1}^k a_i \cdot m_i \right] \cdot \left[ \sum_{j=1}^l b_j \cdot n_j \right] &  \\
        \equiv & \sum_{i=1}^k a_i \cdot m_i \cdot \left[ \sum_{j=1}^l b_j \cdot n_j \right] & \byax{KA-Dist-R} \\
        \equiv & \sum_{i=1}^k a_i \cdot \left[ m_i \cdot \sum_{j=1}^l b_j \cdot n_j \right] & \byax{KA-Seq-Assoc} \\
        \equiv & \sum_{i=1}^k a_i \cdot \left[ \sum_{j=1}^l m_i \cdot  b_j \cdot n_j \right] & \byax{KA-Dist-L} \\
        \equiv & \sum_{i=1}^k a_i \cdot \left[ \sum_{j=1}^l x_{ij} \cdot n_j \right] & \by{IH (\ref{pbdot})} \\
        \equiv & \sum_{i=1}^k \sum_{j=1}^l a_i \cdot x_{ij} \cdot n_j & \byax{KA-Dist-L} \\
      \end{array} \]
    \end{proofcases}

    \paragraph*{Kleene star of normal forms ($x^* \PBJ y$)}
    \begin{proofcases}
    \item[(\rn{StarZero})] We have $0^* \PBstar 1$. We compute:
      \[ \begin{array}{rl@{\qquad}r}
               & 0^* & \\
        \equiv & 1 + 0 \cdot 0^* & \byax{KA-Unroll-L} \\
        \equiv & 1 + 0 & \byax{KA-Zero-Seq} \\
        \equiv & 1 & \byax{KA-Plus-Zero}
      \end{array} \]

    \item[(\rn{Slide})] We are trying to pushback the minimal term $a$
      of $x$ through a star, i.e., we have $(a \cdot x)^*$; by the IH
      on (\ref{pbt}), we know there exists some $y$ such that $x \cdot
      a \equiv y$; by the IH on (\ref{pbstar}), we know that $y^*
      \equiv y'$; and by the IH on (\ref{pbj}), we know that $y' \cdot
      x \equiv z$. We must show that $(a \cdot x)^* \equiv 1 + a \cdot
      z$. We compute:
      \[ \begin{array}{rl@{\qquad}r}
               & (a \cdot x)^* & \\
        \equiv & 1 + a \cdot x \cdot (a \cdot x)^* & \byax{KA-Unroll-L} \\
        \equiv & 1 + a \cdot (x \cdot a)^* \cdot x & \by{sliding with $p = x$ and $q = a$; Lemma~\ref{lem:katsliding}} \\
        \equiv & 1 + a \cdot y^* \cdot x & \by{IH (\ref{pbt})} \\
        \equiv & 1 + a \cdot y' \cdot x & \by{IH (\ref{pbstar})} \\
        \equiv & 1 + a \cdot z & \by{IH (\ref{pbj})}
      \end{array} \]


    \item[(\rn{Expand})] We are trying to pushback the minimal term
      $a$ of $x$ through a star, i.e., we have $(a \cdot x)^*$; by the
      IH on (\ref{pbt}), we know that there exist $t$ and $u$ such
      that $x \cdot a \equiv a \cdot t + u$; by the IH on
      (\ref{pbstar}), we know that there exists a $y$ such that $(t +
      u)^* \equiv y$; and by the IH on (\ref{pbj}), we know that there
      is some $z$ such that $y \cdot x \equiv z$.  We compute:
      \[ \begin{array}{rl@{\qquad}r}
               & (a \cdot x)^* & \\
        \equiv & 1 + a \cdot x + a \cdot x \cdot a \cdot x \cdot (a \cdot x)^* & \byax{KA-Unroll-L} \\
        \equiv & 1 + a \cdot x + a \cdot x \cdot a \cdot (x \cdot a)^* \cdot x & \\
        \multicolumn{3}{r}{\by{sliding with $p = x$ and $q = a$; Lemma~\ref{lem:katsliding}}} \\
        \equiv & 1 + a \cdot x + a \cdot \left[ x \cdot a \cdot (x \cdot a)^* \right] \cdot x & \byax{KA-Seq-Assoc} \\
        \equiv & 1 + a \cdot x + a \cdot \left[ (a \cdot t + u) \cdot (t + u)^* \right] \cdot x & \\
        \multicolumn{3}{r}{\by{expansion using IH (\ref{pbt}); Lemma~\ref{lem:katexpand}}} \\
        \equiv & 1 + a \cdot x + a \cdot (a \cdot t + u) \cdot (t + u)^* \cdot x & \byax{KA-Seq-Assoc} \\
        \equiv & 1 + a \cdot x + (a \cdot a \cdot t + a \cdot u) \cdot (t + u)^* \cdot x & \byax{KA-Dist-L} \\
        \equiv & 1 + a \cdot x + (a  \cdot t + a \cdot u) \cdot (t + u)^* \cdot x & \byax{BA-Seq-Idem} \\
        \equiv & 1 + a \cdot x + a \cdot (t + u) \cdot (t + u)^* \cdot x & \byax{BA-Seq-Idem} \\
        \equiv & 1 + a \cdot 1 \cdot x + a \cdot (t + u) \cdot (t + u)^* \cdot x & \byax{KA-One-Seq} \\
        \equiv & 1 + (a \cdot 1 + a \cdot (t + u) \cdot (t + u)^*) \cdot x & \byax{KA-Dist-R} \\
        \equiv & 1 + a \cdot (1 + (t + u) \cdot (t + u)^*) \cdot x & \byax{KA-Dist-L} \\
        \equiv & 1 + a \cdot (t + u)^* \cdot x & \byax{KA-Unroll-L} \\
        \equiv & 1 + a \cdot y \cdot x & \by{IH (\ref{pbstar})} \\
        \equiv & 1 + a \cdot z & \by{IH (\ref{pbj})}
      \end{array} \]


    \item[(\rn{Denest})] We have a compound normal form $a \cdot x +
      y$ under a star; we will push back the maximal test $a$. By our
      first IH on (\ref{pbstar}) we know that that $y^* \equiv y'$ for
      some $y'$; by our first IH on (\ref{pbj}), we know that $x \cdot
      y' \equiv x'$ for some $x'$; by our second IH on (\ref{pbstar}),
      we know that $(a \cdot x')^* \equiv z$ for some $z$; and by our
      second IH on (\ref{pbj}), we know that $y' \cdot z \equiv z'$
      for some $z'$. We must show that $(a \cdot x + y)^* \equiv z'$.
      We compute:
      \[ \begin{array}{rl@{\qquad}r}
               & (a \cdot x + y)^* & \\
        \equiv & y^* \cdot (a \cdot x \cdot y^*)^* & \by{denesting with $p = a \cdot x$ and $q = y$; Lemma~\ref{lem:katdenesting}} \\
        \equiv & y' \cdot (a \cdot x \cdot y')^* & \by{first IH (\ref{pbstar})} \\
        \equiv & y' \cdot (a \cdot x')^* & \by{first IH (\ref{pbj})} \\
        \equiv & y' \cdot z & \by{second IH (\ref{pbstar})} \\
        \equiv & z' & \by{second IH (\ref{pbj})}
      \end{array} \]        
      

    \end{proofcases}

    \paragraph*{Pushing tests through actions ($m \cdot a \PBdot y$)}
    \begin{proofcases}[leftmargin=2.5cm]
    \item[(\rn{SeqZero})] We are pushing $0$ back through a restricted
      action $m$. We immediately find $m \cdot 0 \equiv 0$ by
      \ax{KA-Seq-Zero}.

    \item[(\rn{SeqOne})] We are pushing $1$ back through a restricted
      action $m$. We find:
      \[ \begin{array}{rl@{\qquad}r}
               & m \cdot 1 & \\
        \equiv & m & \byax{KA-One-Seq} \\
        \equiv & 1 \cdot m & \byax{KA-Seq-One}
      \end{array} \]

    \item[(\rn{SeqSeqTest})] We are pushing the tests $a \cdot b$
      through the restricted action $m$. By our first IH on
      (\ref{pbdot}), we have $m \cdot a \equiv y$; by our second IH on
      (\ref{pbdot}), we have $y \cdot b \equiv z$. We compute:
      \[ \begin{array}{rl@{\qquad}r}
               & m \cdot (a \cdot b) & \\
        \equiv & m \cdot a \cdot b & \byax{KA-Seq-Assoc} \\
        \equiv & y \cdot b & \by{first IH (\ref{pbdot})} \\
        \equiv & z & \by{second IH (\ref{pbdot})}
      \end{array} \]
      
    \item[(\rn{SeqSeqAction})] We are pushing the test $a$ through the
      restricted actions $m \cdot n$. By our IH on (\ref{pbdot}), we have
      $n \cdot a \equiv x$; by our IH on (\ref{pbr}), we have $m \cdot
      x \equiv y$.  We compute:
      \[ \begin{array}{rl@{\qquad}r}
               & (m \cdot n) \cdot a & \\
        \equiv & m \cdot (n \cdot a) & \byax{KA-Seq-Assoc} \\
        \equiv & m \cdot x & \by{IH (\ref{pbdot})} \\
        \equiv & y & \by{IH (\ref{pbr})}
      \end{array} \]
      
    \item[(\rn{SeqParTest})] We are pushing the tests $a + b$ through
      the restricted action $m$. By our first IH on (\ref{pbdot}), we
      have $m \cdot a \equiv x$; by our second IH on (\ref{pbdot}), we
      have $m \cdot b \equiv y$. We compute:
      \[ \begin{array}{rl@{\qquad}r}
               & m \cdot (a + b) & \\
               & m \cdot a + m \cdot b & \byax{KA-Dist-L} \\
        \equiv & x + m \cdot b & \by{first IH (\ref{pbdot})} \\
        \equiv & x + y & \by{second IH (\ref{pbdot})}
      \end{array} \]
      
    \item[(\rn{SeqParAction})] We are pushing the test $a$ through the
      restricted actions $m + n$. By our first IH on (\ref{pbdot}), we
      have $m \cdot a \equiv x$; by our second IH on (\ref{pbdot}), we
      have $n \cdot a \equiv y$. We compute:
      \[ \begin{array}{rl@{\qquad}r}
               & (m + n) \cdot a & \\
               & m \cdot a + n \cdot a & \byax{KA-Dist-R} \\
        \equiv & x + n \cdot a & \by{first IH (\ref{pbdot})} \\
        \equiv & x + y & \by{second IH (\ref{pbdot})}
      \end{array} \]

    \item[(\rn{Prim})] We are pushing a primitive predicate $\alpha$
      through a primitive action $\pi$.  We have, by assumption, that
      $\pi \cdot a \WP \{ a_1, \dots, a_k \}$.  By
      definition of the \WP relation, it must be the case that $\pi
      \cdot \alpha \equiv \sum_{i=1}^k a_i \cdot \pi$

    \item[(\rn{PrimNeg})] We are pushing a negated predicate $\neg a$
      back through a primitive action $\pi$. We have, by assumption,
      that $\pi \cdot a \PBdot \sum_i a_i \cdot pi$ and that
      $\nnf(\neg(\sum_i a_i)) = b$, so $\neg(\sum_i a_i) \equiv b$
      (Lemma~\ref{lem:nnfsound}).
      By the IH, we know that $\pi \cdot a \equiv \sum_i a_i \cdot
      \pi$; we must show that $\pi \cdot \neg a \equiv b \cdot \pi$.
      By our assumptions, we know that $b \cdot \pi \equiv \neg(\sum_i
      a_i) \cdot \pi$, so by pushback negation (\ax{Pushback-Neg}/Lemma~\ref{lem:pushbackneg}).

    \item[(\rn{SeqStarSmaller})] We are pushing the test $a$ through
      the restricted action $m^*$. By our IH on (\ref{pbdot}), we have $m
      \cdot a \equiv x$ for some $x$; by our IH on (\ref{pbr}), we
      have $m^* \cdot x \equiv y$ for some $y$.  We compute:
      \[ \begin{array}{rl@{\qquad}r} 
               & m^* \cdot a & \\
        \equiv & (1 + m^* \cdot m) \cdot a & \byax{KA-Unroll-R} \\
        \equiv & a + m^* \cdot m \cdot a & \byax{KA-Dist-R} \\
        \equiv & a + m^* \cdot (m \cdot a) & \byax{KA-Seq-Assoc} \\
        \equiv & a + m^* \cdot x & \by{IH (\ref{pbdot})} \\
        \equiv & a + y & \by{IH (\ref{pbr})}
      \end{array} \]

    \item[(\rn{SeqStarInv})] We are pushing the test $a$ through the
      restricted action $m^*$. By our IH on (\ref{pbdot}), there exist
      $t$ and $u$ such that $m \cdot a \equiv a \cdot t + u$; by our
      IH on (\ref{pbr}), there exists an $x$ such that $m^* \cdot u
      \equiv x$; by our IH on (\ref{pbstar}), there exists a $y$ such
      that $u^* \equiv y$; and by our IH on (\ref{pbj}), there exists
      a $z$ such that $x \cdot y \equiv z$. We compute:

      m . a = a . t + u
      m* a = (a + m* . u) + t*
      \[ \begin{array}{rl@{\qquad}r} 
               & m^* \cdot a & \\
        \equiv & (a + m^* \cdot u) \cdot t^* & \by{star invariant on IH (\ref{pbdot}); Lemma~\ref{lem:katstarinv}} \\
        \equiv & a \cdot t^* + m^* \cdot u \cdot t^* & \byax{KA-Dist-R} \\
        \equiv & a \cdot t^* + x \cdot t^* & \by{IH (\ref{pbr})} \\
        \equiv & a \cdot y + x \cdot y & \by{IH (\ref{pbstar})} \\
        \equiv & a \cdot y + z & \by{IH (\ref{pbj})}
      \end{array} \]

    \end{proofcases}

    \paragraph*{Pushing normal forms through actions ($m \cdot x \PBR z$)}
    \begin{proofcases}
    \item[(\rn{Restricted})] We have $x = \sum_{i=1}^k a_i \cdot
      n_i$. By the IH on (\ref{pbdot}), $m \cdot a_i \PBdot y_i$. We
      compute:
      \[ \begin{array}{rl@{\qquad}r}
               & m \cdot x & \\
        \equiv & m \cdot \sum_{i=1}^k a_i \cdot n_i & \\
        \equiv & \sum_{i=1}^k m \cdot a_i \cdot n_i & \byax{KA-Dist-L} \\
        \equiv & \sum_{i=1}^k y_i \cdot n_i & \by{IH (\ref{pbdot})}
      \end{array} \]
    \end{proofcases}

    \paragraph*{Pushing tests through normal forms ($x \cdot a \PBT y$)}
    \begin{proofcases}
    \item[(\rn{Test})] We have $x = \sum_{i=1}^k a_i \cdot m_i$. By
      the IH on (\ref{pbdot}), we have $m_i \cdot a \PBdot y_i$
      where $y_i = \sum_{j=1}^l b_{ij} \cdot m_{ij}$.
      We compute:
      \[ \begin{array}{rl@{\qquad}r}
               & x \cdot a & \\
        \equiv & \left[ \sum_{i=1}^k a_i \cdot m_i \right] \cdot a & \\
        \equiv & \sum_{i=1}^k a_i \cdot m_i \cdot a & \byax{KA-Dist-R} \\
        \equiv & \sum_{i=1}^k a_i \cdot (m_i \cdot a) & \byax{KA-Seq-Assoc} \\
        \equiv & \sum_{i=1}^k a_i \cdot y_i & \by{IH (\ref{pbdot})} \\
        \equiv & \sum_{i=1}^k a_i \cdot \sum_{j=1}^l b_{ij} \cdot m_{ij} & \\
        \equiv & \sum_{i=1}^k \sum_{j=1}^l a_i \cdot b_{ij} \cdot m_{ij} & \byax{KA-Dist-L}
      \end{array} \]
    \qedhere

    \end{proofcases}
    \fi}
  \end{proof}
\end{theorem}

\begin{theorem}[Pushback existence]
  \label{fullthm:pushbackexist}
  {\iffull
  For all $x$ and $m$ and $a$:
  \begin{enumerate}
  \item \label{exj} For all $y$ and $z$, if $x \preceq z$ and $y
    \preceq z$ then there exists some $z' \preceq z$ such that $x
    \cdot y \PBJ z'$.
  \item \label{exstar} There exists a $y \preceq x$ such
    that $x^* \PBstar y$.
  \item \label{exdot} There exists some $y \preceq
    a$ such that $m \cdot a \PBdot y$.
  \item \label{exr} There exists a $y \preceq x$
    such that $m \cdot x \PBR y$.
  \item \label{ext} If $x \preceq z$ and $a \preceq z$ then there
    exists a $y \preceq z$ such that $x \cdot a \PBT y$.
  \end{enumerate}
  \else Each \PB relations' left relates to some right that is no larger than the left's parts, e.g., for all $x$ there exists $y \preceq
  x$ such that $x^* \PBstar y$. \fi}
  \begin{proof}
    By induction on the lexicographical order of: 
    the subterm ordering ($\prec$); 
    the size of $x$\iffull\ (for (\ref{exj}), (\ref{exstar}), (\ref{exr}), and (\ref{ext}))\fi; 
    the size of $m$\iffull\ (for (\ref{exdot}) and (\ref{exr}))\else\ (for \PBdot and \PBR)\fi; and
    the size of $a$\iffull\ (for (\ref{exdot}))\else\ (for \PBdot and \PBT)\fi.
    {\iffull\else Cases first split
      (Lemma~\ref{lem:splitting}) to show that derivations exist;
      subterm ordering congruence finds
      orderings to apply the IH. \qedhere \fi}

    {\iffull
    \paragraph*{Sequential composition of normal forms ($x \cdot y \PBJ z$)}

    We have $x = \sum_{i=1}^k a_i \cdot m_i$ and $y = \sum_{j=1}^l b_j
    \cdot n_j$; by the IH on (\ref{exdot}) with the size decreasing on
    $m_i$, we know that $m_i \cdot b_j \PBdot x_{ij}$ for each $i$ and
    $j$ such that $x_{ij} \preceq a_i$, so by \rn{Join}, we know
    that $x \cdot y \PBJ \sum_{i=1}^k \sum_{j=1}^l a_i x_{ij} n_j =
    z'$.

    Given that $x, y \preceq z$, it remains to be seen that $z'
    \preceq z$. We've assumed that $a_i \preceq x \preceq z$. By our
    IH on (\ref{exdot}) we found earlier that $x_{ij} \preceq a_i
    \preceq z$. Therefore, by unpacking $x$ and applying test bounding
    (Lemma~\ref{lem:nforder}), $a_i \cdot x_{ij} \cdot n_j \preceq
    z$. By normal form parallel congruence (Lemma~\ref{lem:nforder}),
    we have $z' \preceq z$.)

    \paragraph*{Kleene star of normal forms ($x^* \PBJ y$)}

    If $x$ is vacuous, we find that $0^* \PBstar 1$ by \rn{StarZero},
    with $1 \preceq 0$ since they have the same maximal terms (just
    $1$).

    If $x$ isn't vacuous, then we have $x \equiv a \cdot x_1 + x_2$
    where $x_1, x_2 \prec x$ and $a \in \mt(x)$ by splitting
    (Lemma~\ref{lem:splitting}). We first consider whether $x_2$ is
    vacuous.

    \begin{proofcases}[leftmargin=2.5cm]
    \item[($x_2$ is vacuous)] We have $x \equiv a \cdot x_1 + 0 \equiv
      a \cdot x_1$. 

      By our IH on (\ref{ext}) with $x_1$ decreasing in size, we
      have $x_1 \cdot a \PBT w$ where $w \preceq x$ (because $x_1
      \prec x$ and $a \preceq x$). By maximal test inequality
      (Lemma~\ref{lem:mtprec}), we have two cases: either $a \in
      \mt(w)$ or $w \prec a \preceq x$.

      \begin{proofcases}
      \item[($a \in \mt(w)$)] By splitting
        (Lemma~\ref{lem:splitting}), we have $w \equiv a \cdot t + u$
        for some normal forms $t, u \prec w$.

        By normal-form parallel congruence (Lemma~\ref{lem:nforder}),
        $t + u \prec x$; so by the IH on (\ref{exstar}) with our
        subterm ordering decreasing on $t + u \prec x$, we find that
        $(t + u)^* \PBstar w'$ for some $w' \preceq (t + u)^* \prec w
        \preceq x$. Since $w' \prec x$, we can apply our IH on
        (\ref{exj}) with our subterm ordering decreasing on $w' \prec
        x$ to find that $w' \cdot x_1 \PBJ z$ such that $z \preceq x_1
        \prec x$ (since $w' \preceq x$ and $x_1 \prec x$).
        
        Finally, we can see by \rn{Expand} that $x = (a \cdot x_1)^*
        \PBstar 1 + a \cdot z = y$. Since each $1, a, z \preceq x$, we
        have $y = 1 + a \cdot z \preceq x$ as needed.


      \item[($w \prec a$)] Since $w \prec a$, we can apply our IH on
        (\ref{pbstar}) with our subterm order decreasing on $w \prec x$ to
        find that $w^* \PBstar w'$ such that $w' \preceq w \prec a
        \preceq x$. By our IH on (\ref{pbj}) with our subterm order
        decreasing on $w' \prec x$ to find that $w' \cdot x_1 \PBJ z$ where $z
        \preceq x$ (because $w' \preceq x$ and $x_1 \prec x$).

        We can now see by \rn{Slide} that $x = (a \cdot x_1)^* \PBstar 1
        + a \cdot z = y$. Since each $1, a, z \preceq x$, we have $y =
        1 + a \cdot z \preceq x$ as needed.


      \end{proofcases}
      
    \item[($x_2$ isn't vacuous)] We have $x \equiv a \cdot x_1 + x_2$
      where $x_i \prec x$ and $a \in mt(x)$. Since $x_2$ isn't
      vacuous, we must have $a \prec x$, not just $a \preceq x$.

      By the IH on (\ref{exstar}) with the subterm ordering decreasing
      on $x_2 \prec x$, we find $x_2 \PBstar w$ such that $w \preceq
      x_2$.  By the IH on (\ref{exj}) with the subterm ordering
      decreasing on $x_1 \prec x$, we have $x_1 \cdot w \PBJ v$ where
      $v \preceq x$ (because $x_1 \preceq x$ and $w \preceq x$). By
      the IH on (\ref{exstar}) with the subterm ordering decreasing on
      $a \cdot v \prec x$, we find $(a \cdot v)^* \PBstar z$ where $z
      \preceq a \cdot v \prec x$. By our IH on (\ref{exj}) with the
      subterm ordering decreasing on $w \prec x$, we find $w \cdot
      z \PBJ y$ where $y \prec x$ (because $w \prec x$ and $z \prec
      x$).

      By \rn{Denest}, we can see that $x \equiv (a \cdot x_1 + x_2)^*
      \PBstar y$, and we've already found that $y \preceq x$ as needed.
      

    \end{proofcases}

    \paragraph*{Pushing tests through actions ($m \cdot a \PBdot y$)}

    We go by cases on $a$ and $m$ to find the $y \preceq a$ such that
    $m \cdot a \PBdot y$.

    \begin{proofcases}

    \item[$(m,0)$] We have $m \cdot 0 \PBdot 0$ by \rn{SeqZero}, and
      $0 \preceq 0$ immediately.

    \item[$(m,1)$] We have $m \cdot 1 \PBdot 1 \cdot m$ by \rn{SeqOne}
      and $1 \preceq 1$ immediately.

    \item[$(m,a \cdot b)$] By the IH on (\ref{exdot}) decreasing in
      size on $a$, we know that $m \cdot a \PBdot x$ where $x \preceq a
      \preceq a \cdot b$. By the IH on (\ref{ext}) decreasing in size
      on $b$, we know that $x \cdot b \PBT y$. Finally, we know by
      \rn{SeqSeqTest} that $m \cdot (a \cdot b) \PBdot y$. Since $x
      \preceq a \cdot b$ and $b \preceq a \cdot b$, we know by the IH
      on (\ref{ext}) earlier that $y \preceq a \cdot b$.

    \item[$(m,a + b)$] By the IH on (\ref{exdot}) decreasing in size
      on $a$, we know that $m \cdot a \PBdot x$ such that $x \preceq a
      \preceq a + b$. Similarly, by the IH on (\ref{exdot}) decreasing
      in size on $b$, we know that $m \cdot b \PBdot z$ such that $z
      \preceq b \preceq a + b$. By \rn{SeqParTest}, we know that $m
      \cdot (a + b) \PBdot x + z = y$; by normal form parallel
      congruence, we know that $y = x + z \preceq a + b$ as needed.

    \item[$(m \cdot n,a)$] By the IH on (\ref{exdot}) decreasing in
      size on $n$, we know that $n \cdot a \PBdot x$ such that $x
      \preceq a$. By the IH on (\ref{exr}) decreasing in size on $m$,
      we know that $m \cdot x \PBR y$ such that $y \preceq x \preceq
      a$ (which are the size bounds on $y$ we needed to show). All
      that remains to be seen is that $(m \cdot n) \cdot a \PBdot y$,
      which we have by \rn{SeqSeqAction}.

    \item[$(m + n,a)$] By the IH on (\ref{exdot}) decreasing in size
      on $m$, we know that $m \cdot a \PBdot x$. Similarly, by the IH
      on (\ref{exdot}) decreasing in size on $n$, we know that $n
      \cdot a \PBdot z$. By \rn{SeqParAction}, we know that $(m + n)
      \cdot a \PBdot x + z = y$. Furthermore, both IHs let us know
      that $x, z \preceq a$, so by normal form parallel congruence, we
      know that $y = x + z \preceq a$.

    \item[$(\pi,\neg a)$] By the IH on (\ref{exdot}) decreasing in
      size on $a$, we can find that $\pi \cdot a \PBdot \sum_i a_i
      \cdot \pi$ where $\sum_i a_i \preceq a$, and $\nnf(\neg(\sum_i
      a_i)) = b$ for some term $b$. It remains to be seen that $b
      \preceq \neg a$, which we have by monotonicity of $\nnf$
      (Lemma~\ref{lem:nnfmonotonic}).

    \item[$(\pi,\alpha)$] In this case, we fall back on the client
      theory's pushback operation (Definition~\ref{def:wp}). We
      have $\pi \cdot \alpha \WP \{ a_1, \dots, a_k \}$
      such that $a_i \preceq \alpha$. By \rn{Prim}, we have $\pi \cdot
      \alpha \PBdot \sum_{i=1}^k a_i \cdot \pi = y$; since each $a_i
      \preceq \alpha$, we find $y \preceq \alpha$ by the monotonicity
      of union (Lemma~\ref{lem:mtuniondist}).

    \item[$(m^*,a)$] We've already ruled out the case where $a = b
      \cdot c$, so it must be the case that $\seqs(a) = \{ a \}$, so
      $\mt(a) = \{ a \}$.

      By the IH on (\ref{exdot}) decreasing in size on $m$, we know that $m
      \cdot a \PBdot x$ such that $x \preceq a$. There are now two
      possibilities: either $x \prec a$ or $a \in \mt(x) = \{ a \}$.
      
      \begin{proofcases}[leftmargin=1cm]

      \item[($x \prec a$)] By the IH on (\ref{exr}) with $x \prec a$,
        we know by \rn{SeqStarSmaller} that $m^* \cdot x \PBR y$ such
        that $y \preceq x \prec a$.

      \item[($a \in \mt(x)$)] By splitting
        (Lemma~\ref{lem:splitting}), we have $x \equiv a \cdot t + u$,
        where $t$ and $u$ are normal forms such that $t, u \prec x
        \preceq a$.

        By the IH on (\ref{exr}) with $t \prec a$, we know that $m^*
        \cdot t \PBR w$ such that $w \preceq t \prec x \preceq a$. By
        the IH on (\ref{exstar}) with $u \prec x \preceq a$, we know
        that $u^* \PBstar z$ such that $z \preceq u \prec x \preceq a$.
        By the IH on (\ref{exj}) with $w \prec a$ and $z \prec a$, we
        find that $w \cdot z \PBJ v$ such that $v \preceq w \prec a$.

        Finally we have our $y$: by \rn{SeqStarInv}, we have $m^*
        \cdot a \PBdot a \cdot z + v = y$. Since $z \preceq a$ and $a
        \preceq a$, we have $a \cdot z \preceq a$ (mixed sequence
        congruence; Lemma~\ref{lem:nforder}) and $v \prec a$. By normal
        form parallel congruence, we have $a \cdot z + v \preceq a$
        (Lemma~\ref{lem:nforder}).

      \end{proofcases}

    \end{proofcases}

    \paragraph*{Pushing normal forms through actions ($m \cdot x \PBR z$)}

    We have $x = \sum_{i=1}^k a_i \cdot n_i$; by the IH on
    (\ref{exdot}) with the size decreasing on $n_i$, we know that $m
    \cdot a_i \PBdot x_i$ for each $i$ such that $x_i \preceq a_i$, so
    by \rn{Restricted}, we know that $m \cdot x \PBR \sum_{i=1}^k
    x_i n_i = y$.

    We must show that $y \preceq x$. By our IH on (\ref{exdot}) we
    found earlier that $x_i \preceq a_i$. By normal form parallel congruence
    (Lemma~\ref{lem:nforder}), we have $y \preceq x$.

    \paragraph*{Pushing tests through normal forms ($x \cdot a \PBT y$)}

    We have $x = \sum_{i=1}^k a_i \cdot m_i$; by the IH on
    (\ref{exdot}) with the size decreasing on $m_i$, we know that $m_i
    \cdot a \PBdot y_i = \sum_{j=1}^l b_{ij} m_{ij}$ where $y_i
    \preceq a$. Therefore, we know that $x \cdot a \PBT \sum_{i=1}^k
    \sum_{j=1}^l a_i \cdot b_{ij} \cdot m_{ij} = y$ by \rn{Test}.

    Given that $x \preceq z$ and $a \preceq z$, We must show that $y
    \preceq z$. We already know that $a_i \preceq x \preceq z$, and we
    found from the IH on (\ref{exdot}) earlier that $b_{ij} \preceq
    y_i \preceq a \preceq z$. By test bounding
    (Lemma~\ref{lem:nforder}), we have $a_i \cdot b_{ij} \preceq z$,
    and therefore $y \preceq z$ by normal form parallel congruence
    (Lemma~\ref{lem:nforder}). \qedhere
 
    \fi}
  \end{proof}
\end{theorem}

\begin{corollary}[Normal forms]
  \label{fullcor:nf}
  For all $p \in \THYkat$, there exists a normal form $x$ such that
  $p \norm x$ and that $p \equiv x$.
  \begin{proof}
    By induction on $p$\iffull\else, using Theorems~\ref{thm:pushbackexist}
    and~\ref{thm:pushbacksound} in the \rn{Seq} and \rn{Star} case\fi.
    \begin{proofcases}
    \item[(\rn{Pred})] We have $a \equiv a$ immediately.
    \item[(\rn{Act})] We have $\pi \equiv 1 \cdot \pi$ by
      \ax{KA-Seq-One}.
    \item[(\rn{Par})] By the IHs and congruence.
    \item[(\rn{Seq})] We have $p = q \cdot r$; by the IHs, we
      know that $q \norm x$ and $r \norm y$. By pushback existence
      (Theorem~\ref{thm:pushbackexist}), we know that $x \cdot y \PBJ
      z$ for some $z$. By pushback soundness
      (Theorem~\ref{thm:pushbacksound}), we know that $x \cdot y
      \equiv z$. By congruence, $p \equiv z$.
    \item[(\rn{Star})] We have $p = q^*$. By the IH, we know
      that $q \norm x$. By pushback existence
      (Theorem~\ref{thm:pushbackexist}), we know that $x^* \PBstar
      y$. By pushback soundness (Theorem~\ref{thm:pushbacksound}), we
      know that $x^* \equiv y$. \qedhere
    \end{proofcases}
  \end{proof}  
\end{corollary}

\section{Completeness proofs} 

\begin{lemma}[Pushback negation (\ax{Pushback-Neg})]
  \label{lem:pushbackneg}
  $p \cdot a \equiv b \cdot p$ iff $p \cdot \neg a \equiv \neg b \cdot p$.
  \begin{proof}
    We show that both sides $p \cdot \neg a$ and $\neg b \cdot p$
    are equivalent to $\neg b \cdot p \cdot \neg a$ by way of \ax{BA-Excl-Mid}:
    \[ \begin{array}{rl@{\quad}r}
               p \cdot \neg a
      \equiv & (b + \neg b) \cdot p \cdot \neg a & \by{\ax{KA-Seq-One}, \ax{BA-Excl-Mid}} \\
      \equiv & b \cdot p \cdot \neg a + \neg b \cdot p \cdot \neg a & \byax{KA-Dist-L} \\
      \equiv & p \cdot a \cdot \neg a + \neg b \cdot p \cdot \neg a & \by{assumption} \\
      \equiv & p \cdot 0 + \neg b \cdot p \cdot \neg a & \byax{BA-Contra} \\
      \equiv & \neg b \cdot p \cdot \neg a & \by{\ax{KA-Plus-Comm}, \ax{KA-Plus-Zero}} \\
      \equiv & 0 \cdot p + \neg b \cdot p \cdot \neg a & \byax{BA-Contra} \\
      \equiv & \neg b \cdot b \cdot p + \neg b \cdot p \cdot \neg a & \by{assumption} \\
      \equiv & \neg b \cdot p \cdot a + \neg b \cdot p \cdot \neg a & \byax{KA-Dist-R} \\
      \equiv & \neg b \cdot p \cdot (a + \neg a) & \by{\ax{KA-One-Seq}, \ax{BA-Excl-Mid}} \\
      \equiv & \neg b \cdot p & \qedhere \\
    \end{array} \] 

    The other direction of the proof is symmetric, with the two terms meeting at $b \cdot p \cdot a$.
    \[ \begin{array}{rl@{\quad}r}
               p \cdot a
      \equiv & (b + \neg b) \cdot p \cdot a & \by{\ax{KA-Seq-One}, \ax{BA-Excl-Mid}} \\
      \equiv & b \cdot p \cdot a + \neg b \cdot p              \cdot a & \byax{KA-Dist-L} \\
      \equiv & b \cdot p \cdot a + \neg b \cdot p \cdot \neg a \cdot a & \by{assumption} \\
      \equiv & b \cdot p \cdot a + \neg b \cdot p \cdot 0 & \byax{BA-Contra} \\
      \equiv & b \cdot p \cdot a & \by{\ax{KA-Plus-Comm}, \ax{KA-Plus-Zero}} \\
      \equiv & b \cdot p \cdot a + 0 \cdot p \cdot \neg a & \byax{BA-Contra} \\
      \equiv & b \cdot p \cdot a + b \cdot \neg b \cdot p \cdot \neg a & \by{assumption} \\
      \equiv & b \cdot p \cdot a + b \cdot              p \cdot \neg a & \byax{KA-Dist-R} \\
      \equiv & b \cdot p \cdot (a + \neg a) & \by{\ax{KA-One-Seq}, \ax{BA-Excl-Mid}} \\
      \equiv & b \cdot p & \qedhere \\
    \end{array} \] 
\end{proof}
\end{lemma}

\begin{figure}[t]\small
\iffull\sidebyside[.55][.44][t]\fi
  {\[ \begin{array}{rcl}
    \R            &:& \RA \rightarrow \mathcal{P}(\Pi_\THY^*) \\
    \R(1)         &=& \{ \epsilon \} \\
    \R(\pi)       &=& \{ \pi \} \\
    \R(m + n)     &=& \R(m) \cup \R(n) \\
    \R(m \cdot n) &=& \{ uv \mid u \in \R(m), v \in R(n) \} \\
    \R(m^*)       &=& \bigcup_{0 \le i} \R(m)^i \\
  \end{array} \]}
  {\[ \begin{array}{rcl}
    \lbl &:& \mathsf{Trace} \rightarrow \Pi_\THY^* \\
    \lbl(\langle \sigma, \bot \rangle) &=& \epsilon \\
    \lbl(t \langle \sigma, \pi \rangle) &=& \lbl(t) \pi \\
    && \\
    \mathcal{L}^0 &=& \{ \epsilon \} \\
    \mathcal{L}^{n+1} &=& \{ uv \mid u \in \mathcal{L}, v \in \mathcal{L}^n \} \\
  \end{array} \]}

\vspace*{-1em}
  \caption{Regular interpretation of restricted actions}
  \label{fig:regularaction}
\end{figure}

\begin{lemma}[Restricted actions are ahistorical]
  \label{lem:restrictedcontext}
  If $\denot{m}(t_1) = t_1,t$ and $\last(t_1) = \last(t_2)$ then $\denot{m}(t_2) = t_2,t$.
  \begin{proof}
    By induction on $m$.
    {\iffull
    \begin{proofcases}
    \item[($m=1$)] Immediate, since $t$ is empty.
    \item[($m=\pi$)] We immediately have $t = \langle \last(t_1), \pi \rangle$.
    \item[($m=m + n$)] We have $\denot{m + n}(t_1) = \denot{m}(t_1)
      \cup \denot{n}(t_1)$ and $\denot{m + n}(t_2) = \denot{m}(t_2)
      \cup \denot{n}(t_2)$. By the IHs.
    \item[($m=m \cdot n$)] We have $\denot{m \cdot n}(t_1) =
      (\denot{m} \bullet \denot{n})(t_1)$ and $\denot{m \cdot n}(t_2)
      = (\denot{m} \bullet \denot{n})(t_2)$. It must be that
      $\denot{m}(t_1) = \{ t_1,t_{mi} \}$, so by the IH we have
      $\denot{m}(t_2) = \{ t_2,t_{mi} \}$. These sets have the same
      last states, so we can apply the IH again for $n$, and we are
      done.
    \item[($m=m^*$)] We have $\denot{m^*}(t_1) = \bigcup_{0 \le i}
      \denot{m}^i(t_1)$. By induction on $i$.
      \begin{proofcases}
      \item[($i=0$)] Immediate, since $\denot{m}^0(t_i) = t_i$ and so $t$ is empty.
      \item[($i=i+1$)] By the IH and the reasoning above for $\cdot$.
      \end{proofcases}
    \end{proofcases}
    \fi}
  \end{proof}
\end{lemma}

\begin{lemma}[Labels are regular]
  \label{lem:regularsound}

  $\{ \lbl(\denot{m}(\langle \sigma, \bot \rangle)) \mid \sigma \in \mathsf{State} \} = \R(m)$
  \begin{proof}
    By induction on the restricted action $m$.
    {\iffull
    \begin{proofcases}
    \item[($m=1$)] We have $\R(1) = \{ \epsilon \}$. For all $\sigma$,
      we find $\denot{1}(\langle \sigma, \bot \rangle) = \{ \langle
      \sigma, \bot \rangle \}$, and $\lbl(\langle \sigma, \bot
      \rangle) = \epsilon$.
    \item[($m=\pi$)] We $\R(\pi) = \{ \pi \}$.  For all $\sigma$, we
      find $\denot{\pi}(\langle \sigma, \bot \rangle) = \{ \langle
      \sigma, \bot \rangle  \langle \act(\pi, \sigma), \pi \rangle
      \}$, and so $\lbl(\langle \sigma, \bot \rangle  \langle
      \act(\pi, \sigma), \pi \rangle) = \pi$.
    \item[($m=m + n$)] We have $\R(m + n) = \R(m) \cup \R(n)$. For all $\sigma$, we have:
      \[ \begin{array}{rl}
          \lbl(\denot{m + n}(\langle \sigma, \bot \rangle)) 
      = & \lbl(\denot{m}(\langle \sigma, \bot \rangle) \cup \denot{n}(\langle \sigma, \bot \rangle)) \\
      = & \lbl(\denot{m}(\langle \sigma, \bot \rangle)) \cup \lbl(\denot{n}(\langle \sigma, \bot \rangle))
      \end{array} \]
      and we are done by the IHs.
    \item[($m=m \cdot n$)] We have $\R(m \cdot n) = \{ uv \mid u \in \R(m), v \in \R(n) \}$. For all $\sigma$, we have:
      \[ \begin{array}{rl}
              \lbl(\denot{m \cdot n}(\langle \sigma, \bot \rangle)) 
          = & \lbl((\denot{m} \bullet \denot{n})(\langle \sigma, \bot \rangle)) \\
          = & \lbl(\bigcup_{t \in \denot{m}(\langle \sigma, \bot \rangle)} \lbl(\denot{n}(t))) \\
          = & \lbl(\bigcup_{t \in \denot{m}(\langle \sigma, \bot \rangle)} \lbl(t \denot{n}(\langle \sigma, \bot \rangle))) \text{\quad by Lemma~\ref{lem:restrictedcontext}} \\
          = & \lbl(\denot{m}(\langle \sigma, \bot \rangle)) \lbl(\denot{n}(\langle \sigma, \bot \rangle))
      \end{array} \]
      and we are done by the IHs.
    \item[($m=m^*$)] We have $\R(m^*) = \bigcup_{0 \le i} \R(m)^i$. For all $\sigma$, we have:
      \[ \begin{array}{rl}
              \lbl(\denot{m^*}(\langle \sigma, \bot \rangle)) 
          = & \lbl(\bigcup_{0 \le i} \denot{m}^i(\langle \sigma, \bot \rangle)) \\
          = & \bigcup_{0 \le i} \lbl(\denot{m}^i(\langle \sigma, \bot \rangle))
      \end{array} \]
      and we are done by the IH.
    \end{proofcases}
    \fi}
  \end{proof}
\end{lemma}

\noindent
Our proof of completeness works by normalizing each side of the
equation, making each side locally unambiguous, making the entire
equation unambiguous, and then using word equality to ensure that
normal forms with equivalent predicates have equivalent actions.
The full proof is in Theorem~\ref{thm:completeness}, in the main body
of the text.

\fi}

\end{document}